\documentclass[letterpaper,english,aps,prb,reprint,amsmath,amssymb,showpacs,onecolumn]{revtex4-1}
\usepackage[T1]{fontenc}
\usepackage[latin9]{inputenc}
\usepackage{color}
\usepackage{babel}
\usepackage{mathtools}
\usepackage{bm}
\usepackage{amsmath}
\usepackage{amssymb}
\usepackage{graphicx}
\usepackage[unicode=true,pdfusetitle,
 bookmarks=true,bookmarksnumbered=true,bookmarksopen=true,bookmarksopenlevel=2,
 breaklinks=false,pdfborder={0 0 1},backref=false,colorlinks=true]
 {hyperref}

\makeatletter

\pdfpageheight\paperheight
\pdfpagewidth\paperwidth

\providecommand{\tabularnewline}{\\}

\usepackage{bm}
\usepackage{graphicx}
\usepackage{braket}
\usepackage{microtype}
\hypersetup{linkcolor = blue,
            urlcolor  = blue,
            citecolor = blue,
            anchorcolor = blue}

\makeatother

\begin{document}

\preprint{0APS/123-QED}

\title{Spin-lattice relaxation of magnetic centers in molecular crystals at low temperature}

\author{Le Tuan Anh Ho}
\email{anh.holetuan@chem.kuleuven.be}

\affiliation{Theory of Nanomaterials Group, Katholieke Universiteit Leuven, Celestijnenlaan 200F, B-3001 Leuven, Belgium}

\affiliation{Institute of Research and Development, Duy Tan University, Da Nang, Viet Nam}

\author{Liviu F. Chibotaru}
\email{liviu.chibotaru@chem.kuleuven.be }

\affiliation{Theory of Nanomaterials Group, Katholieke Universiteit Leuven, Celestijnenlaan 200F, B-3001 Leuven, Belgium}

\date{\today}
\begin{abstract}
We study the spin-phonon relaxation rate of both Kramers and non-Kramers molecular magnets in strongly diluted samples at low temperature. Using the ``rotational'' contribution to the spin-phonon Hamiltonian, universal formulae for the relaxation rate are obtained. Intriguingly, these formulae are all entirely expressed via measurable or \emph{ab initio} computable physical quantities. Moreover, they are also independent of the energy gaps to excited states involved in the relaxation process. These obtained expressions for direct and Raman processes offer an easy way to determine the lowest limit of the spin-phonon relaxation of any spin system based on magnetic properties of the ground doublet only. In addition, some intriguing properties of Raman process are also found. Particularly, Raman process in Kramers system is found dependent on the magnetic field's orientation but independent of its magnitude, meanwhile the same process in non-Kramers system is significantly reduced out of resonance, i.e. for an applied external field. Interestingly, Raman process is demonstrated to vary as $T^{9}$ for both systems. Application of the theory to a recently investigated cobalt(II) complex shows that it can provide a reasonably good description for the relaxation. Based on these findings, a strategy in developing efficient single-molecule magnets by enhancing the mechanical rigidity of the molecular unit is proposed.
\end{abstract}


\maketitle

\section{Introduction}

\global\long\def\chip{\chi^{\prime}}
 \global\long\def\chipp{\chi''}
 \global\long\def\hmt{\mathcal{H}}
 \global\long\def\vt#1{\bm{\mathrm{#1}}}

The spin-lattice interaction plays a major role in a number of application-potential physical phenomena such as magnetic relaxation in single-molecule/single-atom magnets \cite{Gatteschi2006} (SMMs), entanglement of quantum system to the environment \cite{Horodecki2009}, and decoherence of magnetic qubits \cite{Zurek2003}. In particular, for technological applications of single-molecule magnets, a low spin-lattice relaxation rate is a critical factor\cite{Gatteschi2003,Gatteschi2006} and a complete understanding of mechanisms governing this rate is a must. However, since the accurate evaluation of this spin-phonon relaxation rate requires knowledge of corresponding matrix elements of the deformation potential with respect to all phonons in the crystal \cite{VanVleck1940}, a universal form of this rate is hardly achievable.

In the case of spin complexes, such as Mn$_{12}$ac and Fe$_{8}$ \cite{Gatteschi2003}, the magnetic relaxation proceeds via zero-field-split (ZFS) components of the ground state spin $S$. Due to quenched orbital momentum in the ground state of these complexes the energy separations between the ZFS components of the $S$-term do not exceed 10$\div$20 cm$^{-1}$, i.e., they are in the range of frequencies of acoustic phonons even in molecular crystals. The latter are well described by the Debye model resulting in expressions for spin-phonon relaxation rates \cite{Abragam1970} which have been widely employed for the description of magnetization relaxation in $S$-complexes \cite{Villain1994,Hartmann-Boutron1996a,Garanin1997,Leuenberger2000}. Besides the limitation to the acoustic phonons, the description of spin-phonon interaction requires knowledge of derivatives with respect to deformations of two parameters defining the ZFS of a spin manifold \footnote{Besides modifying the ZFS parameters, the deformation also induces the rotation of the main anisotropy axes of the ZFS tensors described by three parameters (Euler angles) \cite{Chibotaru2013} }: 
\begin{equation}
H_{ZFS}=DS_{z}^{2}+E(S_{x}^{2}-S_{y}^{2}),\label{ZFS-S}
\end{equation}
where $S_{\alpha}$, $\alpha=x,y,z$, are spin operators acting on the spin states of the ground $S$ term. Often only the contribution of the parameter of axial anisotropy ($D$) is considered while the contribution of the parameter of rhombic anisotropy ($E$) is usually neglected due to its smallness in SMMs \cite{Garanin1997,Leuenberger2000}. On the same reason are neglected also the quartic terms in the ZFS Hamiltonian \eqref{ZFS-S} and their contribution to the spin-phonon interaction.

Despite this great simplification, the description of spin-phonon relaxation in SMM $S$-complexes still requires the knowledge of derivative of the parameter $D$ and $E$ and directions of the anisotropy axes \cite{Chibotaru2013} with respect to all phonons. In connection with this problem, Chudnovsky \textit{et. al.} \cite{Chudnovsky2005} made an important observation that given the strong dependence of the transition rate on the velocity of sound of acoustic phonons, the most efficient should be the relaxation via the transversal phonon modes. In molecular crystals, the deformations related to the transversal phonons and concomitantly strongly affecting the ZFS interaction \eqref{ZFS-S} are the vibration modes corresponding to small rotational displacements of the undeformed SMM molecule from the equilibrium point. The variation of $H_{\mathrm{ZFS}}$ under these distortions reduces to the rotation of its main anisotropy axes following the rotation of the molecule without modifying the parameters $D$ and $E$. Given the kinematic nature of this ``rotational'' contribution to the spin-phonon interaction, the corresponding expression for the transition rate between two tunneling states is found to be solely defined by the form of the spin Hamiltonian and independent of any spin-phonon phenomenological coupling parameters \cite{Chudnovsky2005}. This remarkable finding allows to evaluate the main contribution to the relaxation rate in $S$-complexes at low temperature without knowledge of the details of spin-phonon interaction. In a general case when other deformations play important role, a calculation of this rotational contribution to the relaxation rate is also meaningful in that it gives the lower boundary for the relaxation rate, which is a crucial insight for the design of efficient SMMs.

In lanthanide-based complexes, which is attracting increasing attention in the last years due to their exceptional magnetization blocking \cite{Layfield2015,Woodruff2013,Ishikawa2003}, the spin-phonon interaction is by far more complex. The reason is that the operator describing the crystal-field (CF) splitting of the ground $J$ atomic-multiplet of the lanthanide ion, 
\begin{equation}
H_{\text{CF}}=\sum_{p=2,4,6}\sum_{q=-p}^{p}B_{pq}O_{p}^{q}(\mathbf{J}),\label{H-CF}
\end{equation}
where $O_{p}^{q}\left(\vt J\right)$ are the extended Stevens operators \cite{Rudowicz1985,Rudowicz2004a,Gatteschi2006}, contains generally 27 parameters instead of five ($D$, $E$, and the directions of anisotropy axes) in the case of $S$-complexes. Taking into account the variations with local distortions of all parameters $B_{pq}$ results in a highly complex form of spin-phonon interaction \cite{Dohm1975}, containing a tremendous number of parameters. On the other hand the arguments given by Chudnovsky \textit{et. al.} \cite{Chudnovsky2005} should be applicable for lanthanide complexes too, for which the derivation of a simple universal expression for spin-phonon transition rate is especially desirable. Additionally, in lanthanides the ground CF doublet is usually well separated from the excited ones. Accordingly, the magnetic relaxation in such complexes takes place preponderantly within the ground doublet in a broad temperature domain and the relaxation reduces to the spin-phonon transition between its two tunnel-split states \cite{Garanin2011}.

Given the general complexity of the spin-phonon interaction and the potential in giving a universal form of the relaxation rate from rotation-only vibration modes, some studies have been done with encouraging results. Specifically, by using the rotational spin-phonon Hamiltonian \cite{Dohm1975,Bonsall1976}, a universal formula of direct transition rate between tunnel-split states in a strong axially anisotropic non-Kramers system described by ZFS in Eq. \eqref{ZFS-S} has been found \cite{Chudnovsky2005}. In the same work, the direct process rate formula for a specific condition $\left[\hmt_{A},S_{z}\right]=0$ was also provided in spite of its questionable applicability. Within the same approach, Raman process was also considered for a specific non-Kramers system with biaxial spin Hamiltonian \cite{Calero2006a}. However, as far as we are aware, both direct and Raman process for arbitrary spin Hamiltonians and type of doublet(Kramers and non-Kramers) have not yet been considered despite their importance for the understanding of magnetic relaxation in a broad class of systems.

In this work, we derive universal expressions for the rate of spin-phonon relaxation at low temperature in Kramers and non-Kramers systems in the presence of applied magnetic field based on the ``rotational'' contribution to the spin-phonon interaction. These include direct process and first- and second-order Raman processes. In order to exclude the effect of hyperfine and dipolar interaction, we restrict the consideration to a strongly diluted systems in which the decoherence is substantially weakened and the system is therefore in the coherent relaxation regime. The article is divided into seven sections. In Section II, spin-phonon relaxation in SMMs at low temperature and the rotational spin-phonon Hamiltonian are introduced; general expressions for the direct and Raman process are also developed. Detailed form of the corresponding relaxation rates for Kramers and non-Kramers system are respectively presented in Section III and IV. Section V is devoted to the comparison between the direct process and Raman processes in different conditions. Section VI shows an example of application of obtained expressions to a real system. A summary of the novel expressions for the relaxation rates and the discussion of their application is given in the last section. 

\section{Spin-phonon relaxation at low temperature and ``rotational'' spin-phonon Hamiltonian\label{sec:Section II}}

The spin-phonon relaxation in a SMM is governed by the Redfield equation which in the interaction picture is\cite{Garanin2011}, 
\begin{equation}
\frac{d\rho_{\alpha\beta}^{\left(I\right)}}{dt}=\sum_{\alpha',\beta'}e^{i\left(\omega_{\alpha\beta}-\omega_{\alpha'\beta'}\right)t}R_{\alpha\beta,\alpha'\beta'}\rho_{\alpha'\beta'}^{\left(I\right)},\label{eq:Redfield equation}
\end{equation}
where $\rho_{\alpha\beta}$ is the density matrix written in the eigenstates of the ZFS of the ground $S$ term or of CF of the atomic $J$ multiplet and $R_{\alpha\beta,\alpha'\beta'}$ are matrix elements of the relaxation Redfield super-operator. Due to the complexity of this equation, some approximations are used, the most known being the secular approximation. It is based on the fact that the time change of the density matrix element due to the relaxation is slow compared to the frequency difference $\left|\omega_{\alpha\beta}-\omega_{\alpha'\beta'}\right|$. As a consequence, the corresponding fast terms on the right-hand side of the equation can be averaged out and ignored. At low temperature, when only the ground doublet eigenstates $\ket{\pm}$ are populated, the off-diagonal density matrix elements $\rho_{+-}$ $\left(\rho_{-+}\right)$ will decay with the rate $\tilde{\Gamma}_{+-}\equiv-R_{+-,+-}=1/2\left(R_{++,--}+R_{--,++}\right)\equiv1/2\left(\Gamma_{+-}+\Gamma_{-+}\right)$ since the other terms on the right-hand side (RHS) of Eq. \eqref{eq:Redfield equation} oscillate rapidly with the frequency $\left|\omega_{+-}\right|$ or 2$\left|\omega_{+-}\right|$ and, therefore, can be neglected within the secular approximation if $\tilde{\Gamma}_{+-}\ll\left|\omega_{+-}\right|$. For the diagonal density matrix elements $\rho_{++}$ $\left(\rho_{--}\right)$, whereas the left-hand side varies with the rate $\Gamma_{\mathrm{relax}}=\Gamma_{+-}+\Gamma_{-+}$ \footnote{The derivation is given in Sec. IV.E2 of Ref. \onlinecite{Garanin2011}}, the terms containing off-diagonal density matrix elements in the RHS oscillate with frequency $\left|\omega_{+-}\right|$ and the condition for ignoring these fast terms turns out to be similar, $\Gamma_{\mathrm{relax}}\ll\left|\omega_{+-}\right|$ , which is thus a joint condition for applying the secular approximation in this regime. This condition seems to be satisfied for most SMMs at sufficiently low temperature. 

Defining the ``low temperature'' in this way, one should have in mind that it also implies that the excited states are sufficiently high that the relaxation proceeds within the ground doublet only. Since all excited doublets are excluded at low temperature, the Orbach relaxation process is  ruled out. As evident from the expression $\Gamma_{\mathrm{relax}}=\Gamma_{+-}+\Gamma_{-+}$, the relaxation of the magnetic system now comes only from the direct and Raman process between two eigenstates of the ground doublet, i.e. between its two tunnel-split states. This is opposite to the incoherent quantum tunneling regime\cite{Leuenberger2000,Garanin1997,Gatteschi2006,Abragam1970} which arises when electron-phonon escape rate to excited states considerably exceeds $\Gamma_{\mathrm{relax}}$. This problem is appropriately described in the natural basis of the two states of the ground doublet, which are $\ket{\pm m}=\frac{1}{\sqrt{2}}\left(\ket{+}\pm\ket{-}\right)|_{\vt H=0}$, and generally requires going beyond the simple secular approximation \cite{Ho2017b}. Lowering the temperature, the electron-phonon escape rate is suppressed, however, incoherent quantum tunneling regime can still be maintained due to interaction with nuclear spins and through spin dipolar interaction between magnetic molecules in the lattice \cite{Vijayaraghavan2009,vijayaraghavan2011tunneling,Prokof'ev2000,Garanin1999,Sinitsyn2003}, provided the tunneling gap is sufficiently small. Since we do not introduce any restriction for its value, hereinafter we keep out the incoherent quantum tunneling effects caused by coupling with nuclear spins (hyperfine field) or other magnetic molecules (dipolar field), by considering strongly diluted systems. Hence, the calculation of the relaxation rate of SMM systems reduces to calculating the relaxation rate between two tunnel-split states of the ground doublet, which consists of the direct and Raman processes. To this end, a knowledge of spin-phonon Hamiltonian is required.

In the following, the treatment is done explicitly for $S$-complexes. However, given their generality, the results are straightforwardly applied also to systems characterized by total angular momentum $J$ as well. That is to say, the results are valid for both transition metal and lanthanide-based magnetic complexes.

As mentioned, a magnetic molecule interacts with the lattice through molecular degrees of freedom including both distortions and rotations. For a rigid magnetic molecule or cluster, the ``rotational'' contribution, which changes the direction of the main anisotropy axis, is apparently the most efficient. In the presence of an external magnetic field, the total Hamiltonian for the system interacting in this way with the phonon bath can be written as 
\begin{equation}
\hmt_{\mathrm{total}}=R\hmt_{A}R^{-1}+\hmt_{Z}+\hmt_{\mathrm{ph}},\label{eq:H-total}
\end{equation}
where $R=e^{-i\vt S\cdot\vt{\delta\varphi}}$ is the rotation operator corresponding to a small rotation angle $\vt{\delta\varphi}$ of the entire complex \cite{Chudnovsky2005,Garanin2011}, $\hmt_{A}$ is the anisotropic crystal (ligand) field Eq. \eqref{ZFS-S} or \eqref{H-CF}, $\hmt_{\mathrm{ph}}$ is the phonon bath Hamiltonian, and $\hmt_{Z}=g\mu_{B}\vt H\cdot\vt S$ ($\hmt_{Z}=g_{J}\mu_{B}\vt H\cdot\vt J$ in the case of lanthanides) is the Zeeman Hamiltonian corresponding to the external magnetic field $\vt H$. Expanding Eq. \eqref{eq:H-total} up to the second order in $\bm{\delta\varphi}$ yields the spin-phonon Hamiltonian \cite{Garanin2011}, 
\begin{flalign}
\text{\ensuremath{\hmt_{\mathrm{sp-ph}}}} & =\hmt_{\mathrm{sp-ph}}^{\left(1\right)}+\hmt_{\mathrm{sp-ph}}^{\left(2\right)},\label{eq:spin-phonon Hamiltonian}\\
\hmt_{\mathrm{sp-ph}}^{\left(1\right)} & =\sum_{\alpha=x,y,z}i\left[\hmt_{A},S_{\alpha}\right]\delta\varphi_{\alpha}\equiv\sum_{\alpha=x,y,z}\hmt^{\left(1\right)\alpha}\delta\varphi_{\alpha},\label{eq:H_sp-ph(1)}\\
\hmt_{\mathrm{sp-ph}}^{\left(2\right)} & =-\frac{1}{2}\sum_{\substack{\alpha=x,y,z\\
\beta=x,y,z
}
}\left[\left[\hmt_{A},S_{\alpha}\right],S_{\beta}\right]\delta\varphi_{\alpha}\delta\varphi_{\beta}\equiv\sum_{\substack{\alpha=x,y,z\\
\beta=x,y,z
}
}\hmt^{\left(2\right)\alpha\beta}\,\delta\varphi_{\alpha}\delta\varphi_{\beta}.\label{eq:H_sp-ph(2)}
\end{flalign}

As discussed in the introduction, we further consider the contribution of acoustic phonons only as being the most efficient in direct and Raman process at relatively low temperatures\cite{Chudnovsky2005}. The angle $\bm{\delta\varphi}$ then can be quantized using the canonical phonon quantization of the lattice displacement vector $\vt u\left(\vt r\right)$ \cite{Leuenberger2000,Garanin2011}, 
\begin{equation}
\delta\vt{\varphi}=\frac{1}{2}\nabla\times\vt u\left(\vt r\right)=\sqrt{\frac{\hbar}{8\rho V}}\sum_{\vt k\lambda_{\vt k}}\frac{\left[i\vt k\times\vt e_{\vt k\lambda_{\vt k}}\right]e^{i\vt k\cdot\vt r}}{\sqrt{\omega_{\vt k\lambda_{\vt k}}}}\left(a_{\vt k\lambda_{\vt k}}+a_{-\vt k\lambda_{\vt k}}^{+}\right).
\end{equation}
Here $\rho$ is the mass density, $V$ is the volume of the crystal, $\omega_{\vt k\lambda_{\vt k}}$ is the phonon frequency, and $\vt e_{\vt k\lambda_{\vt k}}$ is the polarization unit vector corresponding to one longitudinal and two transverse polarizations, $\lambda=t_{1},t_{2},l$. As can be seen from the above expression for $\delta\vt{\varphi}$, the longitudinal mode does not contribute to the spin-phonon Hamiltonian due to the fact that $\vt k\parallel\vt e_{\vt kl}$. Since we consider a low temperature, the Debye model's dispersion relation $\omega_{\vt k\lambda_{\vt k}}=v_{\lambda}k$ will be used throughout this work.

\begin{figure}
\includegraphics[width=0.5\columnwidth]{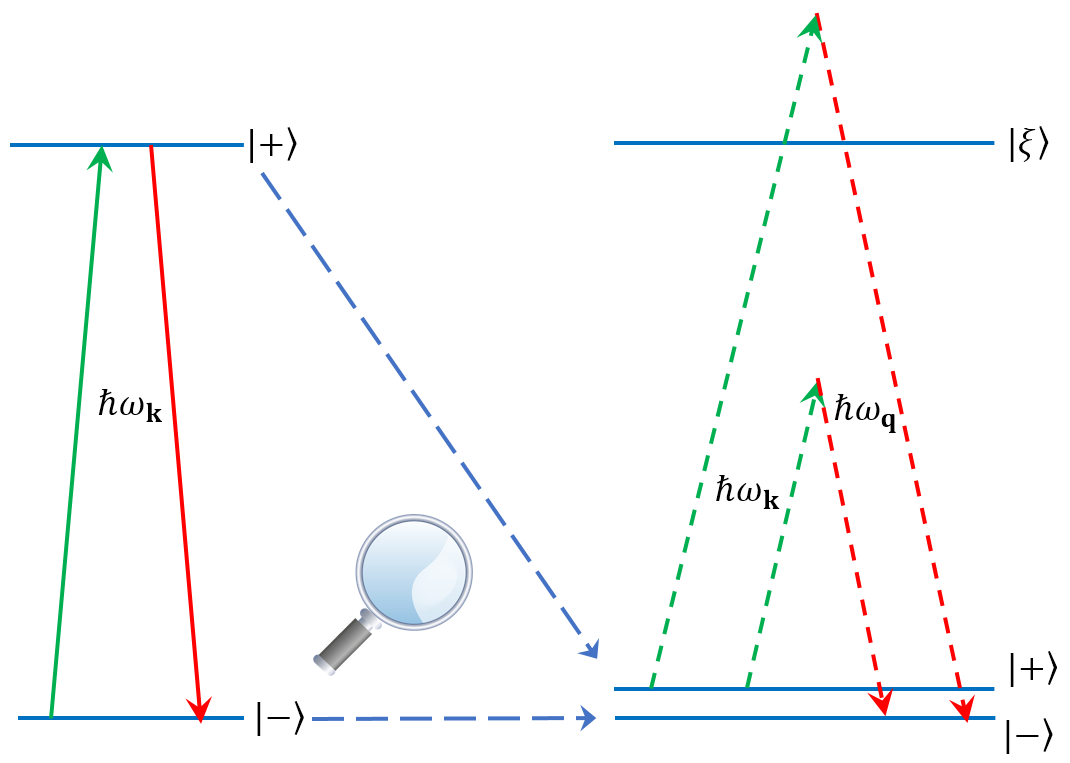}\caption{Phonon-induced Transition between tunnel-split states $\ket{+}$ and $\ket{-}$. (Left) Direct process: spin system absorbs (emits) one resonant phonon with energy $\hbar\omega_{\protect\vt k}$. (Right) Second-order Raman process: spin system virtually absorbs one phonon with energy $\hbar\omega_{\protect\vt k}$ and concomitantly emits one phonon with energy $\hbar\omega_{\protect\vt q}$.\label{fig:Fig. 1}}
\end{figure}

Denoting the states of the thermal bath by the phonon population numbers $n_{\vt k\lambda_{\vt k}}$ for each phonon mode $\vt k\lambda_{\vt k}$, the direct and Raman transition rates between the tunnel-split eigenstate $\ket{+}$ and$\ket{-}$ of the ground doublet can be calculated from the Fermi's golden rule \cite{cohen1977},
\begin{align}
\Gamma_{-+}^{\mathrm{dr}} & =\frac{2\pi}{\hbar}\sum_{\vt k\lambda_{\vt k}}\sum_{n_{\vt k\lambda_{\vt k}}}\left|M^{\left(1\right)}\right|^{2}\delta\left(\Omega-\hbar\omega_{\vt k\lambda_{\vt k}}\right)\frac{e^{-E_{n_{\vt k\lambda_{\vt k}},\ldots}/k_{\mathrm{B}}T}}{Z_{\mathrm{bath}}},\label{eq:direct process rate}\\
\Gamma_{-+}^{\mathrm{Raman}} & =\frac{2\pi}{\hbar}\sum_{\substack{\vt k\lambda_{\vt k}\\
\vt q\lambda_{\vt q}
}
}\sum_{\substack{n_{\vt k\lambda_{\vt k}}\\
n_{\vt q\lambda_{\vt q}}
}
}\left|M^{\left(2\right)}+M^{\left(1+1\right)}\right|^{2}\delta\left(\Omega+\hbar\omega_{\vt k\lambda_{\vt k}}-\hbar\omega_{\vt q\lambda_{\vt q}}\right)\frac{e^{-E_{n_{\vt k\lambda_{\vt k}},n_{\vt q\lambda_{\vt q}},\ldots}/k_{\mathrm{B}}T}}{Z_{\mathrm{bath}}},\label{eq:Raman process rate}
\end{align}
where $\Omega\equiv\varepsilon_{+}-\varepsilon_{-}$ and 
\begin{gather}
M^{\left(1\right)}\equiv\braket{-;n_{\vt k\lambda_{\vt k}}+1|\hmt_{\mathrm{sp-ph}}^{\left(1\right)}|+;n_{\vt k\lambda_{\vt k}}},\label{eq:M(1)}\\
M^{\left(2\right)}\equiv\braket{-;n_{\vt k\lambda_{\vt k}}-1,n_{\vt q\lambda_{\vt q}}+1|\hmt_{\mathrm{sp-ph}}^{\left(2\right)}|+;n_{\vt k\lambda_{\vt k}},n_{\vt q\lambda_{\vt q}}},\label{eq:M(2)}\\
M^{\left(1+1\right)}\equiv\sum_{\xi}\sum_{\substack{\vt k'\lambda_{\vt k'}\\
\vt q'\lambda_{\vt q'}
}
}\sum_{\substack{n_{\vt k'\lambda_{\vt k'}}\\
n_{\vt q'\lambda_{\vt q'}}
}
}\frac{\braket{-;n_{\vt k\lambda_{\vt k}}-1,n_{\vt q\lambda_{\vt q}}+1|\hmt_{\mathrm{sp-ph}}^{\left(1\right)}|\xi;n_{\vt k'\lambda_{\vt k'}},n_{\vt q'\lambda_{\vt q'}}}\braket{\xi;n_{\vt k'\lambda_{\vt k'}},n_{\vt q'\lambda_{\vt q'}}|\hmt_{\mathrm{sp-ph}}^{\left(1\right)}|+;n_{\vt k\lambda_{\vt k}},n_{\vt q\lambda_{\vt q}}}}{\varepsilon_{+}-\varepsilon_{\xi}+E_{n_{\vt k\lambda_{\vt k}},n_{\vt q\lambda_{\vt q}},\ldots}-E_{n_{\vt k'\lambda_{\vt k'}},n_{\vt q'\lambda_{\vt q'}},\ldots}}.\label{eq:M(1+1)}
\end{gather}
Here $\ket{\xi}$ labels eigenstates of the spin $S$. The transition matrix element $M^{\left(1\right)}$ characterizes the direct process where owing to the interaction Hamiltonian $\hmt^{\left(1\right)}$, the spin system emits one phonon to transit from $\ket{+}$ to $\ket{-}$, while $M^{\left(2\right)}$ characterizes the first-order Raman process where the spin system under the effect of $\hmt^{\left(2\right)}$ virtually emits a phonon $\hbar\omega_{\vt q}$ and concomitantly absorbs another one with energy $\hbar\omega_{\vt k}$, their energy difference spanning the states $\ket{+}$ and $\ket{-}$ (Fig. \ref{fig:Fig. 1}); $M^{\left(1+1\right)}$ characterizes the second-order Raman process with the same physical interpretation as the first-order Raman process but deduced from the second order of perturbation of the spin-phonon Hamiltonian $\hmt^{\left(1\right)}$ (Fig. \ref{fig:Fig. 1}).

In order to determine the relaxation rate in Eq. \eqref{eq:direct process rate} and \eqref{eq:Raman process rate}), one should calculate the matrix elements $M^{\left(i\right)}$.

\paragraph*{Direct process:}

From Eq. \eqref{eq:H_sp-ph(1)} and \eqref{eq:M(1)}, we easily reach
\begin{equation}
M^{\left(1\right)}=\sum_{\alpha}\Xi_{\alpha}M_{\mathrm{ph}}^{\alpha},
\end{equation}
where 
\begin{gather}
\Xi_{\alpha}\equiv\braket{-|\hmt^{\left(1\right)\alpha}|+},\\
M_{\mathrm{ph}}^{\alpha}\equiv\sqrt{\frac{\hbar}{8\rho V}}\frac{\left[-i\vt k\times\vt e_{\vt k\lambda_{\vt k}}\right]_{\alpha}e^{-i\vt k\cdot\vt r}}{\sqrt{\omega_{\vt k\lambda_{\vt k}}}}\sqrt{n_{\vt k\lambda_{\vt k}}+1}.
\end{gather}

Substituting $M^{\left(1\right)}$ into Eq. \eqref{eq:direct process rate} of $\Gamma_{-+}^{\mathrm{dr}}$ and averaging over the number of phonon numbers $n_{\vt k\lambda_{\vt k}}$ results in
\begin{align}
\Gamma_{-+}^{\mathrm{dr}} & =\frac{\pi}{4\rho V}\sum_{\vt k\lambda_{\vt k}}\sum_{\alpha,\beta}\Xi_{\alpha}\Xi_{\beta}^{*}\frac{\left[\vt k\times\vt e_{\vt k\lambda_{\vt k}}\right]_{\alpha}\left[\vt k\times\vt e_{\vt k\lambda_{\vt k}}\right]_{\beta}}{\omega_{\vt k\lambda_{\vt k}}}\left(\left\langle n_{\vt k\lambda_{\vt k}}\right\rangle +1\right)\delta\left(\Omega-\hbar\omega_{\vt k\lambda_{\vt k}}\right).\label{eq:general direct process rate}
\end{align}

Using the property $\left[\vt k\times\vt e_{\vt kt}\right]=\pm k\vt e_{\vt kt'}$ where $t=t_{1},t_{2}$ and $t'=t_{2},t_{1}$ respectively and the Debye dispersion relation for the acoustic phonons, then replacing $\sum_{\vt k}$ by $\frac{V}{8\pi^{3}}\int d^{3}k$, one comes up to a simple expression for $\Gamma_{-+}^{\mathrm{dr}}$ (see Appendix \ref{sec:Averaging-phonon-modes} for detailed calculation): 
\begin{align}
\Gamma_{-+}^{\mathrm{dr}} & =\frac{1}{24\pi\hbar^{4}\rho}\left(\frac{1}{v_{t_{1}}^{5}}+\frac{1}{v_{t_{1}}^{5}}\right)\frac{\Omega^{3}}{e^{\Omega/k_{\mathrm{B}}T}-1}e^{\Omega/k_{\mathrm{B}}T}\left|\vt{\Xi}\right|^{2}.\label{eq:General direct process rate}
\end{align}
Here $v_{t_{1}}$, $v_{t_{2}}$ are respectively the phonon velocity for two transverse polarizations, $t_{1}$ and $t_{2}$. The expression for $\vt{\Xi}=\braket{-|\hmt^{\left(1\right)}|+}$ is different for Kramers and non-Kramers doublet. 

\paragraph*{Raman process:}

From Eq. \eqref{eq:M(2)} for $M^{\left(2\right)}$, we obtain 
\begin{align}
M^{\left(2\right)} & =\sum_{\alpha,\beta}\braket{-|\hmt^{\left(2\right)\alpha\beta}|+}\left\{ \braket{n_{\vt q\lambda_{\vt q}}+1|\delta\varphi_{\alpha}|n_{\vt q\lambda_{\vt q}}}\braket{n_{\vt k\lambda_{\vt k}}-1|\delta\varphi_{\beta}|n_{\vt k\lambda_{\vt k}}}+\beta\leftrightarrow\alpha\right\} \nonumber \\
 & =\sum_{\alpha,\beta}\braket{-|\hmt^{\left(2\right)\alpha\beta}+\hmt^{\left(2\right)\beta\alpha}|+}M_{\mathrm{ph}}^{\alpha\beta},
\end{align}
where 
\begin{align}
M_{\mathrm{ph}}^{\alpha\beta} & \equiv\braket{n_{\vt q\lambda_{\vt q}}+1|\delta\varphi_{\alpha}|n_{\vt q\lambda_{\vt q}}}\braket{n_{\vt k\lambda_{\vt k}}-1|\delta\varphi_{\beta}|n_{\vt k\lambda_{\vt k}}}\nonumber \\
 & =\frac{\hbar}{8\rho V}\frac{\left(\vt q\times\vt e_{\vt q\lambda_{\vt q}}\right)_{\alpha}\left(\vt k\times\vt e_{\vt k\lambda_{\vt k}}\right)_{\beta}}{\sqrt{\omega_{\vt q\lambda_{\vt q}}\omega_{\vt k\lambda_{\vt k}}}}\sqrt{\left(n_{\vt q\lambda_{\vt q}}+1\right)n_{\vt k\lambda_{\vt k}}}e^{i\left(\vt k-\vt q\right)\vt r}.\label{eq:M_ph_alpha_beta}
\end{align}

Similar derivation for $M^{\left(1+1\right)}$ yields 
\begin{align}
M^{\left(1+1\right)} & =\sum_{\alpha,\beta}\sum_{\xi}\left(\frac{\hmt_{-\xi}^{\left(1\right)\alpha}\hmt_{\xi+}^{\left(1\right)\beta}}{\hbar\omega_{+\xi}+\hbar\omega_{\vt k\lambda_{\vt k}}}+\frac{\hmt_{-\xi}^{\left(1\right)\beta}\hmt_{\xi+}^{\left(1\right)\alpha}}{\hbar\omega_{+\xi}-\hbar\omega_{\vt q\lambda_{\vt q}}}\right)M_{\mathrm{ph}}^{\alpha\beta}.
\end{align}
During the derivation, the intermediate states $\ket{n_{\vt k\lambda_{\vt k}}-1,n_{\vt q\lambda_{\vt q}}}$ and $\ket{n_{\vt k\lambda_{\vt k}},n_{\vt q\lambda_{\vt q}}+1}$ of the bath, which correspond to the first and second term in the above equation, have been substituted into Eq. \eqref{eq:M(1+1)}. 

Since $M^{\left(1+1\right)}$ connects the ground doublet eigenstate $\ket{+}$ to $\ket{-}$ via the states $\ket{\xi}$, it is not necessary to take into account the admixture of other excited doublets caused by the Zeeman interaction. Denoting this non-admixing part of the eigenstates $\ket{\pm}$ by $\ket{p_{1}\left(m_{1}\right)}$ (short notation for ``plus'' and ``minus'') and the non-admixing part of $\ket{\xi_{m}}$ corresponding to $m^{\mathrm{th}}$ doublet by $\ket{x_{m}}$, $x=p$ or $m$, we have, 
\begin{align}
M^{\left(2\right)}+M^{\left(1+1\right)} & =\sum_{\alpha,\beta}\left[\left(\hmt_{m_{1}p_{1}}^{\left(2\right)\alpha\beta}+\hmt_{m_{1}p_{1}}^{\left(2\right)\beta\alpha}\right)+\delta M_{\mathrm{sp}}^{\alpha\beta}\right.\nonumber \\
 & \qquad\quad\left.+\sum_{x_{m}}\left(\frac{\hmt_{m_{1}x_{m}}^{\left(1\right)\alpha}\hmt_{x_{m}p_{1}}^{\left(1\right)\beta}}{\hbar\omega_{+x_{m}}+\hbar\omega_{\vt k\lambda_{\vt k}}}+\frac{\hmt_{m_{1}x_{m}}^{\left(1\right)\beta}\hmt_{x_{m}p_{1}}^{\left(1\right)\alpha}}{\hbar\omega_{+x_{m}}-\hbar\omega_{\vt q\lambda_{\vt q}}}\right)\right]M_{\mathrm{ph}}^{\alpha\beta}\equiv\sum_{\alpha,\beta}M_{\mathrm{sp}}^{\alpha\beta}M_{\mathrm{ph}}^{\alpha\beta},\label{eq:M(2)+M(1+1)}
\end{align}
where $\delta M_{\mathrm{sp}}^{\alpha\beta}$ results from the excited doublets admixing part of the eigenstates $\ket{\pm}$ in the expression of $M^{\left(2\right)}$. Its detailed form will be shown separately for Kramers and non-Kramers system.

Finally, we set two approximations further used for calculation of the Raman relaxation rate. First, since we are working in low temperature regime, it is assumed that $\hbar\omega_{\vt k\lambda_{\vt k}}\ll\hbar\omega_{\xi1}$, $\ket{\xi}\ne\ket{\pm}$. Second, due to the nature of the second-order Raman process, most of the phonon frequencies involved in the expression for$M^{\left(1+1\right)}$ are supposed to be much larger than the energy gap between $\ket{+}$ and $\ket{-}$, i.e. $\Omega\ll\hbar\omega_{\vt k\lambda_{\vt k}}$. We thus have the combined assumption, 
\begin{equation}
\Omega\ll\hbar\omega_{\vt k\lambda_{\vt k}}\ll\hbar\omega_{\xi1},\ket{\xi}\ne\ket{\pm}.\label{eq:assumptions}
\end{equation}

\section{Spin-phonon relaxation in Kramers doublets}

As mentioned above, at low temperature only the ground doublet is populated and the relaxation rate of the system reduces to the one between the two tunnel-split states. For Kramers system, without an applied magnetic field, the direct and first-order Raman process disappear in virtue of the Kramers' theorem \cite{Abragam1970}. The latter is violated in the presence of an external magnetic field through the admixture of excited Kramers doublets via Zeeman interaction \cite{Abragam1970}. This process thus requires an enlargement of the working Hilbert space to include all excited doublets of $S$ even though these are not thermally populated. In this space, the electronic Hamiltonian in zero external magnetic field can be written as 
\begin{equation}
\hmt_{A}=\sum_{m}\varepsilon_{m}\left(\ket{m^{\left(0\right)}}{\bra{m^{\left(0\right)}}}+\ket{\bar{m}^{\left(0\right)}}{\bra{\bar{m}^{\left(0\right)}}}\right),
\end{equation}
where $\ket{\bar{m}^{\left(0\right)}}=\theta\ket{m^{\left(0\right)}}$ is the Kramers conjugate states of the eigenstate $\ket{m^{\left(0\right)}}$, $\theta$ is the time-reversal operator, and the index $m=1\ldots\left(S+1/2\right)$ indicates the doublets. Since we are interested in the ground doublet, a choice of the reference frame's axes along three main magnetic axes of this doublet will apparently be the most appropriate. Further, the main magnetic $z$-axis of the pseudospin\cite{Chibotaru2012} $\tilde{S}=1/2$ corresponding to the ground doublet will be chosen as the quantization axis. The submatrix of the electronic Hamiltonian in a magnetic field $\vt H=(H_{x},H_{y},H_{z})$ then is 
\begin{align}
\left(\hmt_{A}+\hmt_{Z}\right)_{\left\{ \ket{1^{\left(0\right)}},\ket{\bar{1}^{\left(0\right)}}\right\} } & =\frac{W}{2}\left(\ket{1^{\left(0\right)}}\bra{1^{\left(0\right)}}-\ket{\bar{1}^{\left(0\right)}}\bra{\bar{1}^{\left(0\right)}}\right)+\frac{\Delta}{2}\left(e^{-i\varphi}\ket{1^{\left(0\right)}}\bra{\bar{1}^{\left(0\right)}}+e^{i\varphi}\ket{\bar{1}^{\left(0\right)}}\bra{1^{\left(0\right)}}\right).
\end{align}
where 
\begin{gather}
\Delta_{x}=g_{x}H_{x},\Delta_{y}=g_{y}H_{y},\label{eq:DeltaX and Y - Kramers}\\
\Delta=\sqrt{\Delta_{x}^{2}+\Delta_{y}^{2}},W=g_{z}H_{z},\label{eq:Delta and W - Kramers}\\
\sin\varphi=\frac{\Delta_{y}}{\Delta},\cos\varphi=\frac{\Delta_{x}}{\Delta}.
\end{gather}
Here $g_{x},g_{y},g_{z}$ are the principal values of the $g$-tensor of the ground doublet. For convenience, we set $\mu_{B}=1$ and $\varepsilon_{1}=0$.

To make this Hamiltonian submatrix real, the basis vectors of the ground doublet are rotated by an angle $\varphi/2$ while other basis vectors are kept unchanged:
\begin{align}
\ket{1} & =e^{-i\varphi/2}\ket{1^{\left(0\right)}},\ket{\bar{1}}=\theta\ket{1}=e^{i\varphi/2}\ket{\bar{1}^{\left(0\right)}},\\
\ket{M} & =\ket{M^{\left(0\right)}}\text{ for }M=m,\bar{m};m=2,\ldots,S+1/2.
\end{align}
Hereinafter, we will use $M$ to denote all basis vectors and $m$ to denote states corresponding to $m^{\mathrm{th}}$ doublet.

In this new basis, the submatrix $\left(\hmt_{A}+\hmt_{Z}\right)_{\left\{ \ket{1},\ket{\bar{1}}\right\} }$ becomes real, 
\begin{equation}
\left(\hmt_{A}+\hmt_{Z}\right)_{\left\{ \ket{1},\ket{\bar{1}}\right\} }=\frac{W}{2}\left(\ket{1}\bra{1}-\ket{\bar{1}}\bra{\bar{1}}\right)+\frac{\Delta}{2}\left(\ket{1}\bra{\bar{1}}+\ket{\bar{1}}\bra{1}\right).
\end{equation}

In order to find the spin-phonon coupling matrix element of the Kramers doublet in the presence of an external magnetic field, we need to pass to the new eigenstates of the spin system. These are given by the diagonalization of the submatrix $\left(\hmt_{A}+\hmt_{Z}\right)_{\left\{ \ket{1},\ket{\bar{1}}\right\} }$: 
\begin{gather}
\varepsilon_{\pm}=\pm\frac{1}{2}\sqrt{W^{2}+\Delta^{2}}\equiv\pm\frac{1}{2}\Omega,\\
\ket{p_{1}}=\cos\frac{\vartheta}{2}\ket{1}+\sin\frac{\vartheta}{2}\ket{\bar{1}},\label{eq:ket(p1)}\\
\ket{m_{1}}=\cos\frac{\vartheta}{2}\ket{\bar{1}}-\sin\frac{\vartheta}{2}\ket{1}=\ket{\bar{p}_{1}}.\label{eq:ket(m1)}
\end{gather}
where $\vartheta$ is defined as 
\begin{equation}
\cos\vartheta\equiv\frac{W}{\Omega},\sin\vartheta\equiv\frac{\Delta}{\Omega},0\le\vartheta\le\pi.
\end{equation}

As can be seen, $\ket{p_{1}}$ and $\ket{m_{1}}$ are still Kramers conjugates states. Accordingly, without admixture of excited doublets, the matrix element of the spin-phonon Hamiltonian, which is a time-even operator, between them is still zero. Applying the perturbation theory w.r.t Zeeman Hamiltonian $\hmt_{Z}$, we get new eigenstates (see Appendix \ref{sec: remainder submatrix diagonalization} for details):
\begin{gather}
\ket{+}\approx\ket{p_{1}}+\sum_{M\ne1,\bar{1}}\ket{M}\frac{\hmt_{Mp_{1}}^{Z}}{\hbar\omega_{1M}},\label{eq:Kramers Psi_+}\\
\ket{-}\approx\ket{m_{1}}+\sum_{M\ne1,\bar{1}}\ket{M}\frac{\hmt_{Mm_{1}}^{Z}}{\hbar\omega_{1M}}.\label{eq:Kramers Psi_-}
\end{gather}

For the later use, we also give here the expressions for $\ket{p_{m}}$ and $\ket{m_{m}}$: 
\begin{gather}
\ket{p_{m}}=\cos\frac{\vartheta_{m}}{2}\ket{m}+\sin\frac{\vartheta_{m}}{2}\ket{\bar{m}},\\
\ket{m_{m}}=\cos\frac{\vartheta_{m}}{2}\ket{\bar{m}}-\sin\frac{\vartheta_{m}}{2}\ket{m}=\ket{\bar{p}_{m}}.
\end{gather}

For the calculation of the direct and Raman transition rates, we need to find the matrices of spin operators $\vt S$ (or $\vt J$ for lanthanide complexes). Notice that in the original sub-basis $\left\{ \ket{1^{\left(0\right)}},\ket{\bar{1}^{\left(0\right)}}\right\} $, the matrix of the magnetic moment $\mu_{\alpha}$ can be calculated by either $\mu_{\alpha}^{\left(0\right)}=-gS_{\alpha}^{\left(0\right)}$ or $\mu_{\alpha}^{\left(0\right)}=-g_{\alpha}\tilde{S}_{\alpha}^{\left(0\right)}=-\frac{1}{2}g_{\alpha}\sigma_{\alpha},$ where $g$ is the isotropic electronic $g$-factor of the whole spin multiplet $S$ and $\sigma_{\alpha}$ $(\alpha=x,y,z)$ are Pauli matrices (in the case of $J$-complexes one uses for the first equation $\mu_{\alpha}^{\left(0\right)}=-g_{J}J_{\alpha}^{\left(0\right)},$ where $g_{J}$ is the Landé $g$-factor), we come to a simple formula relating the real-spin $S_{\alpha}$ and pseudospin $\tilde{S}_{\alpha}$ in the original sub-basis $\left\{ \ket{1^{\left(0\right)}},\ket{\bar{1}^{\left(0\right)}}\right\} $: 
\begin{equation}
S_{\alpha}=\frac{1}{2}\frac{g_{\alpha}}{g}\sigma_{\alpha}.
\end{equation}

A change from sub-basis $\left\{ \ket{1^{\left(0\right)}},\ket{\bar{1}^{\left(0\right)}}\right\} $ to $\left\{ \ket{1},\ket{\bar{1}}\right\} $, therefore, results in a rewriting of the $\vt S$ matrices 
\begin{align}
S_{x} & =\frac{1}{2}\frac{g_{x}}{g}\left(e^{-i\varphi}\ket{\bar{1}}\bra{1}+e^{i\varphi}\ket{1}\bra{\bar{1}}\right),\label{eq:Sx-explicit}\\
S_{y} & =\frac{i}{2}\frac{g_{y}}{g}\left(e^{-i\varphi}\ket{\bar{1}}\bra{1}-e^{i\varphi}\ket{1}\bra{\bar{1}}\right),\label{eq:Sy-explicit}\\
S_{z} & =\frac{1}{2}\frac{g_{z}}{g}\left(\ket{1}\bra{1}-\ket{\bar{1}}\bra{\bar{1}}\right).\label{eq:Sz-explicit}
\end{align}

Equipped with these preliminary results, we now proceed to the calculation of the direct and Raman transition rates.

\subsection{The direct process }

For the direct transition rate $\Gamma_{-+}^{\mathrm{dr}}$, the value of $\vt{\Xi}=\braket{-|\vt{\hmt}^{\left(1\right)}|+}$ is required. Substituting the eigenstates $\ket{\pm}$ from Eq. (\ref{eq:Kramers Psi_+},\ref{eq:Kramers Psi_-}) into $\vt{\Xi}$, then expanding it to the 1st order of perturbation theory, yields 
\begin{equation}
\vt{\Xi}\approx\sum_{M\ne1,\bar{1}}\frac{\hmt_{m_{1}M}^{Z}\vt{\hmt}_{Mp_{1}}^{\left(1\right)}+\vt{\hmt}_{m_{1}M}^{\left(1\right)}\hmt_{Mp_{1}}^{Z}}{\omega_{1M}},\label{eq:Xi-first}
\end{equation}
where the property $\braket{m_{1}|\vt{\hmt}^{\left(1\right)}|p_{1}}=0$ ($\vt{\hmt}^{\left(1\right)}$ is a time-even operator) has been used (see Appendix \ref{sec: Time-even and time-odd operators matrix elements}).

Substituting the expressions for $\ket{p_{1}}$ and $\ket{m_{1}}$ from Eq. \eqref{eq:ket(p1)} and \eqref{eq:ket(m1)} into the above equation and making use of the time reversal symmetry in the matrix elements of time-even/time-odd operators (see Appendix \ref{sec: Time-even and time-odd operators matrix elements}), the equation is simplified to 
\begin{align}
\vt{\Xi} & =\sum_{M\ne1,\bar{1}}\frac{1}{\omega_{1M}}\left[\left(\hmt_{\bar{1}M}^{Z}\vt{\hmt}_{M1}^{\left(1\right)}-\vt{\hmt}_{1M}^{\left(1\right)}\hmt_{M\bar{1}}^{Z}\right)-\sin\vartheta\left(\hmt_{1M}^{Z}\vt{\hmt}_{M1}^{\left(1\right)}+\vt{\hmt}_{1M}^{\left(1\right)}\hmt_{M1}^{Z}\right)+\cos\vartheta\left(\hmt_{\bar{1}M}^{Z}\vt{\hmt}_{M1}^{\left(1\right)}+\vt{\hmt}_{1M}^{\left(1\right)}\hmt_{M\bar{1}}^{Z}\right)\right].\label{eq:braket1bar''Hsp1''}
\end{align}

The explicit form of $\vt{\hmt}_{}^{\left(1\right)}=i\left[\hmt_{A},\vt S\right]$ gives a key relation, 
\begin{equation}
\vt{\hmt}_{MN}^{\left(1\right)}=i\omega_{MN}\vt S_{MN}.
\end{equation}

Inserting this into Eq. \eqref{eq:braket1bar''Hsp1''} results in 
\begin{align}
\vt{\Xi} & =-i\sum_{M\ne1,\bar{1}}\left[\left(\hmt_{\bar{1}M}^{Z}\vt S_{M1}+\vt S_{1M}\hmt_{M\bar{1}}^{Z}\right)+\sin\vartheta\left(\vt S_{1M}\hmt_{M1}^{Z}-\hmt_{1M}^{Z}\vt S_{M1}\right)-\cos\vartheta\left(\vt S_{1M}\hmt_{M\bar{1}}^{Z}-\hmt_{\bar{1}M}^{Z}\vt S_{M1}\right)\right]\nonumber \\
 & \equiv-i\left(\vt z_{1}+\sin\vartheta\,\vt z_{2}-\cos\vartheta\,\vt z_{3}\right).\label{eq:Kramers - direct process - spin-phonon coupling matrix element}
\end{align}

The expression for the vector $\vt z_{1}$ can be simplified by using the completeness relation, properties of the time-even operator \cite{Abragam1970}, and the relation $\left[\hmt_{Z},\vt S\right]=ig\left(\vt S\times\vt H\right)$: 
\begin{align}
\vt z_{1} & =\frac{i}{2}g\left(\vt S_{\bar{1}1}-\vt S_{1\bar{1}}\right)\times\vt H-\vt S_{11}\left(\hmt_{1\bar{1}}^{Z}+\hmt_{\bar{1}1}^{Z}\right)+\hmt_{11}^{Z}\left(\vt S_{\bar{1}1}+\vt S_{1\bar{1}}\right).\label{eq:z1}
\end{align}

From the explicit form of $\vt S$ matrices in Sec. \ref{sec:Section II}, we have $\vt S_{11}=-\vt S_{\bar{1}\bar{1}}=\left(0,0,g_{z}\right)/2g$ and $\vt S_{1\bar{1}}=\vt S_{\bar{1}1}^{*}=\left(g_{x}e^{i\varphi},-ig_{y}e^{i\varphi},0\right)/2g$. This results in 
\begin{align}
\vt z_{1} & \equiv\frac{1}{2}\left(-g_{2}H_{z}\cos\varphi,-g_{1}H_{z}\sin\varphi,g_{2}H_{x}\cos\varphi+g_{1}H_{y}\sin\varphi\right),
\end{align}
where 
\begin{align}
g_{1} & \equiv g_{x}-g_{y}g_{z}/g,\label{eq:g1}\\
g_{2} & \equiv g_{y}-g_{z}g_{x}/g,\label{eq:g2}\\
g_{3} & \equiv g_{z}-g_{x}g_{y}/g.\label{eq:g3}
\end{align}

Similarly, we have for $\vt z_{2}$ and $\vt z_{3}$, 
\begin{align}
\vt z_{2} & =\frac{i}{2}\left(g_{3}H_{y},-g_{3}H_{x},0\right),\label{eq:z2}\\
\vt z_{3} & =-\frac{i}{2}\left(g_{2}H_{z}\sin\varphi,-g_{1}H_{z}\cos\varphi,g_{x}H_{y}\cos\varphi-g_{y}H_{x}\sin\varphi\right).\label{eq:z3}
\end{align}

Inserting these vectors into Eq. \eqref{eq:Kramers - direct process - spin-phonon coupling matrix element} and then into Eq. \eqref{eq:General direct process rate}, we obtain the final expression for the direct transition rate between eigenstates of a Kramers doublet, 
\begin{align}
\Gamma_{-+}^{\mathrm{dr}} & =\frac{1}{96\pi\hbar^{4}\rho}\left(\frac{1}{v_{t_{1}}^{5}}+\frac{1}{v_{t_{2}}^{5}}\right)\frac{\Omega e^{\Omega/k_{\mathrm{B}}T}}{e^{\Omega/k_{\mathrm{B}}T}-1}\nonumber \\
 & \qquad\times\left\{ W^{4}\frac{g_{1}^{2}+g_{2}^{2}}{g_{z}^{2}}\right.\nonumber \\
 & \qquad\qquad+W^{2}\Delta^{2}\left[\frac{g_{1}^{2}\Delta_{y}^{2}+g_{2}^{2}\Delta_{x}^{2}}{g_{z}^{2}\Delta^{2}}+2\frac{g_{3}}{g_{z}}\left(\frac{g_{1}}{g_{x}}\frac{\Delta_{x}^{2}}{\Delta^{2}}+\frac{g_{2}}{g_{y}}\frac{\Delta_{y}^{2}}{\Delta^{2}}\right)+\left(\frac{g_{1}}{g_{y}}\frac{\Delta_{y}^{2}}{\Delta^{2}}+\frac{g_{2}}{g_{x}}\frac{\Delta_{x}^{2}}{\Delta^{2}}\right)^{2}+\left(\frac{g_{x}}{g_{y}}-\frac{g_{y}}{g_{x}}\right)^{2}\frac{\Delta_{x}^{2}}{\Delta^{2}}\frac{\Delta_{y}^{2}}{\Delta^{2}}\right]\nonumber \\
 & \qquad\qquad\qquad\qquad\left.+\Delta^{4}\left[\left(\frac{g_{1}}{g_{y}}\frac{\Delta_{y}^{2}}{\Delta^{2}}+\frac{g_{2}}{g_{x}}\frac{\Delta_{x}^{2}}{\Delta^{2}}\right)^{2}+\left(\frac{g_{3}^{2}}{g_{x}^{2}}\frac{\Delta_{x}^{2}}{\Delta^{2}}+\frac{g_{3}^{2}}{g_{y}^{2}}\frac{\Delta_{y}^{2}}{\Delta^{2}}\right)\right]\right\} .\label{eq:General Kramers doublet direct transition rate}
\end{align}

A remarkable feature of this expression is that it does not involve explicitly any parameters characterizing the excited doublets despite the fact that their admixture is indispensable for the existence of the transition. This is due to the special form of the rotational spin-phonon Hamiltonian mentioned earlier and the implicit incorporation of the contributions of the admixed excited doublets in the ground doublet's $g$-tensor, which thus allows a very compact and handy form of $\Gamma_{-+}^{\mathrm{dr}}$. 

Eq. \eqref{eq:General Kramers doublet direct transition rate} shows that the direct transition rate between the eigenstates of the doublet involves three contributions. The first contribution depends solely on the energy bias $W$ between two eigenvectors of the pseudospin operator $\tilde{S}_{z}$ and hence may be considered as the direct process between two Kramers conjugate states $\ket{1}$ and $\ket{\bar{1}}$ of the ground doublets. . The third contribution only depends on the tunneling splitting gap $\Delta$ and can be regarded as the direct process between two tunneling states $\ket{\pm}$ in the absence of the bias. Meanwhile, the second contribution appears as a term mixing these two. 

The $\Omega$-dependence of the relaxation in \eqref{eq:General Kramers doublet direct transition rate} looks surprisingly different from the conventional expression for a direct transition rate between two Kramers components \cite{Abragam1970}. However, in the case of $\Delta\ll W$, only the first contribution survives, we have $\Omega\approx W$ and Eq. \eqref{eq:General Kramers doublet direct transition rate} acquires a conventional prefactor $\Omega^{5}$. Thus the derived direct transition rate, besides being of a universal form (i.e., not containing explicitly the parameters of spin-phonon coupling \cite{Chudnovsky2005}), is also a generalization of a previous expression \cite{Abragam1970} in that it contains distinct contributions of the bias ($W$) and tunneling gap ($\Delta$), which thus cannot be described as depending on only one direct gap $\Omega=\sqrt{\Delta^{2}+W^{2}}$ as was thought before \cite{Abragam1970}.

Since $W,\Delta\propto H$, it is obvious that the derived direct transition rate is proportional to $H^{4}T$ when $\Omega\ll k_{\mathrm{B}}T$. This is thus in agreement with the rough approximation for the field and temperature dependencies of the direct process in Kramers doublet in Ref. {[}\onlinecite{Abragam1970}{]}. However, Eqs. \eqref{eq:DeltaX and Y - Kramers} and \eqref{eq:Delta and W - Kramers} show that all contributions to $\Gamma_{-+}^{\mathrm{dr}}$ in \eqref{eq:General Kramers doublet direct transition rate} depend also on the direction of the applied magnetic field.

In the case $g_{x}=g_{y}\equiv g_{\perp}$ and $\Omega\ll k_{\mathrm{B}}T$, the transition rate is explicitly independent of the magnetic field orientation and proportional to the square of $g_{\perp}$, 
\begin{align}
\Gamma_{-+}^{\mathrm{dr}} & =\frac{g_{\perp}^{2}H^{2}\Omega^{2}k_{\mathrm{B}}T}{96\pi\hbar^{4}\rho}\left(\frac{1}{v_{t_{1}}^{5}}+\frac{1}{v_{t_{2}}^{5}}\right)\left[\left(1-\frac{g_{z}}{g}\right)^{2}+\left(\frac{g_{z}H}{\Omega}-\frac{\Omega}{gH}\right)^{2}\right].\label{eq:Kramers - direct process - gx=00003Dgy}
\end{align}
This physically means that the more axially anisotropic the system is, the smaller is the direct process relaxation rate. 

On the other hand, for the near-isotropic system such that $g_{x}\approx g_{y}\approx g_{z}\approx g$, resulting in $g_{i}\ll g_{\alpha}$ ($i=1,2,3$ and $\alpha=x,y,z$), $\Gamma_{-+}^{\mathrm{dr}}$ being proportional to $g_{i}$ asymptotically approaches zero in the limit of totally isotropic spin system.

\subsection{The Raman process}

For the evaluation of $\Gamma_{-+}^{\mathrm{Raman}}$, we need to calculate $M^{\left(2\right)}+M^{\left(1+1\right)}$. Since $\ket{m_{1}}=\ket{\bar{p}_{1}}$, Eq. \eqref{eq:ket(m1)}, the first term of $M_{\mathrm{sp}}^{\alpha\beta}$ in Eq. \eqref{eq:M(2)+M(1+1)} for $M^{\left(2\right)}+M^{\left(1+1\right)}$ is zero due to time-reversal symmetry. Meanwhile, the third term can be decomposed into two components.The first involves states of the ground doublet only and therefore is zero due to time-reversal symmetry as well. The second component of the third term, dubbed $M_{\mathrm{sp2}}^{\alpha\beta}$, can be simplified using the assumption \eqref{eq:assumptions}: 

\begin{align}
M_{\mathrm{sp2}}^{\alpha\beta} & \approx\sum_{x_{m}\ne p_{1},m_{1}}\frac{\hmt_{m_{1}x_{m}}^{\left(1\right)\alpha}\hmt_{x_{m}p_{1}}^{\left(1\right)\beta}+\hmt_{m_{1}x_{m}}^{\left(1\right)\beta}\hmt_{x_{m}p_{1}}^{\left(1\right)\alpha}}{\hbar\omega_{1m}}+\frac{\hmt_{m_{1}x_{m}}^{\left(1\right)\beta}\hmt_{x_{m}p_{1}}^{\left(1\right)\alpha}-\hmt_{m_{1}x_{m}}^{\left(1\right)\alpha}\hmt_{x_{m}p_{1}}^{\left(1\right)\beta}}{\hbar\omega_{1m}}\frac{\hbar\omega_{\vt k\lambda_{\vt k}}}{\hbar\omega_{1m}},\label{eq:Msp^alphabeta}
\end{align}
where, like in the case of direct process, Eq. \eqref{eq:Xi-first}, $\hbar\omega_{1m}$ denotes the excitation energy between the ground and $m^{\mathrm{th}}$ excited Kramers doublets in the absence of applied field. Here, we made the approximation $\delta\left(\Omega+\hbar\omega_{\vt k\lambda_{\vt k}}-\hbar\omega_{\vt q\lambda_{\vt q}}\right)\approx\delta\left(\hbar\omega_{\vt k\lambda_{\vt k}}-\hbar\omega_{\vt q\lambda_{\vt q}}\right)$ and thus replaced $\omega_{\vt q\lambda_{\vt q}}$ by $\omega_{\vt k\lambda_{\vt k}}$.

It is easy to show that the first term of $M_{\mathrm{sp2}}^{\alpha\beta}$ is zero. Indeed, for any $m^{\mathrm{th}}$ doublet (in which $\ket{m_{m}}=\ket{\bar{p}_{m}}$), the numerator of the first term becomes 
\begin{equation}
\sum_{x_{m}=p_{m},m_{m}}\left(\hmt_{m_{1}x_{m}}^{\left(1\right)\alpha}\hmt_{x_{m}p_{1}}^{\left(1\right)\beta}+\alpha\leftrightarrow\beta\right)=\left(\hmt_{\bar{p}_{1}p_{m}}^{\left(1\right)\alpha}\hmt_{p_{m}p_{1}}^{\left(1\right)\beta}-\hmt_{\bar{p}_{1}p_{m}}^{\left(1\right)\beta}\hmt_{p_{m}p_{1}}^{\left(1\right)\alpha}\right)+\alpha\leftrightarrow\beta=0
\end{equation}

The second term on the r.h.s of Eq. \eqref{eq:Msp^alphabeta} can be easily found by using the property $\hmt_{A}\ket{x_{m}}=\varepsilon_{m}\ket{x_{m}}$ and $\vt{\hmt}^{\left(1\right)}=i\left[\hmt_{A},\vt S\right]$, 
\begin{align}
M_{\mathrm{sp2}}^{\alpha\beta} & \approx\sum_{x_{m}\ne p_{1},m_{1}}\left(S_{m_{1}x_{m}}^{\beta}S_{x_{m}p_{1}}^{\alpha}-S_{m_{1}x_{m}}^{\alpha}S_{x_{m}p_{1}}^{\beta}\right)\hbar\omega_{\vt k\lambda_{\vt k}}\nonumber \\
 & =\left[-i\sum_{\gamma}\varepsilon_{\alpha\beta\gamma}S_{m_{1}p_{1}}^{\gamma}+2\left(S_{m_{1}p_{1}}^{\alpha}S_{p_{1}p_{1}}^{\beta}-\alpha\leftrightarrow\beta\right)\right]\hbar\omega_{\vt k\lambda_{\vt k}}\equiv Q_{\alpha\beta}\hbar\omega_{\vt k\lambda_{\vt k}},\label{eq:Msp^alphabeta-1}
\end{align}
where $\varepsilon_{\alpha\beta\gamma}$ is the Levi-Civita tensor. 

The second term $\delta M_{\mathrm{sp}}^{\alpha\beta}$, in the expression for $M^{\left(2\right)}+M^{\left(1+1\right)}$, Eq. \eqref{eq:M(2)+M(1+1)}, has the explicit form: 
\begin{align}
\delta M_{\mathrm{sp}}^{\alpha\beta} & \approx\sum_{M\ne1,\bar{1}}\frac{\left(\hmt_{m_{1}M}^{Z}\hmt_{Mp_{1}}^{\left(2\right)\alpha\beta}+\hmt_{m_{1}M}^{\left(2\right)\alpha\beta}\hmt_{Mp_{1}}^{Z}\right)+\alpha\leftrightarrow\beta}{\hbar\omega_{1M}},
\end{align}
where again the denominator corresponds to the absence of applied field.

It can be easily seen that $\delta M_{\mathrm{sp}}^{\alpha\beta}$ is of the order $\mathcal{O}\left(\hmt^{Z}S^{2}\right)$, whereas, from Eq. \eqref{eq:Msp^alphabeta-1}, $M_{\mathrm{sp2}}^{\alpha\beta}$ is of the order $\mathcal{O}\left(\hbar\omega_{\vt k\lambda_{\vt k}}S^{2}\right)$. Since the Zeeman interaction is considered to be much smaller than $\hbar\omega_{\vt k\lambda_{\vt k}}$ according to the condition \eqref{eq:assumptions}, $\delta M_{\mathrm{sp}}^{\alpha\beta}$ can be safely neglected in comparison to $M_{\mathrm{sp2}}^{\alpha\beta}$. As a consequence, 
\begin{align}
\Gamma_{-+}^{\mathrm{Raman}} & \approx2\pi\sum_{\substack{\alpha,\beta\\
\alpha',\beta'
}
}\sum_{\substack{\vt k\lambda_{\vt k}\\
\vt q\lambda_{\vt q}
}
}\sum_{\substack{n_{\vt k\lambda_{\vt k}}\\
n_{\vt q\lambda_{\vt q}}
}
}Q_{\alpha\beta}Q_{\alpha'\beta'}^{*}M_{\mathrm{ph}}^{\alpha\beta}M_{\mathrm{ph}}^{\alpha'\beta'*}\omega_{\vt k\lambda_{\vt k}}^{2}\delta\left(\omega_{\vt k\lambda_{\vt k}}-\omega_{\vt q\lambda_{\vt q}}\right)\frac{e^{-E_{n_{\vt k\lambda_{\vt k}},n_{\vt q\lambda_{\vt q}},\ldots}/k_{\mathrm{B}}T}}{Z_{\mathrm{bath}}},\label{eq:Kramers-Raman-rate-form}
\end{align}

Averaging over the phonon number $n_{\vt k\lambda_{\vt k}}$, $n_{\vt q\lambda_{\vt q}}$ and summing up over all phonon modes (see Appendix \ref{sec:Averaging-phonon-modes} for details) as in the calculation of the direct process transition rate, we finally obtain, 
\begin{align}
\Gamma_{-+}^{\mathrm{Raman}} & \approx\frac{\hbar^{2}}{1152\pi^{3}\rho^{2}}\left(\frac{1}{v_{t_{1}}^{5}}+\frac{1}{v_{t_{2}}^{5}}\right)^{2}\sum_{\alpha,\beta}\left|Q_{\alpha\beta}\right|^{2}\int_{0}^{\omega_{D}}\mathrm{d}\omega\,\omega^{8}\left\langle n_{\omega}\right\rangle \left(\left\langle n_{\omega}\right\rangle +1\right)\nonumber \\
 & =\frac{\hbar^{2}}{1152\pi^{3}\rho^{2}}\left(\frac{1}{v_{t_{1}}^{5}}+\frac{1}{v_{t_{2}}^{5}}\right)^{2}\left(\frac{k_{\mathrm{B}}T}{\hbar}\right)^{9}QI_{8},\label{eq:Raman process rate - first}
\end{align}
where $\omega_{D}$ is the Debye frequency and 
\begin{gather}
Q\equiv\sum_{\alpha,\beta}\left|Q_{\alpha\beta}\right|^{2},I_{8}\equiv\int_{0}^{\hbar\omega_{D}/k_{\mathrm{B}}T}\mathrm{d}x\frac{x^{8}e^{x}}{\left(e^{x}-1\right)^{2}}.
\end{gather}

The value of $Q$ can be calculated straightaway from the explicit form of $\ket{p_{1}}$ Eq. \eqref{eq:ket(p1)}, $\ket{m_{1}}$ Eq. \eqref{eq:ket(m1)}, and $S_{\alpha}$ Eq. (\ref{eq:Sx-explicit}-\ref{eq:Sz-explicit}), 
\begin{equation}
Q=\frac{\left(g_{1}^{2}\sin^{2}\varphi+g_{2}^{2}\cos^{2}\varphi\right)+\left(g_{1}^{2}\cos^{2}\varphi+g_{2}^{2}\sin^{2}\varphi\right)\cos^{2}\vartheta+g_{3}^{2}\sin^{2}\vartheta}{2g^{2}},
\end{equation}
where $\sin\varphi=\Delta_{y}/\Delta$, $\cos\varphi=\Delta_{x}/\Delta$, $\sin\vartheta=\Delta/\Omega$, and $\cos\vartheta=W/\Omega$.

This yields, 
\begin{equation}
\Gamma_{-+}^{\mathrm{Raman}}=\frac{\hbar^{2}I_{8}}{2304\pi^{3}\rho^{2}}\left(\frac{1}{v_{t_{1}}^{5}}+\frac{1}{v_{t_{2}}^{5}}\right)^{2}\left(\frac{k_{\mathrm{B}}T}{\hbar}\right)^{9}\left[\left(\frac{g_{1}^{2}}{g^{2}}\frac{\Delta_{y}^{2}}{\Delta^{2}}+\frac{g_{2}^{2}}{g^{2}}\frac{\Delta_{x}^{2}}{\Delta^{2}}\right)+\left(\frac{g_{1}^{2}}{g^{2}}\frac{\Delta_{x}^{2}}{\Delta^{2}}+\frac{g_{2}^{2}}{g^{2}}\frac{\Delta_{y}^{2}}{\Delta^{2}}\right)\frac{W^{2}}{\Omega^{2}}+\frac{g_{3}^{2}}{g^{2}}\frac{\Delta^{2}}{\Omega^{2}}\right].\label{eq:general Kramers doublet Raman transition rate}
\end{equation}

As can be seen, the Raman process transition rate varies as $T^{9}$ as expected. Surprisingly, the Raman process transition rate only shows a dependence on the orientation of the applied field but not on its magnitude. An interesting consequence of this property is that in experiments with powder sample of a rigid Kramers molecules, the Raman process transition rate is independent of the applied magnetic field.  

Interestingly, for the near-isotropic systems with $g_{x}\approx g_{y}\approx g_{z}\approx g$, the Raman process transition rate derived above yields a small value due to its proportionality to $g_{i}^{2}$, which are expected to be very small according to Eqs. (\ref{eq:g1}-\ref{eq:g3}). This is in line with the situation for the direct process. 

If the Debye frequency $\omega_{D}$ is much larger than $T$, $I_{8}\approx8!$ and 
\begin{align}
\Gamma_{-+}^{\mathrm{Raman}} & =\frac{35\hbar^{2}}{2\pi^{3}\rho^{2}}\left(\frac{1}{v_{t_{1}}^{5}}+\frac{1}{v_{t_{2}}^{5}}\right)^{2}\left(\frac{k_{\mathrm{B}}T}{\hbar}\right)^{9}\left[\left(\frac{g_{1}^{2}}{g^{2}}\frac{\Delta_{y}^{2}}{\Delta^{2}}+\frac{g_{2}^{2}}{g^{2}}\frac{\Delta_{x}^{2}}{\Delta^{2}}\right)+\left(\frac{g_{1}^{2}}{g^{2}}\frac{\Delta_{x}^{2}}{\Delta^{2}}+\frac{g_{2}^{2}}{g^{2}}\frac{\Delta_{y}^{2}}{\Delta^{2}}\right)\frac{W^{2}}{\Omega^{2}}+\frac{g_{3}^{2}}{g^{2}}\frac{\Delta^{2}}{\Omega^{2}}\right].
\end{align}

In the case the energy bias $W$ exceeds the tunneling splitting $\Delta$, a simpler expression can be obtained: 
\begin{align}
\Gamma_{-+}^{\mathrm{Raman}} & =\frac{35\hbar^{2}}{2\pi^{3}\rho^{2}}\left(\frac{1}{v_{t_{1}}^{5}}+\frac{1}{v_{t_{2}}^{5}}\right)^{2}\left(\frac{k_{\mathrm{B}}T}{\hbar}\right)^{9}\frac{g_{1}^{2}+g_{2}^{2}}{g^{2}}.
\end{align}

On the other hand, if $g_{x}=g_{y}\equiv g_{\perp}$ and accordingly $g_{1}=g_{2}\equiv g_{12}$, Eq. \eqref{eq:general Kramers doublet Raman transition rate} becomes 
\begin{equation}
\Gamma_{-+}^{\mathrm{Raman}}=\frac{\hbar^{2}I_{8}}{2304\pi^{3}\rho^{2}}\left(\frac{1}{v_{t_{1}}^{5}}+\frac{1}{v_{t_{2}}^{5}}\right)^{2}\left(\frac{k_{\mathrm{B}}T}{\hbar}\right)^{9}\frac{g_{12}^{2}\left(W^{2}+\Omega^{2}\right)+g_{3}^{2}\Delta^{2}}{g^{2}\Omega^{2}}.
\end{equation}

As can be seen, at variance to a similar situation in the case of the direct process, Eq. \eqref{eq:Kramers - direct process - gx=00003Dgy}, $\Gamma_{-+}^{\mathrm{Raman}}$ still depends on the direction of applied field.

\section{Spin-phonon relaxation in non-Kramers doublets}

A similar treatment can also be applied to non-Kramers systems with a little modification. Particularly, for non-Kramers (quasi) doublets, we introduce a new basis $\ket{m^{*}}$ and $\ket{m'^{*}}$ of $m^{\mathrm{th}}$ doublet: 
\begin{align}
\ket{m^{*}} & =\frac{1}{\sqrt{2}}\left(\ket{+_{m}^{\left(0\right)}}-\ket{-_{m}^{\left(0\right)}}\right),\\
\ket{m'^{*}} & =\frac{1}{\sqrt{2}}\left(\ket{+_{m}^{\left(0\right)}}+\ket{-_{m}^{\left(0\right)}}\right),
\end{align}
where $\ket{\pm_{m}^{\left(0\right)}}$ are eigenstates of the anisotropic Hamiltonian $\hmt_{A}$ corresponding to the $m^{\mathrm{th}}$ doublet, and related to the corresponding real-valued eigenstates, denoted $\ket{\pm_{m,r}^{\left(0\right)}}$, as 
\begin{align}
\ket{+_{m}^{\left(0\right)}} & =\ket{+_{m,r}^{\left(0\right)}},\\
\ket{-_{m}^{\left(0\right)}} & =-i\ket{-_{m,r}^{\left(0\right)}}.
\end{align}

Obviously, 
\begin{equation}
\ket{m'^{*}}\equiv\ket{\bar{m}^{*}}=\theta\ket{m^{*}},\text{ and }\ket{m^{*}}=\theta\ket{m'^{*}}=\theta\ket{\bar{m}^{*}}.
\end{equation}

In this basis, the anisotropic Hamiltonian $\hmt_{A}$ is of the form: 
\begin{equation}
\hmt_{A}=\sum_{m}\varepsilon_{m}\left(\ket{m^{*}}{\bra{m^{*}}}+\ket{m'^{*}}{\bra{m'^{*}}}\right)+\frac{\Delta_{m}}{2}\left(\ket{m^{*}}{\bra{m'^{*}}}+\ket{m'^{*}}{\bra{m^{*}}}\right),
\end{equation}
where $\Delta_{m}$ is the tunneling gap of the $m^{\mathrm{th}}$ doublet. Hereinafter, we choose $\varepsilon_{1}=0$ and denote all quantities corresponding to the ground doublet ($m=1$) without the index 1, e.g. $\Delta\equiv\Delta_{1}$ or $\ket{\pm_{1}}\equiv\ket{\pm}$.

Choosing the $z$-axis to lie along the main magnetic axis of the ground doublet, the spin Hamiltonian in the sub-basis $\left\{ \ket{1^{*}},\ket{\bar{1}{}^{*}}\right\} $ of the doublet under an applied magnetic field becomes 
\begin{align}
\left(\hmt_{A}+\hmt_{Z}\right)_{\left\{ \ket{1^{*}},\ket{\bar{1}^{*}}\right\} } & =\frac{W}{2}\left(\ket{1^{*}}\bra{1^{*}}-\ket{\bar{1}^{*}}\bra{\bar{1}^{*}}\right)+\frac{\Delta}{2}\left(\ket{1}\bra{\bar{1}^{*}}+\ket{\bar{1}^{*}}\bra{1^{*}}\right),\\
W & =g_{z}H_{z},
\end{align}
where $g_{z}$ is the principal value along the main magnetic $z$-axis of the ground doublet's $g$-tensor \cite{Griffith1963a}. For non-Kramers doublet, $g_{x}=g_{y}=0$ \cite{Griffith1963a}.

New eigenstates of the ground doublet can be found by first diagonalizing the submatrix $\left(\hmt_{A}+\hmt_{Z}\right)_{\left\{ \ket{1^{*}},\ket{\bar{1}^{*}}\right\} }$ then using the perturbation theory, 
\begin{gather}
\varepsilon_{\pm}=\pm\frac{1}{2}\sqrt{W^{2}+\Delta^{2}}\equiv\pm\frac{1}{2}\Omega,\\
\ket{p_{1}}=\cos\frac{\vartheta}{2}\ket{1^{*}}+\sin\frac{\vartheta}{2}\ket{\bar{1}^{*}},\label{eq:ket(1 prime)-1}\\
\ket{m_{1}}=\cos\frac{\vartheta}{2}\ket{\bar{1}^{*}}-\sin\frac{\vartheta}{2}\ket{1^{*}},\label{eq:ket(1 prime bar)-1}\\
\ket{+}=\ket{p_{1}}+\sum_{M^{*}\ne1^{*},\bar{1}^{*}}\ket{M^{*}}\frac{\hmt_{M^{*}p_{1}}^{Z}}{\hbar\omega_{1M}},\\
\ket{-}=\ket{m_{1}}+\sum_{M^{*}\ne1^{*},\bar{1}^{*}}\ket{M^{*}}\frac{\hmt_{M^{*}m_{1}}^{Z}}{\hbar\omega_{1M}}.
\end{gather}
Here $\hbar\omega_{MN}\equiv\varepsilon_{M}-\varepsilon_{N}$. Further, we will make use of the expressions for $\ket{p_{m}}$ and $\ket{m_{m}}$: 
\begin{gather}
\ket{p_{m}}=\cos\frac{\vartheta_{m}}{2}\ket{m^{*}}+\sin\frac{\vartheta_{m}}{2}\ket{\bar{m}^{*}},\\
\ket{m_{m}}=\cos\frac{\vartheta_{m}}{2}\ket{\bar{m}^{*}}-\sin\frac{\vartheta_{m}}{2}\ket{m^{*}}.
\end{gather}

Similar to the Kramers system, we also need to know the matrix representation of the spin operators $S_{\alpha}$ in the sub-basis $\left\{ \ket{1^{*}},\ket{\bar{1}^{*}}\right\} $. The former choices of the $z$-axis and the basis yield \cite{Griffith1963a}
\begin{gather}
S_{x}=S_{y}=0,\\
S_{z}=\frac{1}{2}\frac{g_{z}}{g}\sigma_{z}.\label{eq:Sz11}
\end{gather}

\subsection{The direct process}

As is clear from Eq. \eqref{eq:General direct process rate}, the only difference in the expression for $\Gamma_{-+}^{\mathrm{dr}}$ for non-Kramers system compared to Kramers one is the expression for $\vt{\Xi}$. Substituting $\ket{\pm}$ into $\vt{\Xi}=\braket{-|\vt{\hmt}^{\left(1\right)}|+}$, we have 
\begin{equation}
\vt{\Xi}\approx\vt{\hmt}_{m_{1}p_{1}}^{\left(1\right)}+\sum_{M^{*}\ne1^{*},\bar{1}^{*}}\frac{\hmt_{m_{1}M^{*}}^{Z}\vt{\hmt}_{M*p_{1}}^{\left(1\right)}+\vt{\hmt}_{m_{1}M^{*}}^{\left(1\right)}\hmt_{M^{*}p_{1}}^{Z}}{\omega_{\pm,M}}.
\end{equation}

Further derivation is practically identical to the Kramers' case. The key difference here is that the term $\braket{m_{1}|\vt{\hmt}^{\left(1\right)}|p_{1}}$ is non-zero for non-Kramers doublet. Taking into account that for non-Kramers system $\theta^{2}=1$ and the matrix elements from time-even and time-odd operators possess the properties presented in Appendix \ref{sec: Time-even and time-odd operators matrix elements}, we come to the following result: 
\begin{equation}
\vt{\Xi}=\vt{\hmt}_{m_{1}p_{1}}^{\left(1\right)}-\sin\vartheta\sum_{M^{*}\ne1^{*},\bar{1}^{*}}\frac{1}{\omega_{\pm,M}}\left(\hmt_{1^{*}M^{*}}^{Z}\vt{\hmt}_{M^{*}1^{*}}^{\left(1\right)}+\vt{\hmt}_{1^{*}M^{*}}^{\left(1\right)}\hmt_{M^{*}1^{*}}^{Z}\right).
\end{equation}

$\vt{\hmt}^{\left(1\right)}=i\left[\hmt_{A},\vt S\right]$ can be expanded as follows: 
\begin{equation}
\vt{\hmt}_{M^{*},N^{*}}^{\left(1\right)}=i\hbar\omega_{MN}\vt S_{M^{*}N^{*}}+\frac{i}{2}\left(\Delta_{M}\vt S_{\bar{M}^{*}N^{*}}-\Delta_{N}\vt S_{M^{*}\bar{N}^{*}}\right).
\end{equation}

Since $\Delta_{M}$ for all $M$ is much smaller than the energy gap between any two doublets, we have 
\begin{align*}
\vt{\Xi} & \approx\vt{\hmt}_{m_{1}p_{1}}^{\left(1\right)}-i\sin\vartheta\sum_{M\ne m{}^{*},\bar{1}^{*}}\left(\vt S_{1^{*}M}\hmt_{M1^{*}}^{Z}-\hmt_{1^{*}M}^{Z}\vt S_{M1^{*}}\right).
\end{align*}

Using the completeness relation and $\left[\hmt_{Z},\vt S\right]=ig\left(\vt S\times\vt H\right)$,  we obtain 
\begin{align}
\sum_{M\ne m{}^{*},\bar{1}^{*}}\left(\vt S_{1^{*}M}\hmt_{M1^{*}}^{Z}-\hmt_{1^{*}M}^{Z}\vt S_{M1^{*}}\right) & =ig\left(\vt H\times\vt S_{1^{*}1^{*}}\right).
\end{align}

The term $\vt{\hmt}_{m_{1}p_{1}}^{\left(1\right)}$ can be found by utilizing the easy-to-prove property,
\begin{equation}
\hmt_{A}\ket{x_{m}}=\varepsilon_{m}\ket{x_{m}}+\frac{\Delta_{m}}{2}\ket{\bar{x}_{m}}\text{ for }x_{m}=p_{m},m_{m},\label{eq:HA_xm}
\end{equation}
 then simplifying, 
\begin{equation}
\vt{\hmt}_{m_{1}p_{1}}^{\left(1\right)}=-i\frac{\Delta}{2}\left(\vt S_{\bar{p}_{1}m_{1}}+\vt S_{m_{1}\bar{p}_{1}}\right)=i\Delta\vt S_{1^{*}1^{*}}.
\end{equation}

Hence, 
\begin{align}
\vt{\Xi} & =i\Delta\vt S_{1^{*}1^{*}}+g\sin\vartheta\left(\vt H\times\vt S_{1^{*}1^{*}}\right),
\end{align}
and 
\begin{align}
\Gamma_{-+}^{\mathrm{dr}} & =\frac{1}{96\pi\hbar^{4}\rho}\left(\frac{1}{v_{t_{1}}^{5}}+\frac{1}{v_{t_{2}}^{5}}\right)\frac{\Omega e^{\Omega/k_{\mathrm{B}}T}}{e^{\Omega/k_{\mathrm{B}}T}-1}\frac{g_{z}^{2}}{g^{2}}\Delta^{2}\left[\Omega^{2}+g^{2}H_{\perp}^{2}\right].\label{eq:General non-Kramers direct process rate}
\end{align}

It is easy to see that when $\Omega\ll k_{\mathrm{B}}T$ and the energy bias $W$ dominates over the tunneling splitting $\Delta$, the direct transition rate $\Gamma_{-+}^{\mathrm{dr}}$ is proportional to $H^{2}T$ as expected for non-Kramers systems \cite{Abragam1970}.

Interestingly, since the rate is proportional to the square of the tunneling splitting $\Delta$, a reduction in the latter will substantially decrease the direct process transition rate. That is to say, the more anisotropic the system, the slower the direct process transition rate. This is not directly evident in the standard expression for the direct relaxation rate \cite{Abragam1970}.

In a highly anisotropic system where $W=g_{z}H_{z}\approx gH_{z}\left(m-m'\right)$ where $m$ ($m'$) is the magnetic quantum numbers corresponding to the ground doublet, the above expression becomes 
\begin{align}
\Gamma_{-+}^{\mathrm{dr}} & =\frac{\left(m'-m\right)^{2}}{96\pi\hbar^{4}\rho}\left(\frac{1}{v_{t_{1}}^{5}}+\frac{1}{v_{t_{2}}^{5}}\right)\frac{\Omega e^{\Omega/k_{\mathrm{B}}T}}{e^{\Omega/k_{\mathrm{B}}T}-1}\Delta^{2}\left(\Omega^{2}+g^{2}H_{\perp}^{2}\right),
\end{align}
which is similar to the result in Ref. {[}\onlinecite{Chudnovsky2005}{]} if $v_{t_{1}}=v_{t_{2}}$. Compared to it, Eq. \ref{eq:General non-Kramers direct process rate} is more general because it does not require extreme uniaxial anisotropy and is applicable not only to $S$ systems (transition metal complexes) but also to $J$ systems (lanthanide complexes) as well.

\subsection{The Raman process}

For non-Kramers system, the term $M_{\mathrm{sp}}^{\alpha\beta}$ in Eq. \eqref{eq:M(2)+M(1+1)} can be expanded as, 
\begin{align}
M_{\mathrm{sp}}^{\alpha\beta} & \approx\left[\frac{\hmt_{m_{1}p_{1}}^{\left(1\right)\alpha}\left(\hmt_{p_{1}p_{1}}^{\left(1\right)\beta}-\hmt_{m_{1}m_{1}}^{\left(1\right)\beta}\right)}{\hbar\omega_{\vt k\lambda_{\vt k}}}+\frac{\hmt_{m_{1}p_{1}}^{\left(1\right)\beta}\left(\hmt_{m_{1}m_{1}}^{\left(1\right)\alpha}-\hmt_{p_{1}p_{1}}^{\left(1\right)\alpha}\right)}{\hbar\omega_{\vt q\lambda_{\vt q}}}\right]\nonumber \\
 & +\left[\sum_{x_{m}\ne p_{1},m_{1}}\frac{\hmt_{m_{1}x_{m}}^{\left(1\right)\alpha}\hmt_{x_{m}p_{1}}^{\left(1\right)\beta}+\hmt_{m_{1}x_{m}}^{\left(1\right)\beta}\hmt_{x_{m}p_{1}}^{\left(1\right)\alpha}}{\hbar\omega_{1m}}\right]+\left[\hmt_{m_{1}p_{1}}^{\left(2\right)\alpha\beta}+\hmt_{m_{1}p_{1}}^{\left(2\right)\beta\alpha}\right]\nonumber \\
 & +\left[\sum_{x_{m}\ne p_{1},m_{1}}\frac{\hmt_{m_{1}x_{m}}^{\left(1\right)\beta}\hmt_{x_{m}p_{1}}^{\left(1\right)\alpha}-\hmt_{m_{1}x_{m}}^{\left(1\right)\alpha}\hmt_{x_{m}p_{1}}^{\left(1\right)\beta}}{\hbar\omega_{1m}}\frac{\hbar\omega_{\vt k\lambda_{\vt k}}}{\hbar\omega_{1x_{m}}}\right]+\delta M_{\mathrm{sp}}^{\alpha\beta}\nonumber \\
 & \equiv M_{\mathrm{sp1}}^{\alpha\beta}+M_{\mathrm{sp2}}^{\alpha\beta}+M_{\mathrm{sp3}}^{\alpha\beta}+M_{\mathrm{sp4}}^{\alpha\beta}+\delta M_{\mathrm{sp}}^{\alpha\beta}.\label{eq:MspAB-non-Kramers}
\end{align}

Using the property \eqref{eq:HA_xm} and $\vt{\hmt}^{\left(1\right)}=i\left[\hmt_{A},\vt S\right]$, it can be easily proved that $\hmt_{p_{1}p_{1}}^{\left(1\right)\alpha}=\hmt_{m_{1}m_{1}}^{\left(1\right)\alpha}=0$ for all $\alpha$. Consequently,
\begin{equation}
M_{\mathrm{sp1}}^{\alpha\beta}=0.
\end{equation}

Similar to the calculation of $M_{\mathrm{sp1}}^{\alpha\beta}$, noticing that all $\Delta_{i}$ can be neglected as being much smaller than $\hbar\omega_{1m}$, $M_{\mathrm{sp2}}^{\alpha\beta}$ becomes \footnote{Even if we keep all terms containing $\Delta_{i}$, the order of magnitude of these terms, which is $\mathcal{O}\left(\Delta_{i}S^{2}\right)$, is still smaller than one of $M_{\mathrm{sp4}}^{\alpha\beta}$, which is $\mathcal{O}\left(\hbar\omega_{\vt k\lambda_{\vt k}}S^{2}\right)$ as we see later. Hence, by any means, these terms can be safely neglected.} 
\begin{align}
M_{\mathrm{sp2}}^{\alpha\beta} & =\sum_{x_{m}\ne p_{1},m_{1}}\hbar\omega_{1m}\left(S_{m_{1}x_{m}}^{\alpha}S_{x_{m}p_{1}}^{\beta}+S_{m_{1}x_{m}}^{\beta}S_{x_{m}p_{1}}^{\alpha}\right).
\end{align}

Likewise, $M_{\mathrm{sp3}}^{\alpha\beta}$ can be simplified into: 
\begin{align}
M_{\mathrm{sp3}}^{\alpha\beta} & =-\sum_{x_{m}\ne p_{1},m_{1}}\hbar\omega_{1m}\left(S_{m_{1}x_{m}}^{\alpha}S_{x_{m}p_{1}}^{\beta}+S_{m_{1}x_{m}}^{\beta}S_{x_{m}p_{1}}^{\alpha}\right)=-M_{\mathrm{sp2}}^{\alpha\beta}.
\end{align}
Here, we have ignored all terms containing $\Delta_{i}$ as before. 

$M_{\mathrm{sp4}}^{\alpha\beta}$ can be determined without difficulty, 
\begin{align}
M_{\mathrm{sp4}}^{\alpha\beta} & =\sum_{x_{m}\ne p_{1},m_{1}}\left(S_{m_{1}x_{m}}^{\beta}S_{x_{m}p_{1}}^{\alpha}-S_{m_{1}x_{m}}^{\alpha}S_{x_{m}p_{1}}^{\beta}\right)\hbar\omega_{\vt k\lambda_{\vt k}}\nonumber \\
 & =\left[-i\sum_{\gamma}\varepsilon_{\alpha\beta\gamma}S_{m_{1}p_{1}}^{\gamma}+S_{m_{1}p_{1}}^{\alpha}\left(S_{p_{1}p_{1}}^{\beta}-S_{m_{1}m_{1}}^{\beta}\right)-S_{m_{1}p_{1}}^{\beta}\left(S_{p_{1}p_{1}}^{\alpha}-S_{m_{1}m_{1}}^{\alpha}\right)\right]\hbar\omega_{\vt k\lambda_{\vt k}}\equiv Q_{\alpha\beta}\hbar\omega_{\vt k\lambda_{\vt k}}.
\end{align}

The term $\delta M_{\mathrm{sp}}^{\alpha\beta}$ for non-Kramers system has the explicit form: 
\begin{align*}
\delta M_{\mathrm{sp}}^{\alpha\beta} & \approx\sum_{M^{*}\ne1^{*},\bar{1}^{*}}\frac{\left(\hmt_{m_{1}M^{*}}^{Z}\hmt_{M^{*}p_{1}}^{\left(2\right)\alpha\beta}+\hmt_{m_{1}M^{*}}^{\left(2\right)\alpha\beta}\hmt_{M^{*},p_{1}}^{Z}\right)+\alpha\leftrightarrow\beta}{\hbar\omega_{1M}}.
\end{align*}
 As in the Kramers' case, this term is of the magnitude order $\mathcal{O}\left(\hmt^{Z}S^{2}\right)$, which is smaller than the order of magnitude of $M_{\mathrm{sp4}}^{\alpha\beta}$, $\mathcal{O}\left(\hbar\omega_{\vt k\lambda_{\vt k}}S^{2}\right)$, since characteristic $\hbar\omega_{\vt k\lambda_{\vt k}}$ is much larger than the Zeeman interaction. As a consequence, $\delta M_{\mathrm{sp}}^{\alpha\beta}$ can be ignored. Then $\Gamma_{-+}^{\mathrm{Raman}}$ for non-Kramers system acquires a similar form, \eqref{eq:Kramers-Raman-rate-form}, with the Kramers case. Performing similar calculations as for the Kramers' case results in the same Eq. \eqref{eq:Raman process rate - first} but with a different value of $Q\equiv\sum_{\alpha,\beta}\left|Q_{\alpha\beta}\right|^{2}$, 
\begin{equation}
Q\equiv\sum_{\alpha,\beta}\left|Q_{\alpha\beta}\right|^{2}=2\left|\vt S_{1^{*}1^{*}}\right|^{2}\sin^{2}\vartheta,
\end{equation}
resulting in 

\begin{equation}
\Gamma_{-+}^{\mathrm{Raman}}=\frac{\hbar^{2}I_{8}}{2304\pi^{3}\rho^{2}}\left(\frac{1}{v_{t_{1}}^{5}}+\frac{1}{v_{t_{2}}^{5}}\right)^{2}\left(\frac{k_{\mathrm{B}}T}{\hbar}\right)^{9}\frac{g_{z}^{2}}{g}\frac{\Delta^{2}}{\Delta^{2}+W^{2}}.\label{eq:General non-Kramers Raman process rate}
\end{equation}

If the Debye frequency $\omega_{D}$ is much larger than $T$, we have 
\begin{align}
\Gamma_{-+}^{\mathrm{Raman}} & =\frac{35\hbar^{2}}{2\pi^{3}\rho^{2}}\left(\frac{1}{v_{t_{1}}^{5}}+\frac{1}{v_{t_{2}}^{5}}\right)^{2}\left(\frac{k_{\mathrm{B}}T}{\hbar}\right)^{9}\left(\frac{g_{z}}{g}\right)^{2}\frac{\Delta^{2}}{\Delta^{2}+W^{2}},
\end{align}
 In a highly anisotropic system where $W=g_{z}H_{z}\approx gH_{z}\left(m-m'\right)$, the above expression becomes 
\begin{equation}
\Gamma_{-+}^{\mathrm{Raman}}=\frac{35\hbar^{2}\left(m'-m\right)^{2}}{2\pi^{3}\rho^{2}}\left(\frac{1}{v_{t_{1}}^{5}}+\frac{1}{v_{t_{2}}^{5}}\right)^{2}\left(\frac{k_{\mathrm{B}}T}{\hbar}\right)^{9}\frac{\Delta^{2}}{\Delta^{2}+W^{2}}.
\end{equation}

The results obtained here differ the previous treatment of the Raman process in non-Kramers doublets within the same rotational approximation for the spin-phonon coupling in several important aspects. Contrary to the present treatment, Ref. {[}\onlinecite{Calero2006a}{]} is limited to a specific axially-anisotropic spin Hamiltonian $\hmt_{A}$, also called biaxial spin Hamiltonian. It is also important to note that the present theory predicts $T^{9}$ dependence of the Raman relaxation rate for non-Kramers doublets (within rotational spin-phonon approximation), while the previous treatment shows a $T^{11}$ temperature dependence \cite{Calero2006a}. The discrepancy comes from the zero value of the last square bracket in the expression \eqref{eq:MspAB-non-Kramers} for $M_{\mathrm{sp}}^{\alpha\beta}$, found in {[}\onlinecite{Calero2006a}{]}, but not supported by the present calculations.

\section{Direct process versus Raman process in magnetic molecules \label{sec:Direct processs vs. Raman process}}

The derived expressions for direct process and Raman process transition rates permit us to assess the relative importance of these processes in the typical case of $\Omega\ll k_{\mathrm{B}}T$ and $\hbar\omega_{D}\gg k_{\mathrm{B}}T$. This analysis is valid for cases when rotational spin-phonon interaction can be considered dominant

\subsection{Kramers SMMs \label{subsec:Direct Process vs. Raman Process in Kramers SMMs}}

Since there are many parameters in the expressions for $\Gamma_{-+}^{\mathrm{dr}}$ Eq. \eqref{eq:General Kramers doublet direct transition rate} and $\Gamma_{-+}^{\mathrm{Raman}}$ Eq. \eqref{eq:general Kramers doublet Raman transition rate}, we will take as example a typical lanthanide SMM with $g_{J}=4/3$ $(\mathrm{Dy}^{3+})$, $g_{x}=g_{y}=g_{\perp}=0.01$, $g_{z}\approx20$ under the supposition of equal $H_{\alpha}$ ($\alpha=x,y,z$). Taking for the common factor in Eqs. \eqref{eq:General Kramers doublet direct transition rate} and \eqref{eq:general Kramers doublet Raman transition rate}, $E_{t}\equiv\left[2\hbar^{3}\rho v_{t_{1}}^{5}v_{t_{2}}^{5}/\left(v_{t_{1}}^{5}+v_{t_{2}}^{5}\right)\right]^{1/4}$, a typical value $E_{t}\approx100\,\mathrm{K}$, $H$ is measured in kOe, we obtain 
\begin{gather}
\Gamma_{-+}^{\mathrm{dr}}\approx5.76\times10^{-4}H^{4}T,\,\,\Gamma_{-+}^{\mathrm{Raman}}\approx6.54\times10^{-7}T^{9},\,\,\frac{\Gamma_{-+}^{\mathrm{Raman}}}{\Gamma_{-+}^{\mathrm{dr}}}\approx\frac{1}{H^{4}}\left(\frac{T}{2.33}\right)^{8},
\end{gather}

\begin{figure}
\begin{tabular}{ll}
{\footnotesize{}(a)} & {\footnotesize{}(b)}\tabularnewline
\includegraphics[width=0.45\textwidth]{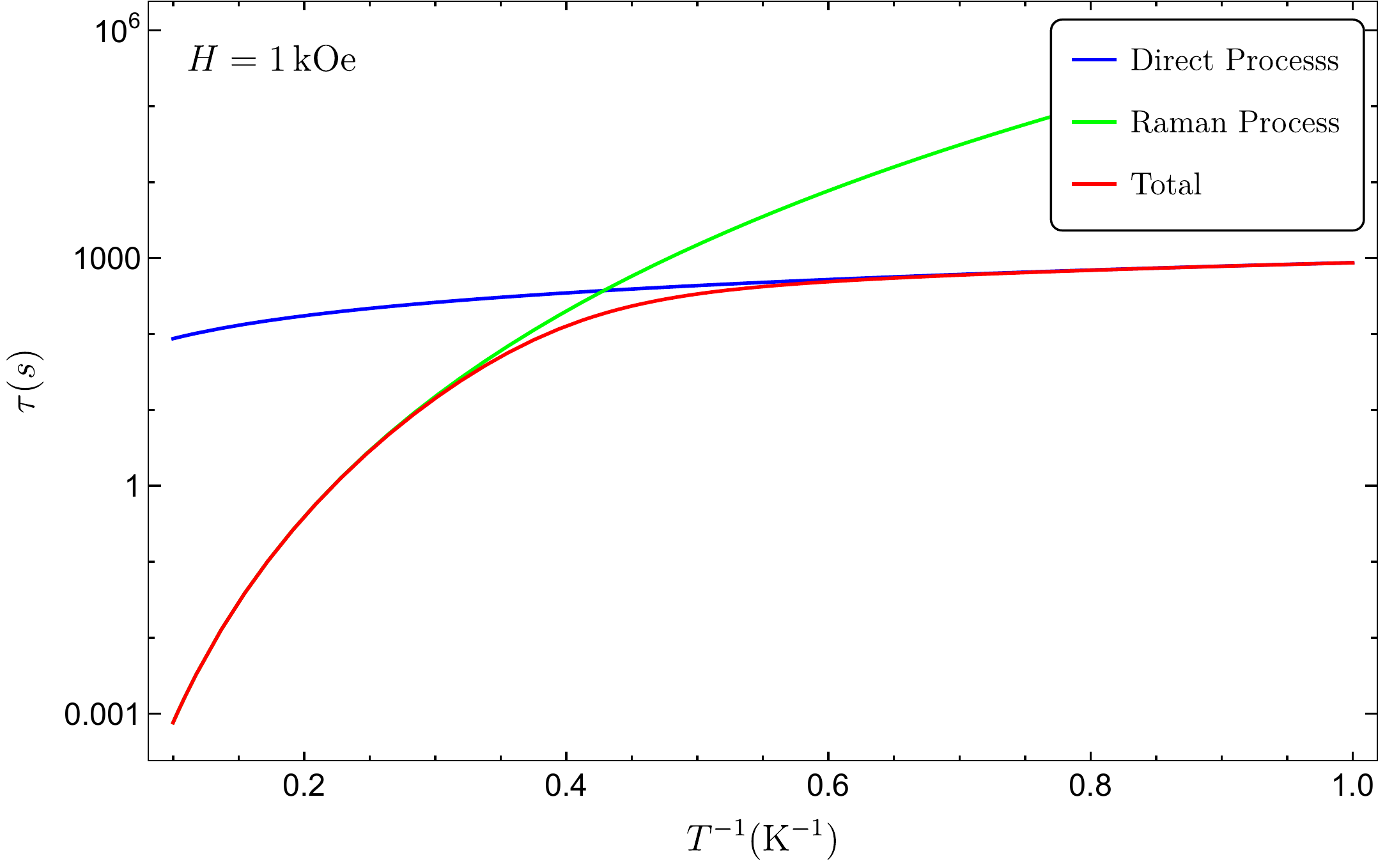} & \includegraphics[width=0.45\textwidth]{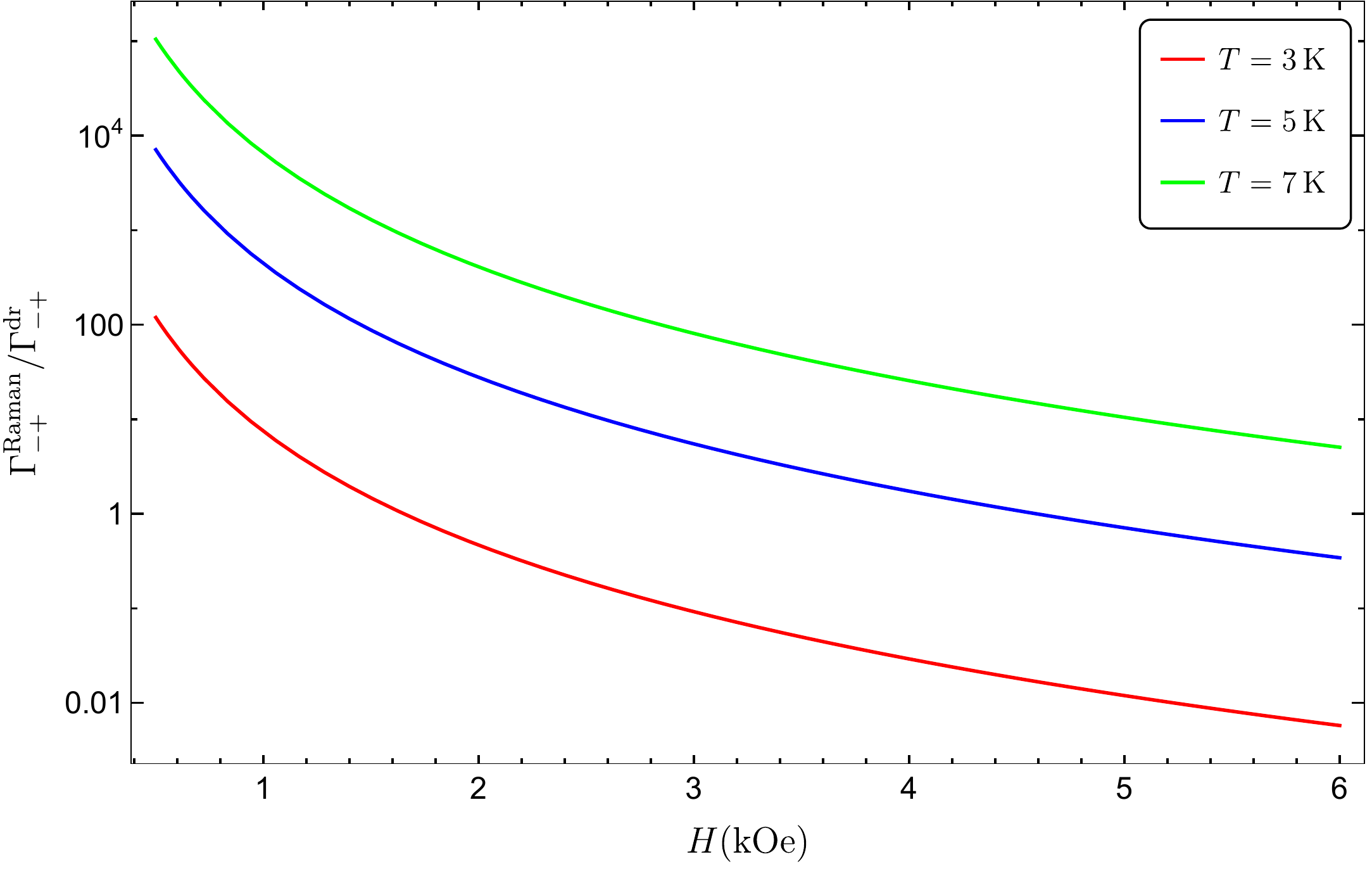}\tabularnewline
\multicolumn{2}{c}{(c)}\tabularnewline
\multicolumn{2}{c}{\includegraphics[width=0.45\textwidth]{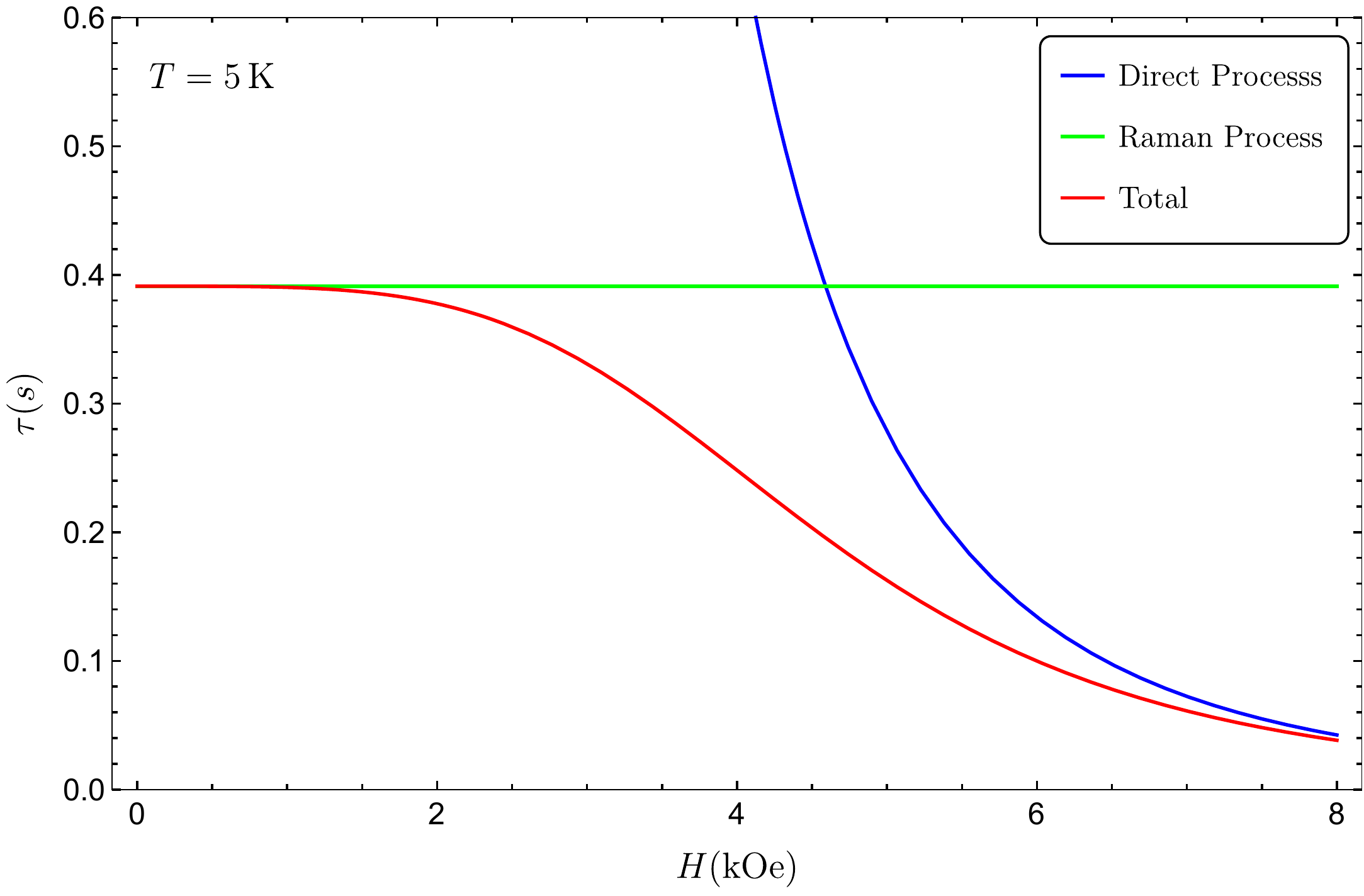}}\tabularnewline
\end{tabular}

\caption{Kramers system: (a) inverse temperature dependence of the relaxation times ($\tau_{\mathrm{direct}}$, $\tau_{\mathrm{Raman}}$, and $\tau_{\mathrm{total}}$) at 1 kOe; (b) magnetic field dependence of $\Gamma_{-+}^{\mathrm{Raman}}/\Gamma_{-+}^{\mathrm{dr}}$ at $T=3$ K, 5 K, and 7 K; (c) Magnetic field dependence of $\tau_{\mathrm{direct}}$, $\tau_{\mathrm{Raman}}$, and $\tau_{\mathrm{total}}$ at $T=5$ K.\label{fig:Kramers rates comparison}}
\end{figure}

Under an applied field of 1 kOe, the Raman transition rate becomes smaller than the direct one only at $T<2.33$ K. Particularly, at $T=5$ K, the Raman transition rate is $1.28\,\mathrm{s^{-1}}$ whereas the direct one is three orders of magnitudes smaller, $0.003\,\mathrm{s^{-1}}$. Fig. \ref{fig:Kramers rates comparison}a shows the dependence of the relaxation times $\tau_{\mathrm{direct}}=\left(2\Gamma_{-+}^{\mathrm{dr}}\right)^{-1}$, $\tau_{\mathrm{Raman}}=\left(2\Gamma_{-+}^{\mathrm{Raman}}\right)^{-1}$, and $\tau_{\mathrm{total}}=\left[2\left(\Gamma_{-+}^{\mathrm{dr}}+\Gamma_{-+}^{\mathrm{Raman}}\right)\right]^{-1}$ as a function of $T^{-1}$ for an applied field $H=1$ kOe, Fig. \ref{fig:Kramers rates comparison}b presents the field dependence of the ratio $\Gamma_{-+}^{\mathrm{Raman}}/\Gamma_{-+}^{\mathrm{dr}}$ for several values of $T$, and Fig. \ref{fig:Kramers rates comparison}c shows the field dependence of these relaxation times at $T=5$ K. 

Several features can be inferred from Fig. \ref{fig:Kramers rates comparison}. First, it can be seen (Fig. \ref{fig:Kramers rates comparison}a) that the switch from one process to another occurs sharply in a small domain of a few Kelvin. This is basically due to the fact that the direct process depends linearly on $T$ while the Raman one shows a $T^{9}$ dependence. Additionally, as expected, the direct process starts gaining control over the Raman process when the field increases (Fig. \ref{fig:Kramers rates comparison}b and \ref{fig:Kramers rates comparison}c). However, for a Kramers system the gain is rather slow. Moreover, the transition from dominant Raman to dominant direct process occurs without passing through an extremum of the total relaxation time/rate, a situation quite opposite to the case of non-Kramers SMMs (see below).

\subsection{Non-Kramers SMMs}

For non-Kramers systems, the number of parameters needed to describe the relaxation is substantially reduced. In fact, in most of cases, we can approximate $\left|\vt S_{1^{*}1^{*}}\right|\approx S$ (or for a lanthanide SMM, $\left|\vt J_{1^{*}1^{*}}\right|\approx J$). Furthermore, since the transverse magnetic field hardly affects the tunneling splitting $\Delta$, only the field projection $H_{z}$ directed along the easy axis of the SMM will affect the relaxation. Then we have 
\begin{gather}
\Gamma_{-+}^{\mathrm{dr}}=\frac{J^{2}}{12\pi}\frac{\Delta^{2}\left(\Delta^{2}+4g_{J}^{2}J^{2}H_{z}^{2}\right)}{E_{t}^{4}}\frac{k_{\mathrm{B}}T}{\hbar},\,\,\Gamma_{-+}^{\mathrm{Raman}}=\frac{280J^{2}}{\pi^{3}}\frac{\Delta^{2}}{\Delta^{2}+4g_{J}^{2}J^{2}H_{z}^{2}}\left(\frac{k_{\mathrm{B}}T}{E_{t}}\right)^{8}\frac{k_{\mathrm{B}}T}{\hbar},\\
\frac{\Gamma_{-+}^{\mathrm{Raman}}}{\Gamma_{-+}^{\mathrm{dr}}}=\frac{3360}{\pi^{2}}\left(\frac{k_{\mathrm{B}}T}{E_{t}}\right)^{4}\left(\frac{k_{\mathrm{B}}T}{\sqrt{\Delta^{2}+4g_{J}^{2}J^{2}H_{z}^{2}}}\right)^{4},
\end{gather}
where $E_{t}$ is defined in the previous subsection.

Accordingly, a $\mathrm{Tb^{3+}}$ SMM with $g_{J}=3/2$, $J=6$, $E_{t}=100$ K, and $\Delta=10^{-2}$ K, will be characterized by the following relaxation rates: 
\begin{gather}
\Gamma_{-+}^{\mathrm{dr}}\approx1.25\times10^{-5}\left(1+1.46\times10^{4}H_{z}^{2}\right)T,\,\,\Gamma_{-+}^{\mathrm{Raman}}\approx\frac{4.25\times10^{-3}T^{9}}{1+1.46\times10^{4}H_{z}^{2}},\,\,\frac{\Gamma_{-+}^{\mathrm{Raman}}}{\Gamma_{-+}^{\mathrm{dr}}}\approx\frac{1}{\left(1+1.46\times10^{4}H_{z}^{2}\right)^{2}}\left(\frac{T}{0.48}\right)^{8},
\end{gather}
where $H_{z}$ is in kOe. In zero field, $\Gamma_{-+}^{\mathrm{Raman}}/\Gamma_{-+}^{\mathrm{dr}}\approx\left(T/0.48\right)^{8}$. As can be seen, even with relatively large tunneling splitting gap of $10^{-2}\,\mathrm{K}$, the Raman process still dominates over the direct process in zero field. At $T$=5 K, the direct relaxation rate is still very small, $6.25\times10^{-5}\mathrm{\,s^{-1}}$, while the Raman process relaxation rate is eight orders of magnitude larger, $\Gamma_{-+}^{\mathrm{Raman}}=8.32\times10^{3}\mathrm{\,s^{-1}}$. However, in a field of 1 kOe, the situation changes significantly, 
\begin{gather}
\Gamma_{-+}^{\mathrm{dr}}\approx0.183T,\,\,\Gamma_{-+}^{\mathrm{Raman}}\approx2.91\times10^{-7}T^{9},\,\,\Gamma_{-+}^{\mathrm{Raman}}/\Gamma_{-+}^{\mathrm{dr}}\approx\left(T/5.31\right)^{8}.
\end{gather}
We can see that the direct process dominates at $T<5.31\,\mathrm{K}$. Thus, at $T$=3 K, $\Gamma_{-+}^{\mathrm{dr}}=5.49\times10^{-1}\,\mathrm{s^{-1}}$ is two orders of magnitude larger than $\Gamma_{-+}^{\mathrm{Raman}}=5.73\times10^{-3}\,\mathrm{s^{-1}}$. This property actually comes from the radically different behavior of the two processes in applied field where the direct process increases fast with the magnetic field but the Raman process is significantly reduced out of resonance. 

\begin{figure}
\begin{tabular}{ll}
{\footnotesize{}(a)} & {\footnotesize{}(b)}\tabularnewline
\includegraphics[width=0.45\textwidth]{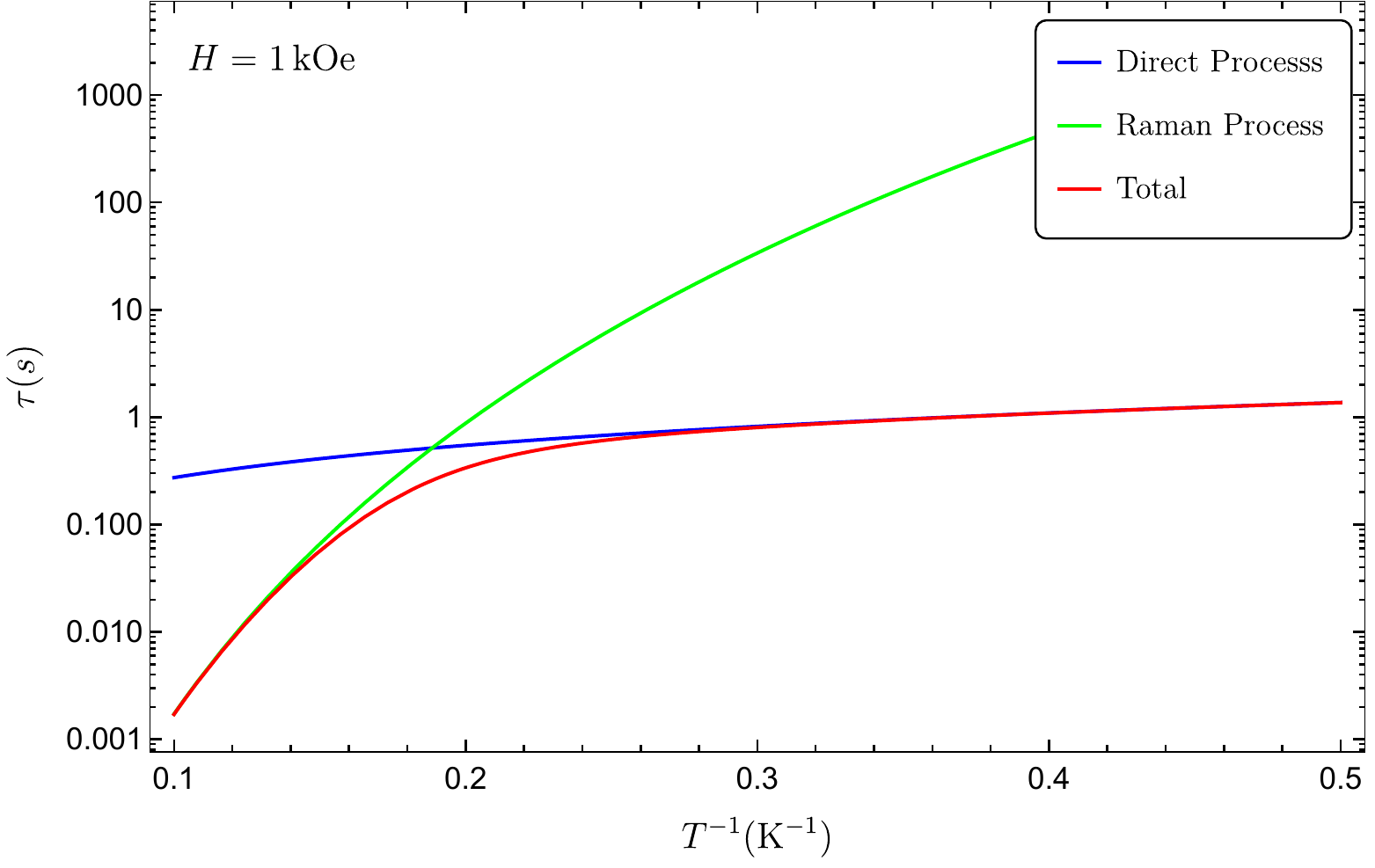} & \includegraphics[width=0.45\textwidth]{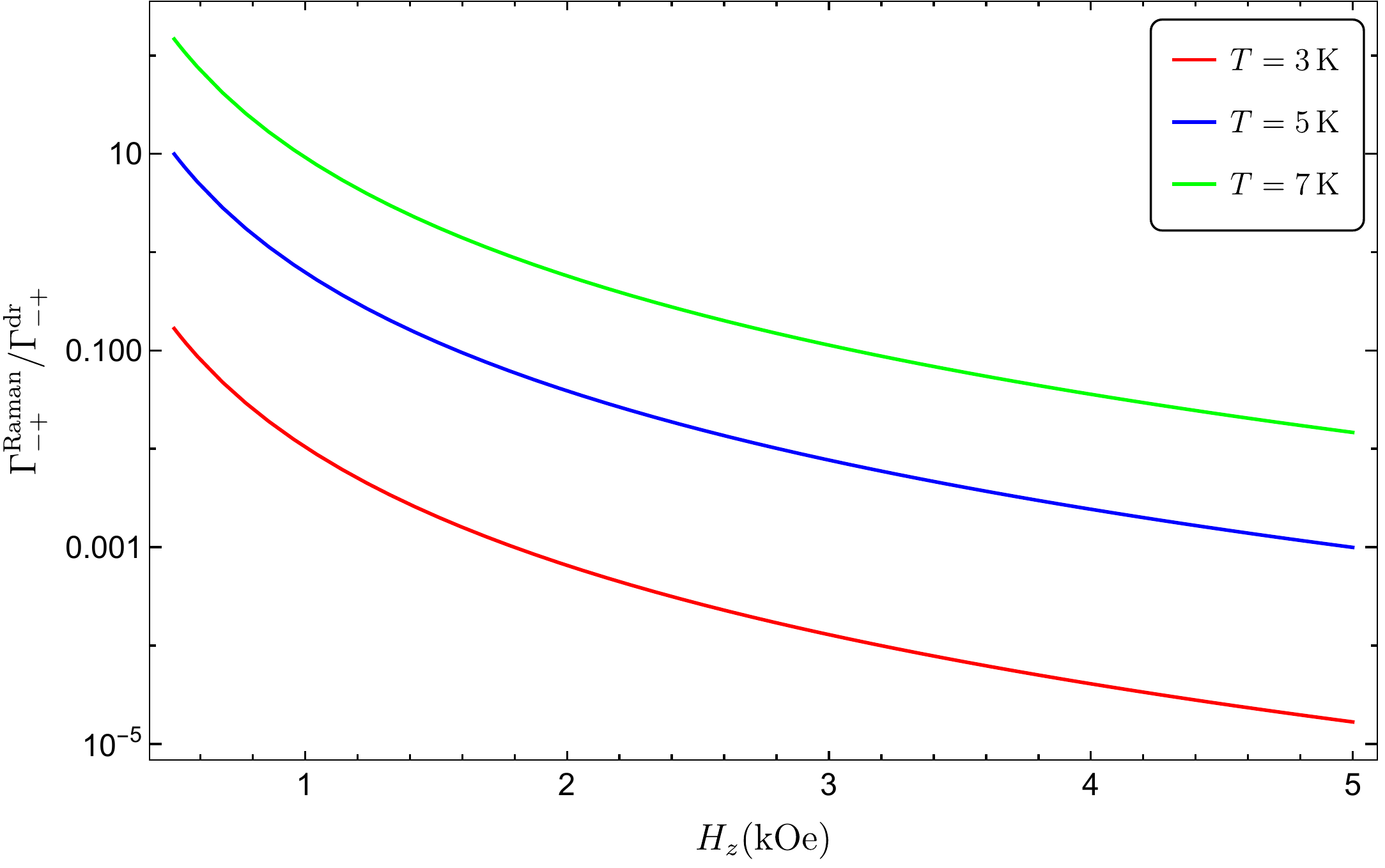}\tabularnewline
\multicolumn{2}{c}{(c)}\tabularnewline
\multicolumn{2}{c}{\includegraphics[width=0.45\textwidth]{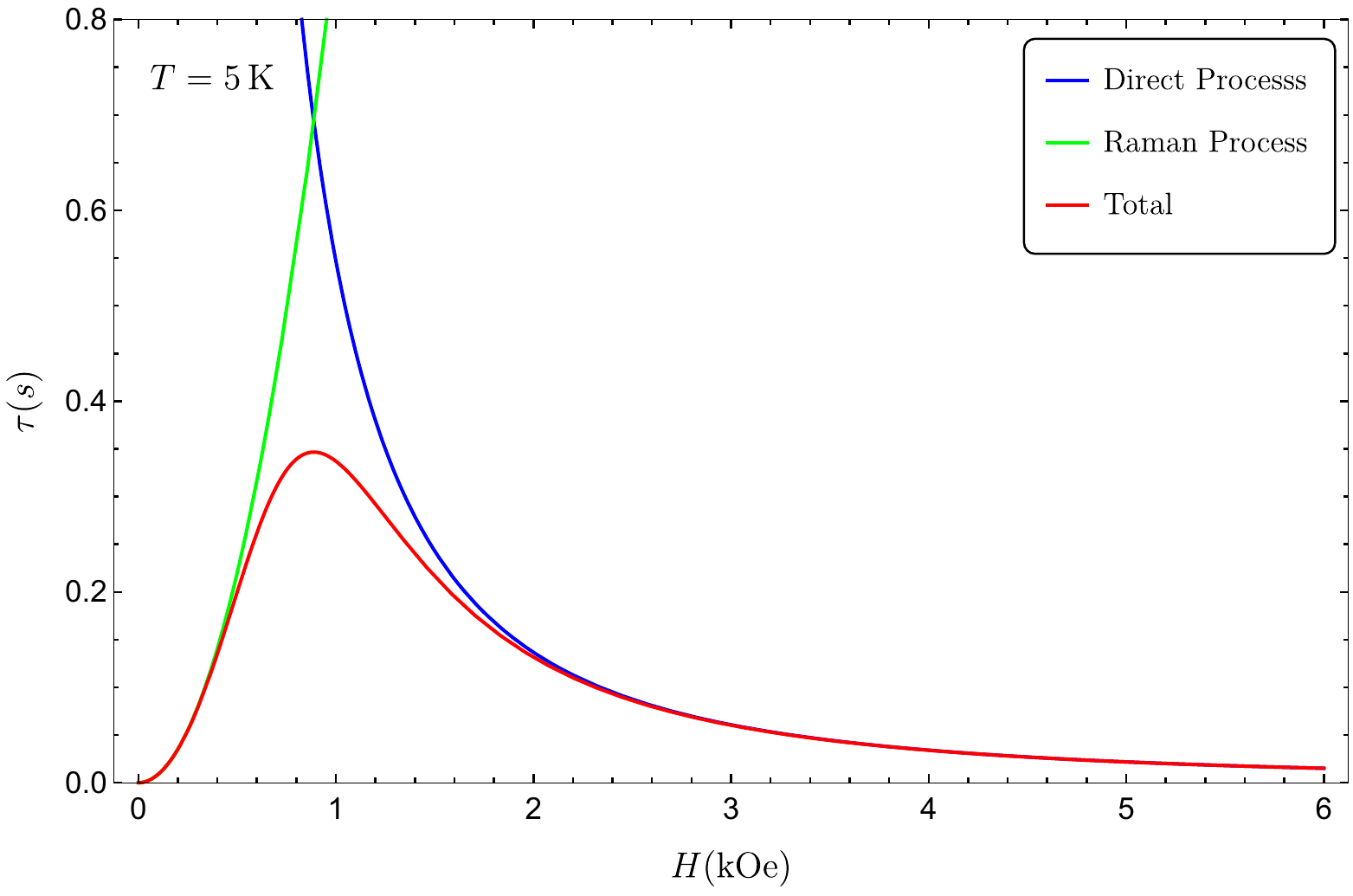}}\tabularnewline
\end{tabular}

\caption{Non-Kramers system: (a) inverse temperature dependence of the relaxation times $\tau_{\mathrm{direct/Raman/total}}$ at 1 kOe; (b) magnetic field dependence of $\Gamma_{-+}^{\mathrm{Raman}}/\Gamma_{-+}^{\mathrm{dr}}$ at $T=3,5,$ and 7 K; (c) Magnetic field dependence of $\tau_{\mathrm{direct/Raman/total}}$ at $T=5$ K.\label{fig:non-Kramers rates comparison}}
\end{figure}

In analogy to Fig. \ref{fig:Kramers rates comparison}, Fig. \ref{fig:non-Kramers rates comparison} shows the field and temperature dependence of the relaxation times $\tau_{\mathrm{direct}}=\left(2\Gamma_{-+}^{\mathrm{dr}}\right)^{-1}$, $\tau_{\mathrm{Raman}}=\left(2\Gamma_{-+}^{\mathrm{Raman}}\right)^{-1}$, and $\tau_{\mathrm{total}}=\left[2\left(\Gamma_{-+}^{\mathrm{dr}}+\Gamma_{-+}^{\mathrm{Raman}}\right)\right]^{-1}$.  As seen in Fig. \ref{fig:non-Kramers rates comparison}a, the switch from one dominant process to another also takes place sharply in a narrow range of temperatures as in the case of Kramers doublets. This also results from the different $T$-dependence of the two relaxation processes. However, in contrast to Kramers systems, the total relaxation time shows an extremum in its $H$-dependence, which is due to the fact that the Raman process transition rate decreases fast with the field. This is not surprising given that the Raman transition rate depends on the ratio $\Delta^{2}/\left(\Delta^{2}+W^{2}\right)$.

Given the typical set of parameters used here, it is likely that for Kramers SMMs, the Raman process will dominate the direct one at temperature $T>5\,\mathrm{K}$ under not too strong magnetic field, $H<2\,\mathrm{kOe}$. For non-Kramers system, it is probable that at $H>1$ kOe and low temperature ($T<5$ K), the direct process takes over the Raman process. However, at temperature $T>10\,\mathrm{K}$ under $H<2\,\mathrm{kOe}$, the Raman process likely suppresses the direct one.

\section{Application: relaxation in a $\mathrm{Co^{2+}}$ complex\label{sec:Theory-application}}

\begin{figure}
\begin{tabular}{ll}
{\footnotesize{}(a)} & {\footnotesize{}(b)}\tabularnewline
\includegraphics[width=0.45\textwidth]{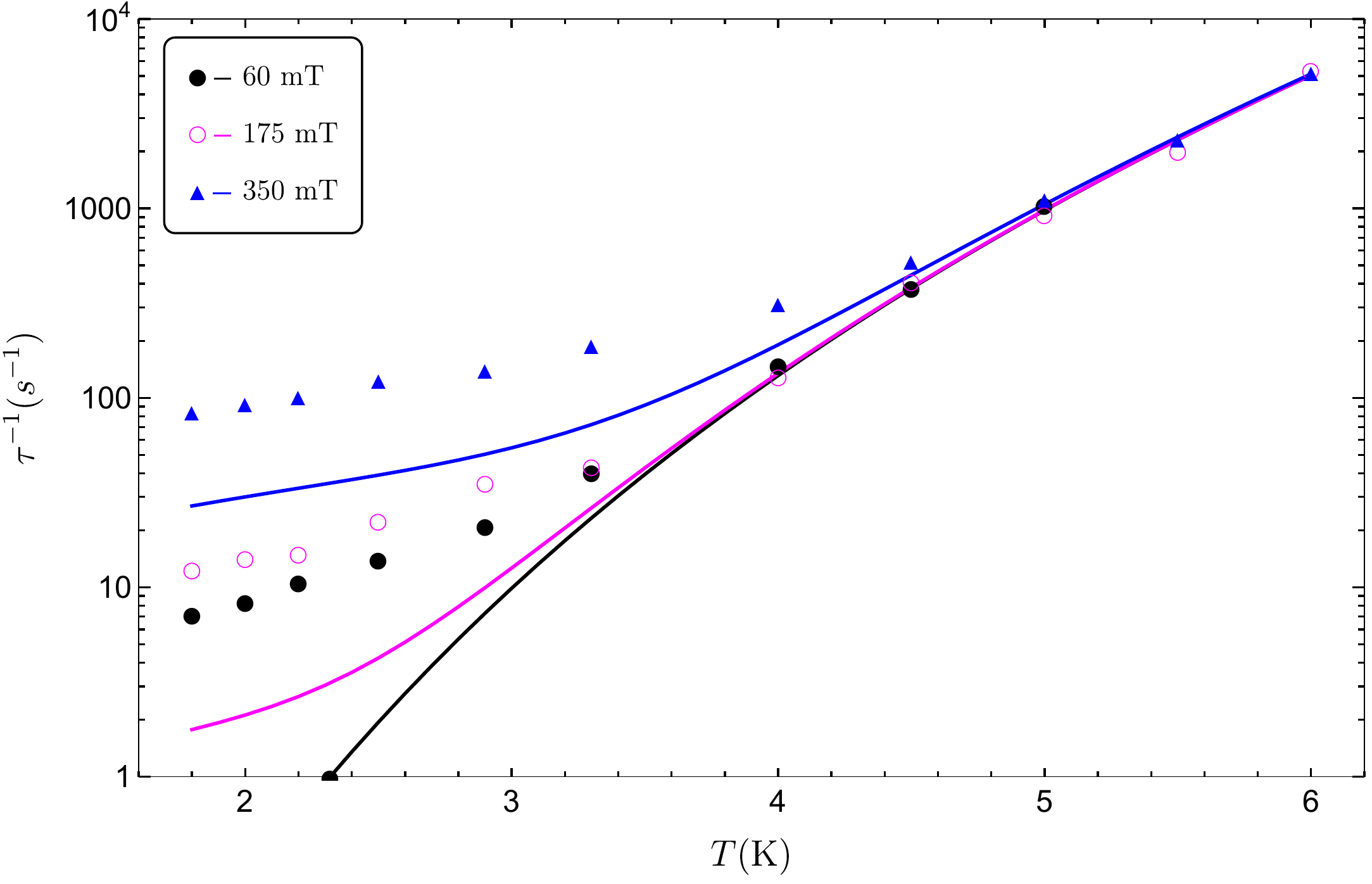} & \includegraphics[width=0.45\textwidth]{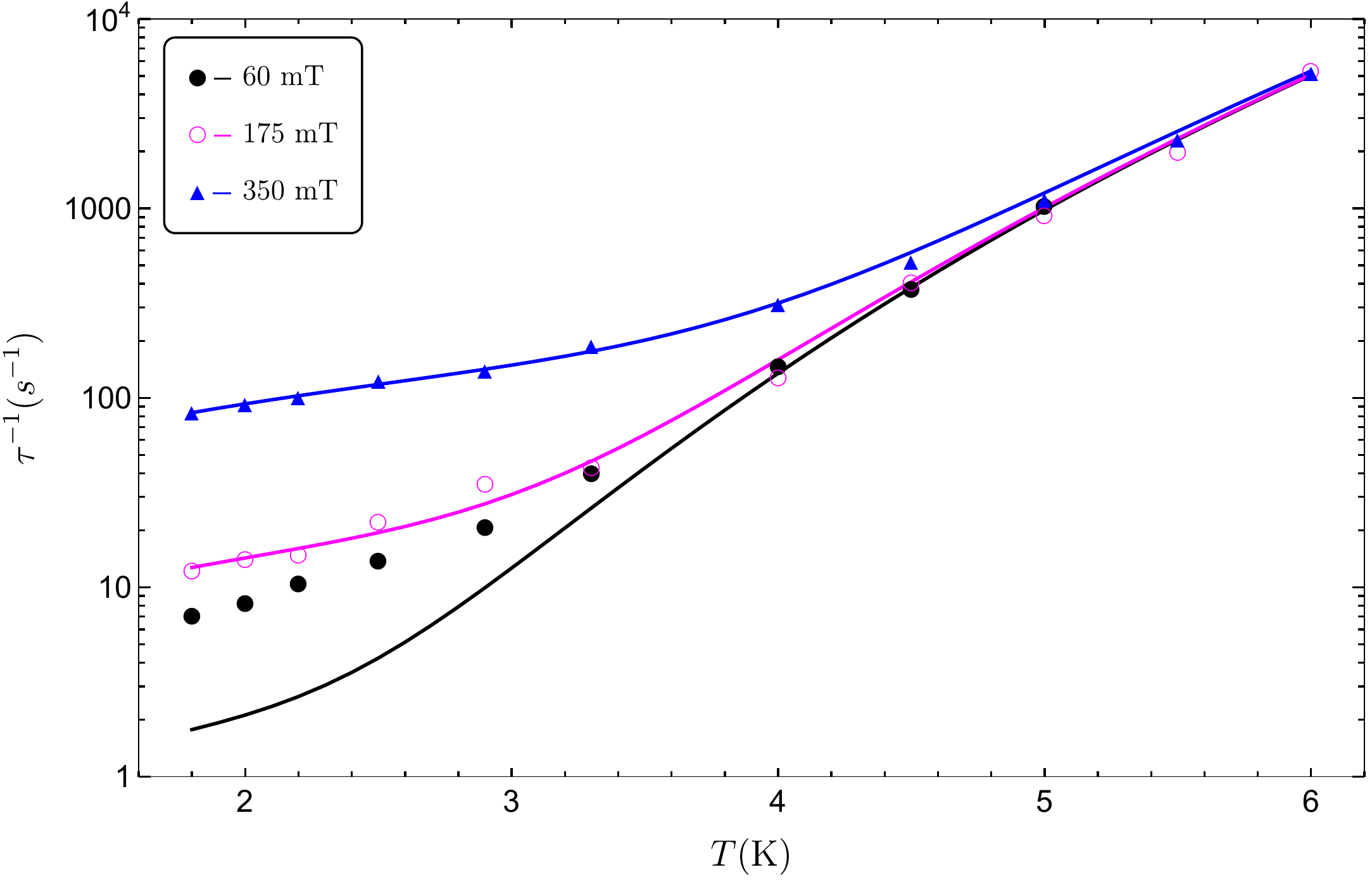}\tabularnewline
\end{tabular}

\caption{Temperature dependence of the total relaxation rate in a $\mathrm{Co}^{2+}$ complex \cite{Gomez-Coca2014}. (a) Calculated data (solid lines) from the present expressions without an additional mean internal field vs. experiment (data points) taken from Ref. {[}\onlinecite{Gomez-Coca2014}{]}; (b) The same as (a) but with an internal field of 115 mT.\label{fig:real_experiment_comparison}}
\end{figure}

As an example of applications to real compounds, in Fig. \ref{fig:real_experiment_comparison}a we show the calculated total relaxation rate with the present expressions and the comparison to experimental data for a cobalt(II) complex\cite{Gomez-Coca2014} showing slow relaxation \footnote{In cases when optical phonons contribute significantly to the spin-lattice relaxation, the Raman relaxation rate for Kramers system may deviate from the $T^{9}$ dependence \cite{Shrivastava1983}. Therefore, in order to have an accurate assessment on the applicability of the rotational spin-phonon approximation, only those experiments which satisfies the following criteria are chosen: (1) Raman process shows a $T^{9}$ dependence, which thus excludes the role of optical phonon; (2) a small number of parameters in the expression of the total relaxation rate, which thus prevents over-parameterization; (3) data and fittings are of high-quality. Ref. {[}\onlinecite{Gomez-Coca2014}{]} satisfies all these criteria.}. The cobalt(II) ion possesses a ground state spin $S=3/2$ with an almost isotropic $g$-factor $g\approx2.5$; the principal values of the ground doublet's $g$-tensor $g_{x}=2.65$, $g_{y}=6.95$, $g_{z}=1.83$, and the mass density $\rho=1528\,\mathrm{kg/m^{3}}$. Since the experiment was conducted on polycrystalline samples, for a simple account of this in the calculation, the field was oriented at an angle $\theta_{\vt H}=\phi_{\vt H}=\pi/4$ w.r.t anisotropic axes $z$ and $x$. A fitting with Eq. \ref{eq:general Kramers doublet Raman transition rate} for $\Gamma_{-+}^{\mathrm{Raman}}$ yields $\Gamma_{\mathrm{Raman}}\approx2\Gamma_{-+}^{\mathrm{Raman}}=5.0\times10^{-4}T^{9}$. The prefactor in this expression corresponds to the value of the parameter $E_{t}=96$ introduced in Sec. \ref{subsec:Direct Process vs. Raman Process in Kramers SMMs}. The latter corresponds to $\bar{v}_{t}\equiv v_{t_{1}}v_{t_{2}}/\sqrt[5]{\left(v_{t_{1}}^{5}+v_{t_{2}}^{5}\right)/2}\approx1.14\times10^{3}\,\mathrm{m/s}$, a value close to the speed of sound extracted from the heat capacity data, $c=2\times10^{3}$ m/s ($c$ is usually higher than the average transverse speed of sound $\bar{v}_{t}$). Using this value, at respectively 175 mT and 350 mT, our formula gives $\Gamma_{\mathrm{direct}}=0.9T$ and $14.9T$, which matches the order of magnitude of experimentally fitted dependencies, $\Gamma_{\mathrm{direct}}=7.2T$ and $47.1T$. In fact, this is a quite good agreement since the effect from other distortions is expected to contribute comparably to the rotational deformations. At 60 mT, our expression gives much smaller value, $\Gamma_{\mathrm{direct}}=0.013T$, than the experimentally fitted one $\Gamma_{\mathrm{direct}}=4.2T$. This discrepancy is easily understood given that at small values of applied field, the hyperfine/dipolar interactions and, accordingly, quantum tunneling effects influence considerably the total relaxation rate. 

Interestingly, an improvement to the calculated results is achieved by a simple addition of a mean internal magnetic field resulting from nuclear spin and dipolar magnetic moments of surroundings molecules. In Fig. \ref{fig:real_experiment_comparison}b, we present the computed total relaxation rate with an additional internal field of 115 mT. As can be seen, taking into account the internal field substantially boosts the overall agreement of the calculated results, $\Gamma_{\mathrm{direct}}=0.9T$, $7.0T$, and $46.3T$, with experimentally fitted $\Gamma_{\mathrm{direct}}=4.2T$, $7.2T$, and $47.1T$ at $H=60$ mT, 175 mT, and 350 mT, respectively. The worst agreement (but still close), as expected, is obtained for 60 mT. As already explained, this may result from the neglected effect of quantum tunneling as well as from the oversimplified mean field approximation, which does not account for the dynamics of the internal field. The latter, as well as the neglected effects from the distortions of the molecule, explains why such a high internal field, exceeding several times the expected one in molecules crystals \cite{Gatteschi2006}, is needed to obtain the the quantitative agreement.

\begin{figure}
\begin{tabular}{ll}
{\footnotesize{}(a)} & {\footnotesize{}(b)}\tabularnewline
\includegraphics[width=0.45\textwidth]{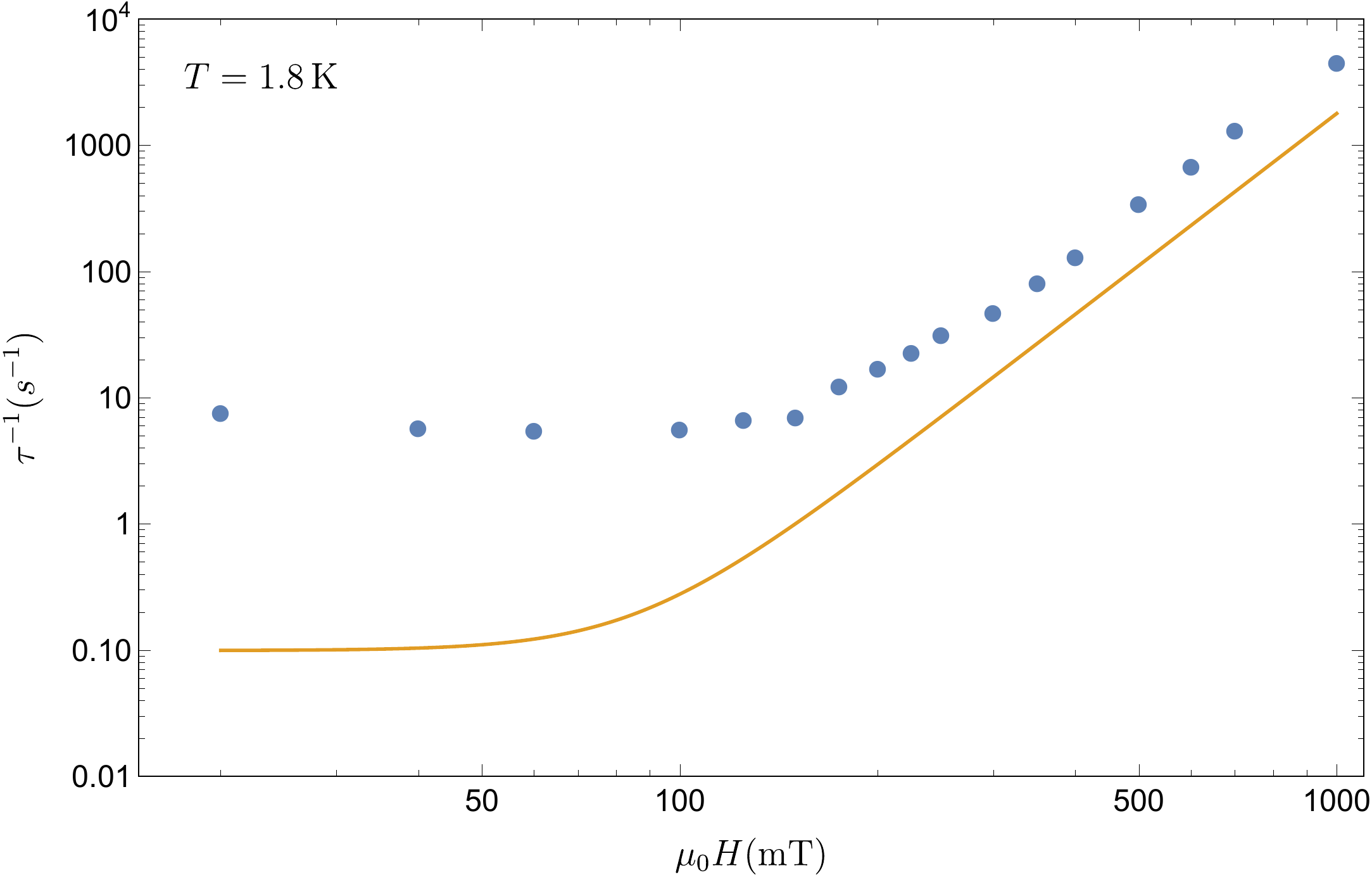} & \includegraphics[width=0.45\textwidth]{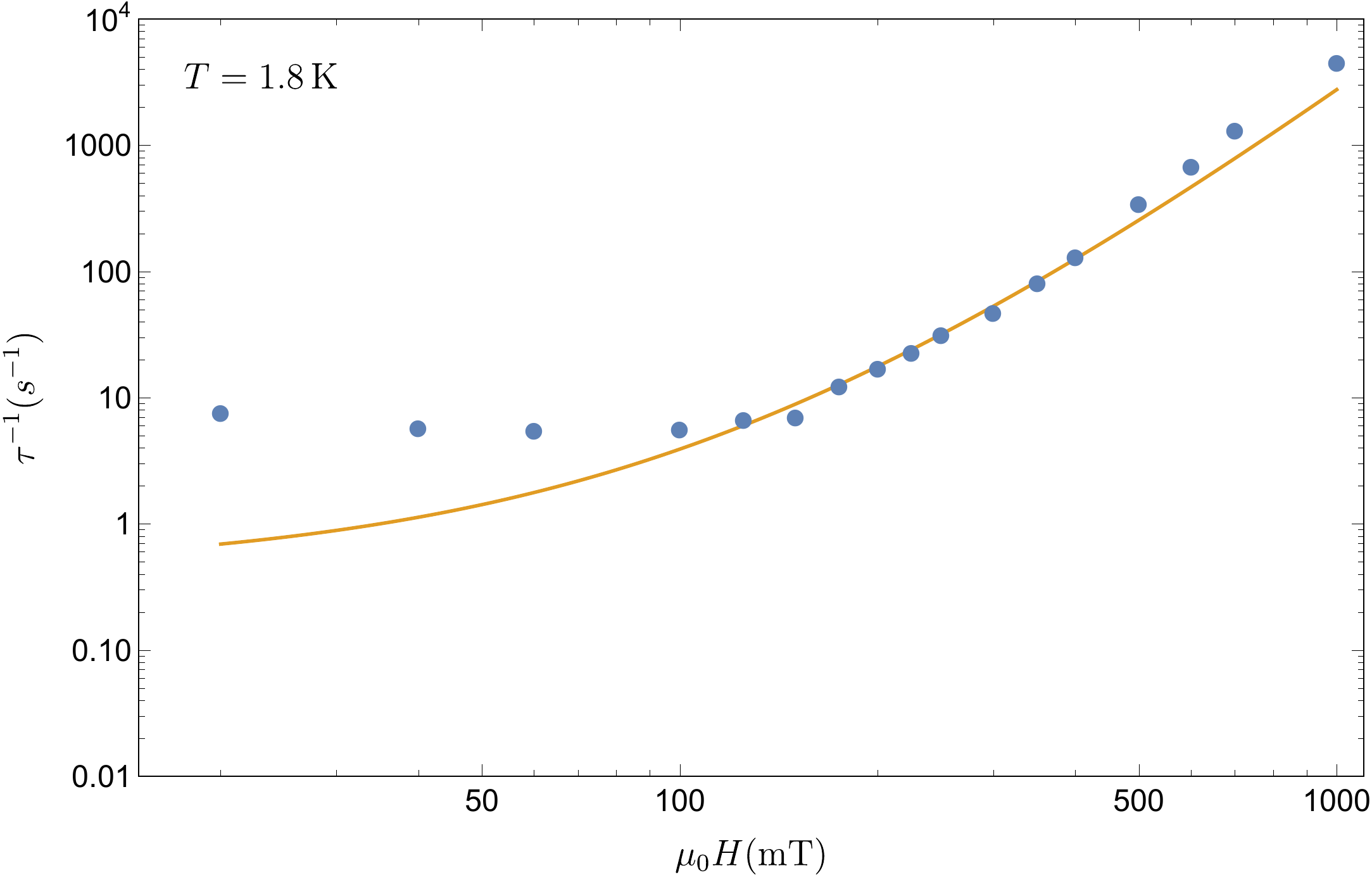}\tabularnewline
\end{tabular}

\caption{Field dependence of the total relaxation rate in a $\mathrm{Co}^{2+}$ SMM \cite{Gomez-Coca2014}. (a) Calculated rate (solid lines) in the absence of internal field vs. experiment (points) taken from Ref. {[}\onlinecite{Gomez-Coca2014}{]}; (b) The same as (a) but with an additional internal field of 115 mT.\label{fig:real_experiment_comparison-H}}
\end{figure}

In order to check the field dependence of the relaxation rate predicted by the present theory, in Fig. \ref{fig:real_experiment_comparison-H} we show this dependence calculated with above expressions vs. experimental data for the same parameters. Quite similarly, without an additional internal magnetic field, the theoretical results at large fields yield a good agreement with the experiment for the orders of magnitude but become worse at small fields. However, with an addition of an internal field of 115 mT , the agreement between theory and experiment becomes striking with only a little difference in the small-field region as expected from the exclusion of the quantum tunneling effects and due to the approximation of a static internal field.

We note that a rationalization of these data was attempted earlier \cite{Gomez-Coca2014} within a mechanism involving the interaction of the electronic $S=3/2$ with the nuclear spin of Co(II). The authors of Ref. {[}\onlinecite{Gomez-Coca2014}{]} advocated in addition the hyperfine-nuclear interaction with an unrealistically high coupling constant ($\alpha_{I}=0.05\,\mathrm{cm^{-1}}$). The obtained agreement with experimental data (Fig. 5b in {[}\onlinecite{Gomez-Coca2014}{]}) is less perfect than in Fig. \ref{fig:real_experiment_comparison-H}b, implying that the relaxation via acoustic phonons is probably more relevant for the presented compound. 

\section{Summary and discussions}

In summary, we have derived the following new expressions for the low-temperature relaxation rate in the rotation approximation for the spin-phonon coupling:
\begin{itemize}
\item The transition rate between two states $\ket{\pm}$ of a ground Kramers doublet
\begin{itemize}
\item The direct process: 
\begin{align}
\Gamma_{-+}^{\mathrm{dr}} & =\frac{1}{96\pi\hbar^{4}\rho}\left(\frac{1}{v_{t_{1}}^{5}}+\frac{1}{v_{t_{2}}^{5}}\right)\frac{\Omega e^{\Omega/k_{\mathrm{B}}T}}{e^{\Omega/k_{\mathrm{B}}T}-1}\nonumber \\
 & \times\left\{ W^{4}\frac{g_{1}^{2}+g_{2}^{2}}{g_{z}^{2}}\right.\nonumber \\
 & \qquad+W^{2}\Delta^{2}\left[\frac{g_{1}^{2}\Delta_{y}^{2}+g_{2}^{2}\Delta_{x}^{2}}{g_{z}^{2}\Delta^{2}}+2\frac{g_{3}}{g_{z}}\left(\frac{g_{1}}{g_{x}}\frac{\Delta_{x}^{2}}{\Delta^{2}}+\frac{g_{2}}{g_{y}}\frac{\Delta_{y}^{2}}{\Delta^{2}}\right)+\left(\frac{g_{1}}{g_{y}}\frac{\Delta_{y}^{2}}{\Delta^{2}}+\frac{g_{2}}{g_{x}}\frac{\Delta_{x}^{2}}{\Delta^{2}}\right)^{2}+\left(\frac{g_{x}}{g_{y}}-\frac{g_{y}}{g_{x}}\right)^{2}\frac{\Delta_{x}^{2}}{\Delta^{2}}\frac{\Delta_{y}^{2}}{\Delta^{2}}\right]\nonumber \\
 & \qquad\qquad\qquad\left.+\Delta^{4}\left[\left(\frac{g_{1}}{g_{y}}\frac{\Delta_{y}^{2}}{\Delta^{2}}+\frac{g_{2}}{g_{x}}\frac{\Delta_{x}^{2}}{\Delta^{2}}\right)^{2}+\left(\frac{g_{3}^{2}}{g_{x}^{2}}\frac{\Delta_{x}^{2}}{\Delta^{2}}+\frac{g_{3}^{2}}{g_{y}^{2}}\frac{\Delta_{y}^{2}}{\Delta^{2}}\right)\right]\right\} .\label{eq:General Kramers doublet direct transition rate-Summary}
\end{align}
\item The Raman process: 
\begin{align}
\Gamma_{-+}^{\mathrm{Raman}} & =\frac{\hbar^{2}I_{8}}{2304\pi^{3}\rho^{2}}\left(\frac{1}{v_{t_{1}}^{5}}+\frac{1}{v_{t_{2}}^{5}}\right)^{2}\left(\frac{k_{\mathrm{B}}T}{\hbar}\right)^{9}\left[\left(\frac{g_{1}^{2}}{g^{2}}\frac{\Delta_{y}^{2}}{\Delta^{2}}+\frac{g_{2}^{2}}{g^{2}}\frac{\Delta_{x}^{2}}{\Delta^{2}}\right)+\left(\frac{g_{1}^{2}}{g^{2}}\frac{\Delta_{x}^{2}}{\Delta^{2}}+\frac{g_{2}^{2}}{g^{2}}\frac{\Delta_{y}^{2}}{\Delta^{2}}\right)\frac{W^{2}}{\Omega^{2}}+\frac{g_{3}^{2}}{g^{2}}\frac{\Delta^{2}}{\Omega^{2}}\right].\label{eq:general Kramers doublet Raman transition rate - summary}
\end{align}
\end{itemize}
\item The transition rate between two tunnel-split states $\ket{\pm}$ of a ground non-Kramers doublet
\begin{itemize}
\item The direct process: 
\begin{align}
\Gamma_{-+}^{\mathrm{dr}} & =\frac{1}{96\pi\hbar^{4}\rho}\left(\frac{1}{v_{t_{1}}^{5}}+\frac{1}{v_{t_{2}}^{5}}\right)\frac{\Omega e^{\Omega/k_{\mathrm{B}}T}}{e^{\Omega/k_{\mathrm{B}}T}-1}\frac{g_{z}^{2}}{g^{2}}\Delta^{2}\left[\Omega^{2}+g^{2}H_{\perp}^{2}\right].\label{eq:General non-Kramers doublet direct transition rate-Summary}
\end{align}
\item The Raman process: 
\begin{align}
\Gamma_{-+}^{\mathrm{Raman}} & =\frac{\hbar^{2}I_{8}}{2304\pi^{3}\rho^{2}}\left(\frac{1}{v_{t_{1}}^{5}}+\frac{1}{v_{t_{2}}^{5}}\right)^{2}\left(\frac{k_{\mathrm{B}}T}{\hbar}\right)^{9}\frac{g_{z}^{2}}{g^{2}}\frac{\Delta^{2}}{\Delta^{2}+W^{2}}.\label{eq:general non-Kramers doublet Raman transition rate - summary}
\end{align}
\end{itemize}
\end{itemize}
In the case of Kramers doublets, all formulae are expressed via the $g$-factor of the ground doublet, $g_{\alpha},\alpha=x,y,z$, and the Zeeman splittings for the field applied in the corresponding direction, $W$ and $\Delta_{x,y}$. Eqs. \eqref{eq:General Kramers doublet direct transition rate-Summary} and \eqref{eq:general Kramers doublet Raman transition rate - summary} deliberately emphasize the $z$-direction, along the main $g$-factor, having in mind application to strongly axial magnetic centers ($g_{z}\gg g_{x},g_{y}$), otherwise being completely symmetric w.r.t $g_{\alpha}$ and $\Delta_{\alpha}$. In this way, the fourth-power dependence on the Zeeman splitting of the relaxation rate for a direct process \cite{Abragam1970} appears in Eq. \eqref{eq:General Kramers doublet direct transition rate-Summary} in the form of three contributions involving the ``bias'' Zeeman splitting $W=g_{z}H_{z}$and the Zeeman ``tunneling gap'' $\Delta=\sqrt{\Delta_{x}^{2}+\Delta_{y}^{2}}.$ In the case of non-Kramers doublets, Eqs. \eqref{eq:General non-Kramers doublet direct transition rate-Summary} and \eqref{eq:general non-Kramers doublet Raman transition rate - summary}, $W$ is the only Zeeman splitting allowed by the Griffith theorem \cite{Griffith1971} and $\Delta$ is the intrinsic tunneling gap. Certainly, all results are independent of our choice of basis as physically expected.

These expressions have been derived under the assumption: $\Omega\ll\hbar\omega_{\vt k\lambda_{\vt k}}\ll\hbar\omega_{\xi1}$ where $\xi$ denotes excited doublets. The remarkable feature of these expressions is their \emph{universal} form for both Kramers and non-Kramers system. Furthermore, these universal formulae are entirely expressed via measurable or \emph{ab initio} computable physical quantities, which facilitates their use.

It is worthy to note that although these results were derived supposing that the $g$-tensor of the multiplet is isotropic, an extension to a general $g$-tensor is trivial. In fact, the main difference now is in the matrix representation of the operator $S_{\alpha}$ (or $J_{\alpha}$), $\alpha=x,y,z$, in the sub-basis of the ground doublet. In particular, for Kramers system, this can be found by simply solving the system of equation $\mu_{\alpha}=-\frac{1}{2}g_{\alpha}\sigma_{\alpha}=-\sum_{\beta}g_{\alpha\beta}S_{\beta}$ where $g_{\alpha}$ are the principal values of the ground doublet's $g$-tensor and $g_{\alpha\beta}$ are component of the multiplet's $g$-tensor in the chosen reference frame. Notice that a new derivation also requires a change in the Zeeman Hamiltonian to $\hmt_{Z}=\sum_{\alpha,\beta}H_{\alpha}g_{\alpha\beta}S_{\beta}$ \footnote{We consider multiplets $S$ or $J$ not far from isotropic case, when the Zeeman operator contains the first-rank contribution only \cite{Chibotaru2013}. This is by far the most common case.}and accordingly the use of new expression $\left[\hmt_{Z},\vt S\right]=i\left(\vt S\times\vt H'\right)$ where $H_{\alpha}'=\sum_{\beta}H_{\beta}g_{\beta\alpha}$. In practice, SMMs are usually characterized by a spin $S$ ($J$) with an almost isotropic $g$-tensor. Therefore, for the sake of simplicity we do not show here the results for this general case.

The total relaxation rate is then simply, 
\begin{equation}
\Gamma_{\mathrm{relax}}=\left(\Gamma_{-+}^{\mathrm{dr}}+\Gamma_{-+}^{\mathrm{Raman}}\right)\left(1+e^{-\Omega/k_{\mathrm{B}}T}\right).
\end{equation}
Since the total relaxation rate was derived in the eigenstates basis, the phonon-induced quantum tunneling effect has been included from the start \cite{Ho2017b}. Hyperfine and dipolar interaction have been excluded from the present treatment, so the obtained results are strictly applicable to diluted samples.

The obtained expressions for direct process transition rate can be applied to tunnel-split states of any excited doublet at any temperature as well. This is because the derivation of the direct process transition rate for both Kramers and non-Kramers doublets was in fact not restricted by low temperature. However, the same cannot be said about the Raman process since its derivation was essentially done under the condition $\hbar\omega_{\vt k\lambda_{\vt k}}\ll\hbar\omega_{\xi1}\,\forall\ket{\xi}\ne\ket{\pm}$, which is typically satisfied at low temperature but generally not fulfilled at high temperature.

Within the rotational approximation to the spin-phonon coupling, one of the notable findings about the Raman transition rates is that the first-order process is always dominated by the second-order one for both Kramers and non-Kramers systems. Furthermore, whereas the Raman process in a Kramers system, as usual, is subject to the $T^{9}$ dependence emerging from the degeneracy of the doublet the Raman process in a non-Kramers system does not show the expected $T^{7}$ dependence but the $T^{9}$ one. Our derivation in fact demonstrates that this property results from the destructive interference of the main contributions to the  $T^{7}$ dependence of the first-order and second-order Raman process, which are together left with the $T^{9}$ dependent terms only.

Moreover, contrary to the traditional belief that the Raman process in non-Kramers system is independent of the field \cite{Abragam1970}, we reveal that in fact, at least within the contribution from rotation spin-phonon coupling, the Raman process is strongly dependent on the applied field. Furthermore, it behaves in a similar manner as the incoherent quantum tunneling rate in the sense that it is significantly reduced when the system is out of resonance \cite{Gatteschi2006,Prokof'ev1998,Prokof'ev2000,Vijayaraghavan2009,Garanin1997,Leuenberger2000}. Not less interesting, for Kramers rigid molecules, the Raman process only depends on the magnetic field orientation but independent of its magnitude. Consequently, the Raman process keeps unchanged under a variation of the magnetic field magnitude. These properties, luckily, are easy verifiable in crystals of rigid SMMs.  It should be noted that these results are only valid in the rotational approximation for the spin-phonon coupling. Although this approximation is sufficient to describe qualitatively the temperature and field dependence of the relaxation rate in the low-$T$ domain, to achieve a quantitative level of description, the contribution of intramolecular distortions, of the nuclear spins and dipolar interactions should be included. In some situation, when the latter become important, they can modify qualitatively the predictions of the present model. For instance, the field dependence of the relaxation rate for the Kramers doublets, predicted here as a monotonic function (Fig. \ref{fig:Kramers rates comparison}c), does show a maximum in the Er-trensal complex \cite{Pedersen2013}, due to an important contribution from the incoherent tunneling induced by nuclear spin and dipolar interactions

Regarding the question of the temperature dependence of the total spin-lattice relaxation rate, the present study shows that the spin-phonon interaction at low temperature always gives rise to a \emph{$T$- }dependent relaxation. This is physically obvious given the fact that the phonon density varies strongly with energy.  As a result, our finding rules out the common belief of a $T$-independent phonon-induced quantum tunneling rate at low temperature and zero field, and hence leaves the explanation of $T$-independence of the relaxation rate observed experimentally to other relaxation mechanisms, such as hyperfine \cite{Prokofev1995,Prokof'ev2000,Garanin1999} or dipolar interaction \cite{Prokof'ev1998,Fernandez2003,Cuccoli1999}.

A new strategy in designing SMMs with long relaxation times may also be drawn from our estimation on the direct and Raman process relaxation rates shown in Sec. \ref{sec:Direct processs vs. Raman process}. Practically, it appears that the relaxation rate caused by the rotational deformations is quite small, even when the system's anisotropy is not high in the case of Kramers system, or the intrinsic tunneling splitting is considerable in the case of non-Kramers system. This observation thus suggests that besides focusing on systems with strong anisotropy, it would be equally efficient to suppress the vibrational deformations by devising \emph{mechanically rigid} SMMs, provided the quantum tunneling caused by hyperfine/dipolar interaction is suppressed. 

In conclusion, we have derived universal ``parameter-free'' analytical expressions for the spin-phonon relaxation rate at low temperature for crystals of rigid molecules (both Kramers and non-Kramers) with arbitrary ZFS or crystal-field Hamiltonian, using a rotational approximation for spin-phonon interaction. These expressions offer a handy and quick estimation of the relaxation rate in arbitrary systems. Since the rotation deformation is always present in spin-phonon interaction, the relaxation rates found here set a lower bound on the total relaxation rates. 
\begin{acknowledgments}
L. T. A. H. is grateful to Dr. N. Iwahara and Dr. L. Ungur for helpful discussions and would like to acknowledge financial support from the Flemish Science Foundation (FWO). 
\end{acknowledgments}

\appendix

\section{Averaging relaxation process rates over phonon modes \label{sec:Averaging-phonon-modes}}

\subsection{Direct process}

We start by simplifying the expression \eqref{eq:general direct process rate} for the direct process transition rate $\Gamma_{-+}^{\mathrm{dr}}$ using the Debye dispersion relation and the fact that $\left[\vt k\times\vt e_{\vt kt}\right]=\pm k\vt e_{\vt kt'}$, where $t,t'$ can be $t_{1}$ or $t_{2}$, to obtain 
\begin{align}
\Gamma_{-+}^{\mathrm{dr}} & =\frac{\pi}{4\rho V\hbar}\sum_{t=t_{1},t_{2}}\sum_{\vt k}\sum_{\alpha,\beta}\Xi_{\alpha}\Xi_{\beta}^{*}e_{\vt kt,\alpha}e_{\vt kt,\beta}\frac{\omega_{\vt kt}}{v_{t}^{2}}\frac{e^{\Omega/k_{\mathrm{B}}T}}{e^{\Omega/k_{\mathrm{B}}T}-1}\delta\left(\Omega/\hbar-\omega_{\vt kt}\right).
\end{align}

Replacing $\sum_{\vt k}$ by $\frac{V}{8\pi^{3}}\int d^{3}k$, then substituting $k=\omega_{\vt kt}/v_{t}$, gives rise to 
\begin{align}
\Gamma_{-+}^{\mathrm{dr}} & =\frac{1}{32\pi^{2}\rho\hbar}\sum_{t=t_{1},t_{2}}\int\mathrm{d}\varphi_{\vt k}\int\sin\theta_{\vt k}\,\mathrm{d}\theta_{\vt k}\int\mathrm{d}\omega_{kt}\frac{\omega_{kt}^{3}}{v_{t}^{5}}\left(\frac{1}{e^{\Omega/k_{\mathrm{B}}T}-1}+1\right)\delta\left(\Omega/\hbar-\omega_{kt}\right)\sum_{\alpha,\beta}\Xi_{\alpha}\Xi_{\beta}^{*}e_{\vt kt,\alpha}e_{\vt kt,\beta}\nonumber \\
 & =\frac{1}{32\pi^{2}\hbar^{4}\rho}\frac{\Omega^{3}e^{\Omega/k_{\mathrm{B}}T}}{e^{\Omega/k_{\mathrm{B}}T}-1}\sum_{t=t_{1},t_{2}}\left[\frac{1}{v_{t}^{5}}\int\mathrm{d}\varphi_{\vt k}\int\sin\theta_{\vt k}\,\mathrm{d}\theta_{\vt k}\sum_{\alpha,\beta}\Xi_{\alpha}\Xi_{\beta}^{*}e_{\vt kt,\alpha}e_{\vt kt,\beta}\right].\label{appendix:eq-direct transition rate}
\end{align}

The term inside the square bracket is the average of  $\left(\vt{\Xi}\cdot\vt e_{\vt kt}\right)\left(\vt{\Xi}\cdot\vt e_{\vt kt}\right)^{*}$ over the unit $k-$sphere. Given the uniform distribution of the polarization vector $\vt e_{\vt kt_{1}}$ ($\vt e_{\vt kt_{2}}$) within the $k$-sphere, the integration over the spherical coordinates $\left(\theta_{\vt k},\varphi_{\vt k}\right)$ of $\vt k$ can be changed to the spherical coordinates of vector $\vt e_{\vt kt_{1}}$ ($\vt e_{\vt kt_{2}}$) itself. The above integral then transforms into 
\begin{equation}
\int\mathrm{d}\varphi_{\vt k}\int\sin\theta_{\vt k}\,\mathrm{d}\theta_{\vt k}\sum_{\alpha,\beta}\Xi_{\alpha}\Xi_{\beta}^{*}e_{\vt kt,\alpha}e_{\vt kt,\beta}=4\pi\sum_{\alpha,\beta}\Xi_{\alpha}\Xi_{\beta}^{*}\left\langle n_{\alpha}n_{\beta}\right\rangle _{\mathrm{sphere}}=\frac{4\pi}{3}\left|\vt{\Xi}\right|^{2},
\end{equation}
\label{appendix:eq - average2} where $\vt n$ is a normal unit vector on a sphere's surface. Substituting Eq. \eqref{appendix:eq - average2} into Eq. \eqref{appendix:eq-direct transition rate} yield Eq. \eqref{eq:General direct process rate}.

\subsection{Raman process}

Starting with the form \eqref{eq:Kramers-Raman-rate-form} for $\Gamma_{-+}^{\mathrm{Raman}}$, valid for both Kramers and non-Kramers systems, averaging over all transversal phonons modes for this form can be done by first taking average over the phonon numbers $n_{\vt k\lambda_{\vt k}}$ and $n_{\vt q\lambda_{\vt q}}$, then substituting the Eq. \eqref{eq:M_ph_alpha_beta} for $M_{\mathrm{ph}}^{\alpha\beta}$ into Eq. \eqref{eq:Kramers-Raman-rate-form} gives: 
\begin{align}
\Gamma_{-+}^{\mathrm{Raman}} & =\frac{\pi\hbar^{2}}{32\rho^{2}V^{2}}\sum_{\substack{\alpha,\beta\\
\alpha',\beta'
}
}\sum_{\substack{\vt k\lambda_{\vt k}\\
\vt q\lambda_{\vt q}
}
}Q_{\alpha\beta}Q_{\alpha'\beta'}^{*}\frac{\left(\vt q\times\vt e_{\vt q\lambda_{\vt q}}\right)_{\alpha}\left(\vt q\times\vt e_{\vt q\lambda_{\vt q}}\right)_{\alpha'}\left(\vt k\times\vt e_{\vt k\lambda_{\vt k}}\right)_{\beta}\left(\vt k\times\vt e_{\vt k\lambda_{\vt k}}\right)_{\beta'}}{\omega_{\vt q\lambda_{\vt q}}\omega_{\vt k\lambda_{\vt k}}}\nonumber \\
 & \qquad\qquad\qquad\qquad\qquad\times\left(\left\langle n_{\vt q\lambda_{\vt q}}\right\rangle +1\right)\left\langle n_{\vt k\lambda_{\vt k}}\right\rangle \omega_{\vt k\lambda_{\vt k}}^{2}\delta\left(\omega_{\vt k\lambda_{\vt k}}-\omega_{\vt q\lambda_{\vt q}}\right).
\end{align}

In analogy to the direct process, Eq. \eqref{appendix:eq-direct transition rate}, using the Debye dispersion relation and the property of transverse phonons $\left[\vt k\times\vt e_{\vt kt}\right]=\pm k\vt e_{\vt kt'}$, we have 
\begin{align*}
\Gamma_{-+}^{\mathrm{Raman}} & =\frac{\pi\hbar^{2}}{32\rho^{2}V^{2}}\sum_{\substack{t=t_{1},t_{2}\\
t'=t_{1},t_{2}
}
}\sum_{\vt k,\vt q}\sum_{\substack{\alpha,\beta\\
\alpha',\beta'
}
}Q_{\alpha\beta}Q_{\alpha'\beta'}^{*}\frac{e_{\vt qt,\alpha}e_{\vt qt,\alpha'}}{v_{t}^{2}}\frac{e_{\vt kt',\beta}e_{\vt kt',\beta'}}{v_{t'}^{2}}\left(\left\langle n_{\vt kt'}\right\rangle +1\right)\left\langle n_{\vt kt'}\right\rangle \omega_{\vt kt'}^{4}\delta\left(\omega_{\vt kt'}-\omega_{\vt qt}\right)\\
 & =\frac{\hbar^{2}}{2048\pi^{5}\rho^{2}}\sum_{\substack{t=t_{1},t_{2}\\
t'=t_{1},t_{2}
}
}\int\mathrm{d}\varphi_{\vt k}\,\mathrm{d}\theta_{\vt k}\,\sin\theta_{\vt k}\int\mathrm{d}\varphi_{\vt q}\,\mathrm{d}\theta_{\vt q}\,\sin\theta_{\vt q}\\
 & \qquad\times\int\mathrm{d}\omega_{\vt kt'}\,\mathrm{d}\omega_{\vt qt}\,\sum_{\substack{\alpha,\beta\\
\alpha',\beta'
}
}Q_{\alpha\beta}Q_{\alpha'\beta'}^{*}\frac{e_{\vt qt,\alpha}e_{\vt qt,\alpha'}}{v_{t}^{5}}\frac{e_{\vt kt',\beta}e_{\vt kt',\beta'}}{v_{t'}^{5}}\left(\left\langle n_{\vt kt'}\right\rangle +1\right)\left\langle n_{\vt kt'}\right\rangle \omega_{\vt kt'}^{8}\delta\left(\omega_{\vt kt'}-\omega_{\vt qt}\right)\\
 & =\frac{\hbar^{2}}{2048\pi^{5}\rho^{2}}\int_{0}^{\omega_{D}}\mathrm{d}\omega_{\vt kt'}\,\frac{e^{\hbar\omega_{\vt kt'}/k_{\mathrm{B}}T}}{\left(e^{\hbar\omega_{\vt kt'}/k_{\mathrm{B}}T}-1\right)^{2}}\omega_{\vt kt'}^{8}\\
 & \qquad\times\sum_{\substack{t=t_{1},t_{2}\\
t'=t_{1},t_{2}
}
}\sum_{\substack{\alpha,\beta\\
\alpha',\beta'
}
}Q_{\alpha\beta}Q_{\alpha'\beta'}^{*}\int\mathrm{d}\varphi_{\vt q}\,\mathrm{d}\theta_{\vt q}\sin\theta_{\vt q}\frac{e_{\vt qt,\alpha}e_{\vt qt,\alpha'}}{v_{t}^{5}}\int\mathrm{d}\varphi_{\vt k}\,\mathrm{d}\theta_{\vt k}\,\sin\theta_{\vt k}\frac{e_{\vt kt',\beta}ke_{\vt kt',\beta'}}{v_{t'}^{5}}\\
 & =\frac{\hbar^{2}}{1152\pi^{3}\rho^{2}}\left(\frac{k_{\mathrm{B}}T}{\hbar}\right)^{9}I_{8}\sum_{\substack{t=t_{1},t_{2}\\
t'=t_{1},t_{2}
}
}\sum_{\alpha,\beta}Q_{\alpha\beta}Q_{\alpha\beta}^{*}\frac{1}{v_{t}^{5}v_{t'}^{5}},
\end{align*}
 which is Eq. \eqref{eq:Raman process rate - first}.

\section{$\left(2S-1\right)\times\left(2S-1\right)$ submatrix diagonalization\label{sec: remainder submatrix diagonalization}}

Here we prove that a diagonalization of the submatrix of dimension $\left(2S-1\right)\times\left(2S-1\right)$ corresponding to the excited doublets $m^{\mathrm{th}}$ $\left(m\ne1\right)$ can be avoided for the perturbed states $\ket{+}$ and $\ket{-}$. Since $\hmt_{Z}$ is small, a diagonalization of this matrix will lead to the following correction for $\ket{m}$, 
\begin{align}
\ket{+_{m}} & =\left(0^{\mathrm{th}}\text{-order const }c_{m}\right)\ket{m}+\left(0^{\mathrm{th}}\text{-order const }c_{\bar{m}}\right)\ket{\bar{m}}+\sum_{M\ne1,\bar{1},m,\bar{m}}\ket{M}\left(1^{\mathrm{st}}\text{-order const }c_{M}\right)\nonumber \\
 & \equiv\ket{m'}+\sum_{M\ne1,\bar{1},m,\bar{m}}\ket{M}\left(1^{\mathrm{st}}\text{-order const }c_{M}\right),\label{eq:m-doubleprime}
\end{align}
and similarly for $\ket{-_{m}}$, where $\ket{\pm_{m}}$ are eigenstates of the $m^{\mathrm{th}}$ doublet in the presence of the field. Since only the zeroth order term plays a role in the expression for  $\ket{\pm}$ of the ground doublet, we can neglect the third term in the r.h.s. of Eq. \eqref{eq:m-doubleprime}. The corrections to the states $\ket{\pm}$ then become:
\begin{gather}
\ket{+}=\ket{p_{1}}+\sum_{m\ne1}\ket{p_{m}}\frac{\hmt_{p_{m}p_{1}}^{Z}}{\omega_{p_{1}p_{m}}}+\ket{m_{m}}\frac{\hmt_{m_{m}p_{1}}^{Z}}{\omega_{p_{1}m_{m}}}\approx\ket{p_{1}}+\sum_{m\ne1}\frac{\left(\ket{p_{m}}\bra{p_{m}}+\ket{m_{m}}\bra{m_{m}}\right)\hmt_{Z}\ket{p_{1}}}{\omega_{1m}},\\
\ket{-}=\ket{m_{1}}+\sum_{m\ne1}\ket{p_{m}}\frac{\hmt_{p_{m}m_{1}}^{Z}}{\omega_{m_{1}p_{m}}}+\ket{m_{m}}\frac{\hmt_{m_{m}m_{1}}^{Z}}{\omega_{m_{1}m_{m}}}\approx\ket{m_{1}}+\sum_{m\ne1}\frac{\left(\ket{p_{m}}\bra{p_{m}}+\ket{m_{m}}\bra{m_{m}}\right)\hmt_{Z}\ket{m_{1}}}{\omega_{1m}}.
\end{gather}

The diagonalization of the submatrix block $\left(2\times2\right)$ corresponding to the $m^{\mathrm{th}}$ doublet keeps the completeness of the basis. In other words, we have, 
\begin{equation}
\ket{p_{m}}\bra{p_{m}}+\ket{m_{m}}\bra{m_{m}}=\ket{m}\bra{m}+\ket{\bar{m}}\bra{\bar{m}}.
\end{equation}

$\ket{\pm}$ then become,

\begin{gather}
\ket{+}=\ket{p_{1}}+\sum_{M\ne1,\bar{1}}\ket{M}\frac{\hmt_{Mp_{1}}^{Z}}{\omega_{1M}},\\
\ket{-}=\ket{m_{1}}+\sum_{M\ne1,\bar{1}}\ket{M}\frac{\hmt_{Mm_{1}}^{Z}}{\omega_{1M}}.
\end{gather}

\section{Time-reversal symmetry and matrix elements of time-even/time-odd operators \label{sec: Time-even and time-odd operators matrix elements}}

The derivation of the relaxation rate in the main text has made use of the following relations between matrix elements imposed by the time reversal symmetry: 
\begin{itemize}
\item $\vt{\hmt}_{}^{\left(1\right)}\equiv i\left[\hmt_{A},\vt S\right]$ is a time-even operator. 
\item Time-even operators matrix elements relation in a Kramers system: $\hmt_{\alpha\bar{\beta}}=-\hmt_{\beta\bar{\alpha}}$ , $\hmt_{\bar{\alpha}\beta}=-\hmt_{\bar{\beta}\alpha}$, $\hmt_{\bar{\alpha}\bar{\beta}}=\hmt_{\beta\alpha}$, and $\hmt_{\bar{\alpha}\alpha}=\hmt_{\alpha\bar{\alpha}}=0$. 
\item Time-odd operators matrix elements relation in a Kramers system: $O_{\alpha\bar{\beta}}=O_{\beta\bar{\alpha}}$, $O_{\bar{\alpha}\beta}=O_{\bar{\beta}\alpha}$, $O_{\bar{\alpha}\bar{\beta}}=-O_{\beta\alpha}$, and $O_{\bar{\alpha}\bar{\alpha}}=-O_{\alpha\alpha}$. 
\item Time-even operators matrix elements relation in a non-Kramers system: $\hmt_{\alpha\bar{\beta}}=\hmt_{\beta\bar{\alpha}}$, $\hmt_{\bar{\alpha}\beta}=\hmt_{\bar{\beta}\alpha}$, $\hmt_{\bar{\alpha}\bar{\beta}}=\hmt_{\beta\alpha}$, and $\hmt_{\bar{\alpha}\bar{\alpha}}=\hmt_{\alpha\alpha}$. 
\item Time-odd operators matrix elements relation in a non-Kramers system: $O_{\alpha\bar{\beta}}=-O_{\beta\bar{\alpha}}$, $O_{\bar{\alpha}\beta}=-O_{\bar{\beta}\alpha}$, $O_{\bar{\alpha}\bar{\beta}}=-O_{\beta\alpha}$, $O_{\bar{\alpha}\bar{\alpha}}=-O_{\alpha\alpha}$, and $O_{\alpha\bar{\alpha}}=O_{\bar{\alpha}\alpha}=0$. 
\end{itemize}
These relations are easily proved by following the steps described in Chapter 15 of Abragam and Bleaney's book \cite{Abragam1970}.

\bibliographystyle{apsrev4-1}

\begin{thebibliography}{42}%
	\makeatletter
	\providecommand \@ifxundefined [1]{%
		\@ifx{#1\undefined}
	}%
	\providecommand \@ifnum [1]{%
		\ifnum #1\expandafter \@firstoftwo
		\else \expandafter \@secondoftwo
		\fi
	}%
	\providecommand \@ifx [1]{%
		\ifx #1\expandafter \@firstoftwo
		\else \expandafter \@secondoftwo
		\fi
	}%
	\providecommand \natexlab [1]{#1}%
	\providecommand \enquote  [1]{``#1''}%
	\providecommand \bibnamefont  [1]{#1}%
	\providecommand \bibfnamefont [1]{#1}%
	\providecommand \citenamefont [1]{#1}%
	\providecommand \href@noop [0]{\@secondoftwo}%
	\providecommand \href [0]{\begingroup \@sanitize@url \@href}%
	\providecommand \@href[1]{\@@startlink{#1}\@@href}%
	\providecommand \@@href[1]{\endgroup#1\@@endlink}%
	\providecommand \@sanitize@url [0]{\catcode `\\12\catcode `\$12\catcode
		`\&12\catcode `\#12\catcode `\^12\catcode `\_12\catcode `\%12\relax}%
	\providecommand \@@startlink[1]{}%
	\providecommand \@@endlink[0]{}%
	\providecommand \url  [0]{\begingroup\@sanitize@url \@url }%
	\providecommand \@url [1]{\endgroup\@href {#1}{\urlprefix }}%
	\providecommand \urlprefix  [0]{URL }%
	\providecommand \Eprint [0]{\href }%
	\providecommand \doibase [0]{http://dx.doi.org/}%
	\providecommand \selectlanguage [0]{\@gobble}%
	\providecommand \bibinfo  [0]{\@secondoftwo}%
	\providecommand \bibfield  [0]{\@secondoftwo}%
	\providecommand \translation [1]{[#1]}%
	\providecommand \BibitemOpen [0]{}%
	\providecommand \bibitemStop [0]{}%
	\providecommand \bibitemNoStop [0]{.\EOS\space}%
	\providecommand \EOS [0]{\spacefactor3000\relax}%
	\providecommand \BibitemShut  [1]{\csname bibitem#1\endcsname}%
	\let\auto@bib@innerbib\@empty
	%</preamble>
	\bibitem [{\citenamefont {Gatteschi}\ \emph {et~al.}(2006)\citenamefont
		{Gatteschi}, \citenamefont {Sessoli},\ and\ \citenamefont
		{Villain}}]{Gatteschi2006}%
	\BibitemOpen
	\bibfield  {author} {\bibinfo {author} {\bibfnamefont {D.}~\bibnamefont
			{Gatteschi}}, \bibinfo {author} {\bibfnamefont {R.}~\bibnamefont {Sessoli}},
		\ and\ \bibinfo {author} {\bibfnamefont {J.}~\bibnamefont {Villain}},\ }\href
	{\doibase 10.1093/acprof:oso/9780198567530.001.0001} {\emph {\bibinfo {title}
			{{Molecular Nanomagnets}}}}\ (\bibinfo  {publisher} {Oxford University
		Press},\ \bibinfo {year} {2006})\ pp.\ \bibinfo {pages} {1--408}\BibitemShut
	{NoStop}%
	\bibitem [{\citenamefont {Horodecki}\ \emph {et~al.}(2009)\citenamefont
		{Horodecki}, \citenamefont {Horodecki}, \citenamefont {Horodecki},\ and\
		\citenamefont {Horodecki}}]{Horodecki2009}%
	\BibitemOpen
	\bibfield  {author} {\bibinfo {author} {\bibfnamefont {R.}~\bibnamefont
			{Horodecki}}, \bibinfo {author} {\bibfnamefont {P.}~\bibnamefont
			{Horodecki}}, \bibinfo {author} {\bibfnamefont {M.}~\bibnamefont
			{Horodecki}}, \ and\ \bibinfo {author} {\bibfnamefont {K.}~\bibnamefont
			{Horodecki}},\ }\href {\doibase 10.1103/RevModPhys.81.865} {\bibfield
		{journal} {\bibinfo  {journal} {Rev. Mod. Phys.}\ }\textbf {\bibinfo {volume}
			{81}},\ \bibinfo {pages} {865} (\bibinfo {year} {2009})},\ \Eprint
	{http://arxiv.org/abs/0702225} {arXiv:0702225 [quant-ph]} \BibitemShut
	{NoStop}%
	\bibitem [{\citenamefont {Zurek}(2003)}]{Zurek2003}%
	\BibitemOpen
	\bibfield  {author} {\bibinfo {author} {\bibfnamefont {W.~H.}\ \bibnamefont
			{Zurek}},\ }\href {\doibase 10.1103/RevModPhys.75.715} {\bibfield  {journal}
		{\bibinfo  {journal} {Rev. Mod. Phys.}\ }\textbf {\bibinfo {volume} {75}},\
		\bibinfo {pages} {715} (\bibinfo {year} {2003})},\ \Eprint
	{http://arxiv.org/abs/0105127} {arXiv:0105127 [quant-ph]} \BibitemShut
	{NoStop}%
	\bibitem [{\citenamefont {Gatteschi}\ and\ \citenamefont
		{Sessoli}(2003)}]{Gatteschi2003}%
	\BibitemOpen
	\bibfield  {author} {\bibinfo {author} {\bibfnamefont {D.}~\bibnamefont
			{Gatteschi}}\ and\ \bibinfo {author} {\bibfnamefont {R.}~\bibnamefont
			{Sessoli}},\ }\href {\doibase 10.1002/anie.200390099} {\bibfield  {journal}
		{\bibinfo  {journal} {Angew. Chemie Int. Ed.}\ }\textbf {\bibinfo {volume}
			{42}},\ \bibinfo {pages} {268} (\bibinfo {year} {2003})}\BibitemShut
	{NoStop}%
	\bibitem [{\citenamefont {{Van Vleck}}(1940)}]{VanVleck1940}%
	\BibitemOpen
	\bibfield  {author} {\bibinfo {author} {\bibfnamefont {J.~H.}\ \bibnamefont
			{{Van Vleck}}},\ }\href {\doibase 10.1103/PhysRev.57.426} {\bibfield
		{journal} {\bibinfo  {journal} {Phys. Rev.}\ }\textbf {\bibinfo {volume}
			{57}},\ \bibinfo {pages} {426} (\bibinfo {year} {1940})}\BibitemShut
	{NoStop}%
	\bibitem [{\citenamefont {Abragam}\ and\ \citenamefont
		{Bleaney}(1970)}]{Abragam1970}%
	\BibitemOpen
	\bibfield  {author} {\bibinfo {author} {\bibfnamefont {A.}~\bibnamefont
			{Abragam}}\ and\ \bibinfo {author} {\bibfnamefont {B.}~\bibnamefont
			{Bleaney}},\ }\href@noop {} {\emph {\bibinfo {title} {{Electron paramagnetic
					resonance of transition ions}}}}\ (\bibinfo  {publisher} {Clarendon Press},\
	\bibinfo {year} {1970})\BibitemShut {NoStop}%
	\bibitem [{\citenamefont {Villain}\ \emph {et~al.}(1994)\citenamefont
		{Villain}, \citenamefont {Hartman-Boutron}, \citenamefont {Sessoli},\ and\
		\citenamefont {Rettori}}]{Villain1994}%
	\BibitemOpen
	\bibfield  {author} {\bibinfo {author} {\bibfnamefont {J.}~\bibnamefont
			{Villain}}, \bibinfo {author} {\bibfnamefont {F.}~\bibnamefont
			{Hartman-Boutron}}, \bibinfo {author} {\bibfnamefont {R.}~\bibnamefont
			{Sessoli}}, \ and\ \bibinfo {author} {\bibfnamefont {A.}~\bibnamefont
			{Rettori}},\ }\href {\doibase 10.1209/0295-5075/27/2/014} {\bibfield
		{journal} {\bibinfo  {journal} {Europhys. Lett.}\ }\textbf {\bibinfo {volume}
			{27}},\ \bibinfo {pages} {159} (\bibinfo {year} {1994})}\BibitemShut
	{NoStop}%
	\bibitem [{\citenamefont {Hartmann-Boutron}\ \emph {et~al.}(1996)\citenamefont
		{Hartmann-Boutron}, \citenamefont {Politi},\ and\ \citenamefont
		{Villain}}]{Hartmann-Boutron1996a}%
	\BibitemOpen
	\bibfield  {author} {\bibinfo {author} {\bibfnamefont {F.}~\bibnamefont
			{Hartmann-Boutron}}, \bibinfo {author} {\bibfnamefont {P.}~\bibnamefont
			{Politi}}, \ and\ \bibinfo {author} {\bibfnamefont {J.}~\bibnamefont
			{Villain}},\ }\href {\doibase 10.1142/S0217979296001148} {\bibfield
		{journal} {\bibinfo  {journal} {Int. J. Mod. Phys. B}\ }\textbf {\bibinfo
			{volume} {10}},\ \bibinfo {pages} {2577} (\bibinfo {year}
		{1996})}\BibitemShut {NoStop}%
	\bibitem [{\citenamefont {Garanin}\ and\ \citenamefont
		{Chudnovsky}(1997)}]{Garanin1997}%
	\BibitemOpen
	\bibfield  {author} {\bibinfo {author} {\bibfnamefont {D.~A.}\ \bibnamefont
			{Garanin}}\ and\ \bibinfo {author} {\bibfnamefont {E.~M.}\ \bibnamefont
			{Chudnovsky}},\ }\href {http://prb.aps.org/abstract/PRB/v56/i17/p11102{\_}1}
	{\bibfield  {journal} {\bibinfo  {journal} {Phys. Rev. B}\ }\textbf {\bibinfo
			{volume} {56}},\ \bibinfo {pages} {102} (\bibinfo {year} {1997})}\BibitemShut
	{NoStop}%
	\bibitem [{\citenamefont {Leuenberger}\ and\ \citenamefont
		{Loss}(2000)}]{Leuenberger2000}%
	\BibitemOpen
	\bibfield  {author} {\bibinfo {author} {\bibfnamefont {M.~N.}\ \bibnamefont
			{Leuenberger}}\ and\ \bibinfo {author} {\bibfnamefont {D.}~\bibnamefont
			{Loss}},\ }\href {\doibase 10.1103/PhysRevB.61.1286} {\bibfield  {journal}
		{\bibinfo  {journal} {Phys. Rev. B}\ }\textbf {\bibinfo {volume} {61}},\
		\bibinfo {pages} {1286} (\bibinfo {year} {2000})}\BibitemShut {NoStop}%
	\bibitem [{Note1()}]{Note1}%
	\BibitemOpen
	\bibinfo {note} {Besides modifying the ZFS parameters, the deformation also
		induces the rotation of the main anisotropy axes of the ZFS tensors described
		by three parameters (Euler angles) \cite {Chibotaru2013}}\BibitemShut
	{NoStop}%
	\bibitem [{\citenamefont {Chibotaru}(2013)}]{Chibotaru2013}%
	\BibitemOpen
	\bibfield  {author} {\bibinfo {author} {\bibfnamefont {L.~F.}\ \bibnamefont
			{Chibotaru}},\ }\href {\doibase 10.1002/9781118571767.ch6} {\bibfield
		{journal} {\bibinfo  {journal} {Adv. Chem. Phys.}\ }\textbf {\bibinfo
			{volume} {153}},\ \bibinfo {pages} {397} (\bibinfo {year}
		{2013})}\BibitemShut {NoStop}%
	\bibitem [{\citenamefont {Chudnovsky}\ \emph {et~al.}(2005)\citenamefont
		{Chudnovsky}, \citenamefont {Garanin},\ and\ \citenamefont
		{Schilling}}]{Chudnovsky2005}%
	\BibitemOpen
	\bibfield  {author} {\bibinfo {author} {\bibfnamefont {E.~M.}\ \bibnamefont
			{Chudnovsky}}, \bibinfo {author} {\bibfnamefont {D.~A.}\ \bibnamefont
			{Garanin}}, \ and\ \bibinfo {author} {\bibfnamefont {R.}~\bibnamefont
			{Schilling}},\ }\href {\doibase 10.1103/PhysRevB.72.094426} {\bibfield
		{journal} {\bibinfo  {journal} {Phys. Rev. B}\ }\textbf {\bibinfo {volume}
			{72}},\ \bibinfo {pages} {094426} (\bibinfo {year} {2005})}\BibitemShut
	{NoStop}%
	\bibitem [{\citenamefont {Layfield}\ and\ \citenamefont
		{Murugesu}(2015)}]{Layfield2015}%
	\BibitemOpen
	\bibfield  {author} {\bibinfo {author} {\bibfnamefont {R.~A.}\ \bibnamefont
			{Layfield}}\ and\ \bibinfo {author} {\bibfnamefont {M.}~\bibnamefont
			{Murugesu}},\ }\href {\doibase 10.1002/9783527673476} {\emph {\bibinfo
			{title} {{Lanthanides and Actinides in Molecular Magnetism}}}}\ (\bibinfo
	{publisher} {Wiley-VCH Verlag},\ \bibinfo {year} {2015})\ pp.\ \bibinfo
	{pages} {1--368}\BibitemShut {NoStop}%
	\bibitem [{\citenamefont {Woodruff}\ \emph {et~al.}(2013)\citenamefont
		{Woodruff}, \citenamefont {Winpenny},\ and\ \citenamefont
		{Layfield}}]{Woodruff2013}%
	\BibitemOpen
	\bibfield  {author} {\bibinfo {author} {\bibfnamefont {D.~N.}\ \bibnamefont
			{Woodruff}}, \bibinfo {author} {\bibfnamefont {R.~E.~P.}\ \bibnamefont
			{Winpenny}}, \ and\ \bibinfo {author} {\bibfnamefont {R.~A.}\ \bibnamefont
			{Layfield}},\ }\href {\doibase 10.1021/cr400018q} {\bibfield  {journal}
		{\bibinfo  {journal} {Chem. Rev.}\ }\textbf {\bibinfo {volume} {113}},\
		\bibinfo {pages} {5110} (\bibinfo {year} {2013})}\BibitemShut {NoStop}%
	\bibitem [{\citenamefont {Ishikawa}\ \emph {et~al.}(2003)\citenamefont
		{Ishikawa}, \citenamefont {Sugita}, \citenamefont {Ishikawa}, \citenamefont
		{Koshihara},\ and\ \citenamefont {Kaizu}}]{Ishikawa2003}%
	\BibitemOpen
	\bibfield  {author} {\bibinfo {author} {\bibfnamefont {N.}~\bibnamefont
			{Ishikawa}}, \bibinfo {author} {\bibfnamefont {M.}~\bibnamefont {Sugita}},
		\bibinfo {author} {\bibfnamefont {T.}~\bibnamefont {Ishikawa}}, \bibinfo
		{author} {\bibfnamefont {S.-Y.}\ \bibnamefont {Koshihara}}, \ and\ \bibinfo
		{author} {\bibfnamefont {Y.}~\bibnamefont {Kaizu}},\ }\href {\doibase
		10.1021/ja029629n} {\bibfield  {journal} {\bibinfo  {journal} {J. Am. Chem.
				Soc.}\ }\textbf {\bibinfo {volume} {125}},\ \bibinfo {pages} {8694} (\bibinfo
		{year} {2003})}\BibitemShut {NoStop}%
	\bibitem [{\citenamefont {Rudowicz}(1985)}]{Rudowicz1985}%
	\BibitemOpen
	\bibfield  {author} {\bibinfo {author} {\bibfnamefont {C.}~\bibnamefont
			{Rudowicz}},\ }\href {\doibase 10.1088/0022-3719/18/7/009} {\bibfield
		{journal} {\bibinfo  {journal} {J. Phys. C Solid State Phys.}\ }\textbf
		{\bibinfo {volume} {18}},\ \bibinfo {pages} {1415} (\bibinfo {year}
		{1985})}\BibitemShut {NoStop}%
	\bibitem [{\citenamefont {Rudowicz}\ and\ \citenamefont
		{Chung}(2004)}]{Rudowicz2004a}%
	\BibitemOpen
	\bibfield  {author} {\bibinfo {author} {\bibfnamefont {C.}~\bibnamefont
			{Rudowicz}}\ and\ \bibinfo {author} {\bibfnamefont {C.~Y.}\ \bibnamefont
			{Chung}},\ }\href {\doibase 10.1088/0953-8984/16/32/018} {\bibfield
		{journal} {\bibinfo  {journal} {J. Phys. Condens. Matter}\ }\textbf {\bibinfo
			{volume} {16}},\ \bibinfo {pages} {5825} (\bibinfo {year}
		{2004})}\BibitemShut {NoStop}%
	\bibitem [{\citenamefont {Dohm}\ and\ \citenamefont {Fulde}(1975)}]{Dohm1975}%
	\BibitemOpen
	\bibfield  {author} {\bibinfo {author} {\bibfnamefont {V.}~\bibnamefont
			{Dohm}}\ and\ \bibinfo {author} {\bibfnamefont {P.}~\bibnamefont {Fulde}},\
	}\href {\doibase 10.1007/PL00020764} {\bibfield  {journal} {\bibinfo
		{journal} {Zeitschrift f{\"{u}}r Phys. B Condens. Matter}\ }\textbf {\bibinfo
		{volume} {21}},\ \bibinfo {pages} {369} (\bibinfo {year} {1975})}\BibitemShut
{NoStop}%
\bibitem [{\citenamefont {Garanin}(2011)}]{Garanin2011}%
\BibitemOpen
\bibfield  {author} {\bibinfo {author} {\bibfnamefont {D.~A.}\ \bibnamefont
		{Garanin}},\ }in\ \href {\doibase 10.1002/9781118135242.ch4} {\emph {\bibinfo
		{booktitle} {Adv. Chem. Phys.}}},\ Vol.\ \bibinfo {volume} {147}\ (\bibinfo
{publisher} {Wiley},\ \bibinfo {year} {2011})\ Chap.~\bibinfo {chapter} {4},
pp.\ \bibinfo {pages} {213--277},\ \Eprint {http://arxiv.org/abs/0805.0391}
{arXiv:0805.0391} \BibitemShut {NoStop}%
\bibitem [{\citenamefont {Bonsall}\ and\ \citenamefont
	{Melcher}(1976)}]{Bonsall1976}%
\BibitemOpen
\bibfield  {author} {\bibinfo {author} {\bibfnamefont {L.}~\bibnamefont
		{Bonsall}}\ and\ \bibinfo {author} {\bibfnamefont {R.~L.}\ \bibnamefont
		{Melcher}},\ }\href {\doibase 10.1103/PhysRevB.14.1128} {\bibfield  {journal}
	{\bibinfo  {journal} {Phys. Rev. B}\ }\textbf {\bibinfo {volume} {14}},\
	\bibinfo {pages} {1128} (\bibinfo {year} {1976})}\BibitemShut {NoStop}%
\bibitem [{\citenamefont {Calero}\ \emph {et~al.}(2006)\citenamefont {Calero},
	\citenamefont {Chudnovsky},\ and\ \citenamefont {Garanin}}]{Calero2006a}%
\BibitemOpen
\bibfield  {author} {\bibinfo {author} {\bibfnamefont {C.}~\bibnamefont
		{Calero}}, \bibinfo {author} {\bibfnamefont {E.~M.}\ \bibnamefont
		{Chudnovsky}}, \ and\ \bibinfo {author} {\bibfnamefont {D.~A.}\ \bibnamefont
		{Garanin}},\ }\href {\doibase 10.1103/PhysRevB.74.094428} {\bibfield
	{journal} {\bibinfo  {journal} {Phys. Rev. B}\ }\textbf {\bibinfo {volume}
		{74}},\ \bibinfo {pages} {094428} (\bibinfo {year} {2006})},\ \Eprint
{http://arxiv.org/abs/0605668} {arXiv:0605668 [cond-mat]} \BibitemShut
{NoStop}%
\bibitem [{Note2()}]{Note2}%
\BibitemOpen
\bibinfo {note} {The derivation is given in Sec. IV.E2 of Ref. \protect
	\rev@citealpnum {Garanin2011}}\BibitemShut {NoStop}%
\bibitem [{\citenamefont {{Ho}}\ and\ \citenamefont
	{{Chibotaru}}(2017)}]{Ho2017b}%
\BibitemOpen
\bibfield  {author} {\bibinfo {author} {\bibfnamefont {L.~T.~A.}\
		\bibnamefont {{Ho}}}\ and\ \bibinfo {author} {\bibfnamefont {L.~F.}\
		\bibnamefont {{Chibotaru}}},\ }\href@noop {} {\bibfield  {journal} {\bibinfo
		{journal} {ArXiv e-prints}\ } (\bibinfo {year} {2017})},\ \Eprint
{http://arxiv.org/abs/1710.02053} {arXiv:1710.02053 [quant-ph]} \BibitemShut
{NoStop}%
\bibitem [{\citenamefont {Vijayaraghavan}\ and\ \citenamefont
	{Garg}(2009)}]{Vijayaraghavan2009}%
\BibitemOpen
\bibfield  {author} {\bibinfo {author} {\bibfnamefont {A.}~\bibnamefont
		{Vijayaraghavan}}\ and\ \bibinfo {author} {\bibfnamefont {A.}~\bibnamefont
		{Garg}},\ }\href {\doibase 10.1103/PhysRevB.79.104423} {\bibfield  {journal}
	{\bibinfo  {journal} {Phys. Rev. B}\ }\textbf {\bibinfo {volume} {79}},\
	\bibinfo {pages} {104423} (\bibinfo {year} {2009})}\BibitemShut {NoStop}%
\bibitem [{\citenamefont {Vijayaraghavan}(2011)}]{vijayaraghavan2011tunneling}%
\BibitemOpen
\bibfield  {author} {\bibinfo {author} {\bibfnamefont {A.}~\bibnamefont
		{Vijayaraghavan}},\ }\emph {\bibinfo {title} {Tunneling in Molecular
		Magnets}},\ \href@noop {} {Ph.D. thesis},\ \bibinfo  {school} {Northwestern
	University} (\bibinfo {year} {2011})\BibitemShut {NoStop}%
\bibitem [{\citenamefont {Prokof'ev}\ and\ \citenamefont
	{Stamp}(2000)}]{Prokof'ev2000}%
\BibitemOpen
\bibfield  {author} {\bibinfo {author} {\bibfnamefont {N.}~\bibnamefont
		{Prokof'ev}}\ and\ \bibinfo {author} {\bibfnamefont {P.}~\bibnamefont
		{Stamp}},\ }\href {http://arxiv.org/abs/cond-mat/0001080} {\bibfield
	{journal} {\bibinfo  {journal} {Reports Prog. Phys.}\ }\textbf {\bibinfo
		{volume} {63}},\ \bibinfo {pages} {669} (\bibinfo {year} {2000})}\BibitemShut
{NoStop}%
\bibitem [{\citenamefont {Garanin}\ \emph {et~al.}(2000)\citenamefont
	{Garanin}, \citenamefont {Chudnovsky},\ and\ \citenamefont
	{Schilling}}]{Garanin1999}%
\BibitemOpen
\bibfield  {author} {\bibinfo {author} {\bibfnamefont {D.~A.}\ \bibnamefont
		{Garanin}}, \bibinfo {author} {\bibfnamefont {E.~M.}\ \bibnamefont
		{Chudnovsky}}, \ and\ \bibinfo {author} {\bibfnamefont {R.}~\bibnamefont
		{Schilling}},\ }\href {\doibase 10.1103/PhysRevB.61.12204} {\bibfield
	{journal} {\bibinfo  {journal} {Phys. Rev. B}\ }\textbf {\bibinfo {volume}
		{61}},\ \bibinfo {pages} {12204} (\bibinfo {year} {2000})},\ \Eprint
{http://arxiv.org/abs/9911055} {arXiv:9911055 [cond-mat]} \BibitemShut
{NoStop}%
\bibitem [{\citenamefont {Sinitsyn}\ and\ \citenamefont
	{Prokof'ev}(2003)}]{Sinitsyn2003}%
\BibitemOpen
\bibfield  {author} {\bibinfo {author} {\bibfnamefont {N. A.}~\bibnamefont
		{Sinitsyn}}\ and\ \bibinfo {author} {\bibfnamefont {N.}~\bibnamefont
		{Prokof'ev}},\ }\href {\doibase 10.1103/PhysRevB.67.134403} {\bibfield
	{journal} {\bibinfo  {journal} {Phys. Rev. B}\ }\textbf {\bibinfo {volume}
		{67}},\ \bibinfo {pages} {134403} (\bibinfo {year} {2003})}\BibitemShut
{NoStop}%
\bibitem [{\citenamefont {Cohen-Tannoudji}\ \emph {et~al.}(1977)\citenamefont
	{Cohen-Tannoudji}, \citenamefont {Diu},\ and\ \citenamefont
	{Lalo{\"e}}}]{cohen1977}%
\BibitemOpen
\bibfield  {author} {\bibinfo {author} {\bibfnamefont {C.}~\bibnamefont
		{Cohen-Tannoudji}}, \bibinfo {author} {\bibfnamefont {B.}~\bibnamefont
		{Diu}}, \ and\ \bibinfo {author} {\bibfnamefont {F.}~\bibnamefont
		{Lalo{\"e}}},\ }\href {https://books.google.be/books?id=2KjvAAAAMAAJ} {\emph
	{\bibinfo {title} {Quantum mechanics}}}\ (\bibinfo  {publisher} {Wiley},\
\bibinfo {year} {1977})\ pp.\ \bibinfo {pages} {1--408}\BibitemShut {NoStop}%
\bibitem [{\citenamefont {Chibotaru}\ and\ \citenamefont
	{Ungur}(2012)}]{Chibotaru2012}%
\BibitemOpen
\bibfield  {author} {\bibinfo {author} {\bibfnamefont {L.~F.}\ \bibnamefont
		{Chibotaru}}\ and\ \bibinfo {author} {\bibfnamefont {L.}~\bibnamefont
		{Ungur}},\ }\href {\doibase 10.1063/1.4739763} {\bibfield  {journal}
	{\bibinfo  {journal} {J. Chem. Phys.}\ }\textbf {\bibinfo {volume} {137}},\
	\bibinfo {pages} {064112} (\bibinfo {year} {2012})}\BibitemShut {NoStop}%
\bibitem [{\citenamefont {Griffith}(1963)}]{Griffith1963a}%
\BibitemOpen
\bibfield  {author} {\bibinfo {author} {\bibfnamefont {J.~S.}\ \bibnamefont
		{Griffith}},\ }\href {\doibase 10.1103/PhysRev.132.316} {\bibfield  {journal}
	{\bibinfo  {journal} {Phys. Rev.}\ }\textbf {\bibinfo {volume} {132}},\
	\bibinfo {pages} {316} (\bibinfo {year} {1963})},\ \Eprint
{http://arxiv.org/abs/arXiv:1011.1669v3} {arXiv:arXiv:1011.1669v3}
\BibitemShut {NoStop}%
\bibitem [{Note3()}]{Note3}%
\BibitemOpen
\bibinfo {note} {Even if we keep all terms containing $\Delta _{i}$, the
	order of magnitude of these terms, which is $\protect \mathcal {O}\left
	(\Delta _{i}S^{2}\right )$, is still smaller than one of $M_{\protect \mathrm
		{sp4}}^{\alpha \beta }$, which is $\protect \mathcal {O}\left (\hbar \omega
	_{\protect \bm {\protect \mathrm {k}}\lambda _{\protect \bm {\protect \mathrm
				{k}}}}S^{2}\right )$ as we see later. Hence, by any means, these terms can be
	safely neglected.}\BibitemShut {Stop}%
\bibitem [{\citenamefont {G{\'{o}}mez-Coca}\ \emph {et~al.}(2014)\citenamefont
	{G{\'{o}}mez-Coca}, \citenamefont {Urtizberea}, \citenamefont {Cremades},
	\citenamefont {Alonso}, \citenamefont {Cam{\'{o}}n}, \citenamefont {Ruiz},\
	and\ \citenamefont {Luis}}]{Gomez-Coca2014}%
\BibitemOpen
\bibfield  {author} {\bibinfo {author} {\bibfnamefont {S.}~\bibnamefont
		{G{\'{o}}mez-Coca}}, \bibinfo {author} {\bibfnamefont {A.}~\bibnamefont
		{Urtizberea}}, \bibinfo {author} {\bibfnamefont {E.}~\bibnamefont
		{Cremades}}, \bibinfo {author} {\bibfnamefont {P.~J.}\ \bibnamefont
		{Alonso}}, \bibinfo {author} {\bibfnamefont {A.}~\bibnamefont {Cam{\'{o}}n}},
	\bibinfo {author} {\bibfnamefont {E.}~\bibnamefont {Ruiz}}, \ and\ \bibinfo
	{author} {\bibfnamefont {F.}~\bibnamefont {Luis}},\ }\href {\doibase
	10.1038/ncomms5300} {\bibfield  {journal} {\bibinfo  {journal} {Nat.
			Commun.}\ }\textbf {\bibinfo {volume} {5}},\ \bibinfo {pages} {4300}
	(\bibinfo {year} {2014})}\BibitemShut {NoStop}%
\bibitem [{Note4()}]{Note4}%
\BibitemOpen
\bibinfo {note} {In cases when optical phonons contribute significantly to
	the spin-lattice relaxation, the Raman relaxation rate for Kramers system may
	deviate from the $T^{9}$ dependence \cite {Shrivastava1983}. Therefore, in
	order to have an accurate assessment on the applicability of the rotational
	spin-phonon approximation, only those experiments which satisfies the
	following criteria are chosen: (1) Raman process shows a $T^{9}$ dependence,
	which thus excludes the role of optical phonon; (2) a small number of
	parameters in the expression of the total relaxation rate, which thus
	prevents over-parameterization; (3) data and fittings are of high-quality.
	Ref. {[}\protect \rev@citealpnum {Gomez-Coca2014}{]} satisfies all these
	criteria.}\BibitemShut {Stop}%
\bibitem [{\citenamefont {Griffith}(1971)}]{Griffith1971}%
\BibitemOpen
\bibfield  {author} {\bibinfo {author} {\bibfnamefont {J.~S.}\ \bibnamefont
		{Griffith}},\ }\href@noop {} {\emph {\bibinfo {title} {The theory of
			transition-metal ions}}}\ (\bibinfo  {publisher} {Cambridge University
	Press},\ \bibinfo {year} {1971})\BibitemShut {NoStop}%
\bibitem [{Note5()}]{Note5}%
\BibitemOpen
\bibinfo {note} {We consider multiplets $S$ or $J$ not far from isotropic
	case, when the Zeeman operator contains the first-rank contribution only
	\cite {Chibotaru2013}. This is by far the most common case.}\BibitemShut
{Stop}%
\bibitem [{\citenamefont {Prokof'ev}\ and\ \citenamefont
	{Stamp}(1998)}]{Prokof'ev1998}%
\BibitemOpen
\bibfield  {author} {\bibinfo {author} {\bibfnamefont {N.~V.}\ \bibnamefont
		{Prokof'ev}}\ and\ \bibinfo {author} {\bibfnamefont {P.~C.~E.}\ \bibnamefont
		{Stamp}},\ }\href {\doibase 10.1103/PhysRevLett.80.5794} {\bibfield
	{journal} {\bibinfo  {journal} {Phys. Rev. Lett.}\ }\textbf {\bibinfo
		{volume} {80}},\ \bibinfo {pages} {5794} (\bibinfo {year}
	{1998})}\BibitemShut {NoStop}%
\bibitem [{\citenamefont {Pedersen}\ \emph {et~al.}(2014)\citenamefont
	{Pedersen}, \citenamefont {Ungur}, \citenamefont {Sigrist}, \citenamefont
	{Sundt}, \citenamefont {Schau-Magnussen}, \citenamefont {Vieru},
	\citenamefont {Mutka}, \citenamefont {Rols}, \citenamefont {Weihe},
	\citenamefont {Waldmann}, \citenamefont {Chibotaru}, \citenamefont {Bendix},\
	and\ \citenamefont {Dreiser}}]{Pedersen2013}%
\BibitemOpen
\bibfield  {author} {\bibinfo {author} {\bibfnamefont {K.~S.}\ \bibnamefont
		{Pedersen}}, \bibinfo {author} {\bibfnamefont {L.}~\bibnamefont {Ungur}},
	\bibinfo {author} {\bibfnamefont {M.}~\bibnamefont {Sigrist}}, \bibinfo
	{author} {\bibfnamefont {A.}~\bibnamefont {Sundt}}, \bibinfo {author}
	{\bibfnamefont {M.}~\bibnamefont {Schau-Magnussen}}, \bibinfo {author}
	{\bibfnamefont {V.}~\bibnamefont {Vieru}}, \bibinfo {author} {\bibfnamefont
		{H.}~\bibnamefont {Mutka}}, \bibinfo {author} {\bibfnamefont
		{S.}~\bibnamefont {Rols}}, \bibinfo {author} {\bibfnamefont {H.}~\bibnamefont
		{Weihe}}, \bibinfo {author} {\bibfnamefont {O.}~\bibnamefont {Waldmann}},
	\bibinfo {author} {\bibfnamefont {L.~F.}\ \bibnamefont {Chibotaru}}, \bibinfo
	{author} {\bibfnamefont {J.}~\bibnamefont {Bendix}}, \ and\ \bibinfo {author}
	{\bibfnamefont {J.}~\bibnamefont {Dreiser}},\ }\href {\doibase
	10.1039/C3SC53044B} {\bibfield  {journal} {\bibinfo  {journal} {Chem. Sci.}\
	}\textbf {\bibinfo {volume} {5}},\ \bibinfo {pages} {1650} (\bibinfo {year}
	{2014})}\BibitemShut {NoStop}%
\bibitem [{\citenamefont {Prokof'ev}\ and\ \citenamefont
	{Stamp}(1996)}]{Prokofev1995}%
\BibitemOpen
\bibfield  {author} {\bibinfo {author} {\bibfnamefont {N.~V.}\ \bibnamefont
		{Prokof'ev}}\ and\ \bibinfo {author} {\bibfnamefont {P.~C.~E.}\ \bibnamefont
		{Stamp}},\ }\href {\doibase 10.1007/BF00754094} {\bibfield  {journal}
	{\bibinfo  {journal} {J. Low Temp. Phys.}\ }\textbf {\bibinfo {volume}
		{104}},\ \bibinfo {pages} {143} (\bibinfo {year} {1996})},\ \Eprint
{http://arxiv.org/abs/9511016} {arXiv:9511016 [cond-mat]} \BibitemShut
{NoStop}%
\bibitem [{\citenamefont {Fern{\'{a}}ndez}\ and\ \citenamefont
	{Alonso}(2003)}]{Fernandez2003}%
\BibitemOpen
\bibfield  {author} {\bibinfo {author} {\bibfnamefont {J. F.}~\bibnamefont
		{Fern{\'{a}}ndez}}\ and\ \bibinfo {author} {\bibfnamefont {J. J.}~\bibnamefont
		{Alonso}},\ }\href {\doibase 10.1103/PhysRevLett.91.047202} {\bibfield
	{journal} {\bibinfo  {journal} {Phys. Rev. Lett.}\ }\textbf {\bibinfo
		{volume} {91}},\ \bibinfo {pages} {047202} (\bibinfo {year}
	{2003})}\BibitemShut {NoStop}%
\bibitem [{\citenamefont {Cuccoli}\ \emph {et~al.}(1999)\citenamefont
	{Cuccoli}, \citenamefont {Fort}, \citenamefont {Rettori}, \citenamefont
	{Adam},\ and\ \citenamefont {Villain}}]{Cuccoli1999}%
\BibitemOpen
\bibfield  {author} {\bibinfo {author} {\bibfnamefont {A.}~\bibnamefont
		{Cuccoli}}, \bibinfo {author} {\bibfnamefont {A.}~\bibnamefont {Fort}},
	\bibinfo {author} {\bibfnamefont {A.}~\bibnamefont {Rettori}}, \bibinfo
	{author} {\bibfnamefont {E.}~\bibnamefont {Adam}}, \ and\ \bibinfo {author}
	{\bibfnamefont {J.}~\bibnamefont {Villain}},\ }\href {\doibase
	10.1007/s100510050974} {\bibfield  {journal} {\bibinfo  {journal} {Eur. Phys.
			J. B}\ }\textbf {\bibinfo {volume} {12}},\ \bibinfo {pages} {39} (\bibinfo
	{year} {1999})}\BibitemShut {NoStop}%
\end{thebibliography}

\end{document}